# Valleytronics in 2D Materials Roadmap

Kyle L. Seyler[1*], Giancarlo Soavi[2*], Bent Weber[3*], Sunit Das[4], Amit Agarwal[4], Ioannis Paradisanos[5,6], Mikhail M. Glazov[7], Oleg Dogadov[8,9], Francesco Gucci[8], Giulio Cerullo[8,10], Stefano Dal Conte[8], Shubhadeep Biswas[11], Jan Wilhelm[12], Igor Žutić[13], Konstantin S. Denisov[13], Tong Zhou[14], Huiyuan Zheng[15], Wang Yao[16,17], Hongyi Yu[18, 19], Ting Cao[20], Dacen Waters[21], Matthew Yankowitz[20,22], Guido Burkard[23], Artem Denisov[24], Thomas Ihn[24], Klaus Ensslin[24], Louis Gaudreau[25], Justin Boddison-Chouinard[26], Zlata Fedorova[27,28], Isabelle Staude[27,28], Kuan Eng Johnson Goh[29,30,31,32], Zhichao Zhou[33], Xiao Li[33].

[1]Wyant College of Optical Sciences, University of Arizona, Tucson, United States
[2]Institute of Solid State Physics, Friedrich Schiller University Jena, Jena, Germany
[3]School of Physical and Mathematical Sciences, Nanyang Technological University, Singapore, Singapore
[4]Department of Physics, Indian Institute of Technology, Kanpur 208016, India
[5]Department of Materials Science and Engineering, University of Crete, Heraklion, 70013, Greece
[6]Institute of Electronic Structure and Laser, Foundation for Research and Technology-Hellas, Heraklion, 70013, Greece
[7]orcid.org/0000-0003-4462-0749
[8]Dipartimento di Fisica, Politecnico di Milano, Piazza L. da Vinci 32, 20133 Milano, Italy
[9]Department of Physical Chemistry, Fritz Haber Institute of the Max Planck Society, 14195 Berlin, Germany
[10]IFN-CNR, Department of Physics, Politecnico di Milano, Piazza L. da Vinci 32, Milano 20133, Italy
[11]Department of Physics, Indian Institute of Science, Bangalore-560012, India
[12]Institute of Theoretical Physics and Regensburg Center for Ultrafast Nanoscopy (RUN), University of Regensburg, 93059 Regensburg, Germany
[13]Department of Physics, University at Buffalo, State University of New York, Buffalo, New York 14260, USA
[14]Eastern Institute for Advanced Study, Eastern Institute of Technology, Ningbo, Zhejiang 315200, China
[15]Department of Materials Science and Engineering, University of Washington, Seattle, Washington 98195, USA
[16]New Cornerstone Science Lab, Department of Physics, The University of Hong Kong, Hong Kong Special Administrative Region of China, People's Republic of China
[17]HK Institute of Quantum Science & Technology, The University of Hong Kong, Hong Kong Special Administrative Region of China, People's Republic of China
[18]Guangdong Provincial Key Laboratory of Quantum Metrology and Sensing & School of Physics and Astronomy, Sun Yat-Sen University (Zhuhai Campus), Zhuhai 519082, People's Republic of China
[19]State Key Laboratory of Optoelectronic Materials and Technologies, Sun Yat-Sen University (Guangzhou Campus), Guangzhou 510275, People's Republic of China
[20]Department of Materials Science and Engineering, University of Washington, Seattle, Washington, USA
[21]Department of Physics and Astronomy, University of Denver, Denver, CO, USA
[22]Department of Physics, University of Washington, Seattle, Washington, USA
[23]Department of Physics and IQST, University of Konstanz, D-78457 Konstanz, Germany
[24]Laboratory for Solid State Physics, ETH Zurich, Switzerland
[25]Quantum and Nanotechnologies Research Centre, National Research Council Canada, Ottawa, Canada
[26]Department of Physics, University of Ottawa, Ottawa, Canada

[27]Institute of Applied Physics, Friedrich Schiller University Jena, Albert-Einstein-Str. 15, 07745 Jena, Germany
[28]Abbe Center of Photonics, Friedrich Schiller University Jena, Albert-Einstein-Str. 6, 07745, Jena, Germany
[29]Institute of Materials Research & Engineering (IMRE), Agency for Science, Technology, and Research (A*STAR), 2 Fusionopolis Way, #08-03 Innovis, Singapore 138634, Republic of Singapore
[30]Centre for Quantum Technologies, National University of Singapore, 3 Science Drive 2, Singapore 117543, Singapore
[31]Department of Physics, National University of Singapore, Republic of Singapore 2 Science Drive 3, Singapore 117551, Singapore
[32]Division of Physics and Applied Physics, School of Physical and Mathematical Sciences, Nanyang Technological University 21 Nanyang Link, 637371, Singapore
[33]Center for Quantum Transport and Thermal Energy Science, Institute of Physics Frontiers and Interdisciplinary Sciences, School of Physics and Technology, Nanjing Normal University, Nanjing 210023, China

E-mails: klseyler@arizona.edu, giancarlo.soavi@uni-jena.de, b.weber@ntu.edu.sg

**Abstract**

Valleytronics exploits non-equivalent energy extrema in the electronic band structure of crystalline solids—the valley degree of freedom—to encode, manipulate, and read out information. The advent of 2D materials, first graphene and then transition-metal dichalcogenides, made valley control practical through optical, electrical, and magnetic routes. This foundation has enabled remarkable progress in recent years spanning established frontiers, such as valley exciton physics and valley Hall effects, as well as emerging directions including lightwave valleytronics, nanophotonic integration, flat-band valleytronics, and spin-valley qubits. In parallel, there are sustained efforts to scale up valleytronic materials and to predict new valleytronic platforms. This Roadmap brings together perspectives from leading experts to chart the key opportunities and challenges at the forefront of 2D material valleytronics. Each section captures a snapshot of progress in a key research area, identifies critical open challenges, and outlines pathways toward future valleytronics breakthroughs.

## Table of Contents

# 1. Introduction

**Kyle L. Seyler[1], Giancarlo Soavi[2] and Bent Weber[3]**

[1] Wyant College of Optical Sciences, University of Arizona, Tucson, United States
[2] Institute of Solid State Physics, Friedrich Schiller University Jena, Jena, Germany
[3] School of Physical and Mathematical Sciences, Nanyang Technological University, Singapore, Singapore

E-mail: klseyler@arizona.edu, giancarlo.soavi@uni-jena.de, b.weber@ntu.edu.sg

Quasiparticles in a crystal can possess internal quantum degrees of freedom, such as spin, that enable encoding and processing information. Interest in exploiting such degrees of freedom naturally extended to the valley pseudospin, which labels degenerate but non-equivalent energy extrema in momentum space. The ideas of "valleytronics" were seeded in studies of silicon in the 1970s [1] and AlAs quantum wells in early 2000s [2], but it did not develop as a distinct field until the arrival of 2D materials, particularly graphene [3] and transition metal dichalcogenides (TMDs). An important theoretical breakthrough came in 2007 with the recognition that inversion symmetry breaking in 2D hexagonal lattices endows the ±K valleys (at the corners of the hexagonal Brillouin zone) with contrasting Berry curvature and intrinsic orbital magnetic moment, enabling the valley Hall effect and magnetic field coupling [4]. The field accelerated dramatically with the study of monolayer group-6 TMD semiconductors ($MoS_2$, $WS_2$, $MoSe_2$, $WSe_2$, $MoTe_2$) in the early 2010s [5]. These hexagonal-lattice materials not only possess intrinsic broken inversion symmetry, but also exhibit strong spin-orbit coupling that locks spin and valley degrees of freedom [6], direct bandgaps at their ±K points [7, 8], and valley-dependent optical selection rules for circularly polarized light [6, 9]. By the mid-2010s, foundational valleytronic functionalities were demonstrated, including optical initialization (absorption) and readout (photoluminescence) of valley polarization and coherence [10-13], valley Hall transport in both TMDs and graphene heterostructures [14-18], and valley manipulation via magnetic fields [19-22] and Floquet engineering (optical Stark and Bloch-Siegert effects) [23-25]. The scope of valley physics also began to expand beyond intrinsic monolayers and bilayers to other systems such as TMD heterostructures [26] and individual electron (or hole) spins confined to quantum dot potentials [27]. These achievements, comprehensively reviewed elsewhere [5, 28-32], established TMDs and graphene as practical platforms for valleytronics. The 2D material valleytronics landscape has since evolved along a few distinct but interconnected trajectories.

First, there continued to be important valleytronics-specific advances that enabled new manipulation pathways and deepened our understanding of valley phenomena. Established areas such as valley Hall effects (section 2) and exciton valley physics (section 3) have matured significantly. For example, valley Hall transport has expanded into nonlinear and quantum regimes, while valley exciton studies have achieved deeper understanding of depolarization mechanisms while expanding control through dielectric engineering, strain, moiré superlattices, and cavity coupling.

Ultrafast valleytronics (section 4) has worked to understand and address the challenge of rapid valley depolarization in complementary ways: extending valley lifetimes through heterostructures and strain engineering, while also developing protocols to manipulate valley states with phase-locked few-cycle laser pulses before the polarization decays. This latter paradigm, lightwave valleytronics, has demonstrated all-optical valley initialization, switching, and readout even in centrosymmetric crystals and on few femtosecond timescales, using symmetry-engineered strong light fields such as trefoil waveforms (section 5). Furthermore, nonlinear optical spectroscopy has emerged as an ideal ultrafast and non-invasive probe of valley properties, with second-harmonic generation and third-harmonic Faraday/Kerr rotation being able to detect broken time-reversal symmetry in both centrosymmetric and non-centrosymmetric atomically thin crystals (section 6).

While these all-optical valleytronics advances were unfolding, improvements in sample quality (particularly through hexagonal boron nitride encapsulation) and the emergence of multilayer devices transformed these materials into premier platforms for studying broader condensed matter phenomena. Interactions across van der Waals interfaces induce proximity effects that profoundly alter valleytronic properties, from lifting the valley degeneracy to stabilizing robust quantum spin-valley Hall kink states (section 7). The flat electronic bands of moiré superlattices as well as rhombohedral graphene further unveiled strong correlation phenomena such as superconductivity, magnetism, and quantum anomalous Hall states. While the valley structure of course remains central to understanding such materials, the focus of many studies shifted to emergent many-body phenomena rather than valley control per se, with valley pseudospin becoming one of several important coupled degrees of freedom such as layer pseudospin and moiré sublattice sites (sections 8-9).

Moving towards technological applications, the strong coupling of spin and valley degrees of freedom in individual electron spin qubits in graphene and TMD quantum dots can lead to a dramatic enhancement of qubit lifetimes (sections 10-12), which is paramount for devices. Finally, interfacing 2D materials with nanophotonic structures serves to initialize, manipulate, route, and read out valley information (section 13).

While these developments have significantly expanded the possibilities for valley-based classical and quantum information processing, most functionalities are still only demonstrated on small devices with mechanically exfoliated flakes. Therefore, parallel efforts to realize high-quality large area materials and address other scaling issues are essential (section 14). Moreover, experimental research has been largely limited to TMDs and graphene, leaving a wide variety of theoretically predicted 2D valleytronic materials awaiting experimental realization (section 15).

This Roadmap brings together leading experts to chart paths forward across this diverse landscape. The contributions span fundamental valley physics, advanced manipulation techniques, quantum information applications, emerging directions, and practical considerations. Each section assesses current progress and identifies key challenges and opportunities. Rather than providing a comprehensive review, this Roadmap is a snapshot of the state of the field and is resolutely forward-looking. It remains to be seen whether valleytronics can become a viable pathway to classical and quantum information processing and storage as compared to other established approaches. Regardless, it is clear that the scientific insights gained from 2D valleytronic materials will resonate across multiple fields for years to come.

**Acknowledgements**

KLS acknowledges support by the U.S. Department of Energy, Office of Science, Basic Energy Sciences under Award #DE-SC0025619. BW acknowledges support from the Singapore Ministry of Education (MOE) Academic Research Fund Tier 3 grant (MOE-MOET32023-0003) “Geometrical Quantum Materials” and the US Air Force Office of Scientific Research under award number FA2386-24-1-4064. GS acknowledges funding from the Deutsche Forschungsgemeinschaft (DFG, German Research Foundation) – Project-ID 398816777 – SFB 1375.

## References

[1] Ando T, Fowler A B and Stern F 1982 Electronic properties of two-dimensional systems *Rev. Mod. Phys.* **54** 437–672
[2] Shkolnikov Y, De Poortere E, Tutuc E and Shayegan M 2002 Valley splitting of AlAs two-dimensional electrons in a perpendicular magnetic field *Phys. Rev. Lett.* **89** 226805
[3] Rycerz A, Tworzydło J and Beenakker C W J 2007 Valley filter and valley valve in graphene *Nat. Phys.* **3** 172–175
[4] Xiao D, Yao W and Niu Q 2007 Valley-contrasting physics in graphene: magnetic moment and topological transport *Phys. Rev. Lett.* **99** 236809
[5] Xu X, Yao W, Xiao D and Heinz T F 2014 Spin and pseudospins in layered transition metal dichalcogenides *Nat. Phys.* **10** 343–350
[6] Xiao D, Liu G-B, Feng W, Xu X and Yao W 2012 Coupled spin and valley physics in monolayers of $MoS_2$ and other group-VI dichalcogenides *Phys. Rev. Lett.* **108** 196802
[7] Splendiani A et al. 2010 Emerging photoluminescence in monolayer $MoS_2$ *Nano Lett.* **10** 1271–1275
[8] Mak K F, Lee C, Hone J, Shan J and Heinz T F 2010 Atomically thin $MoS_2$: a new direct-gap semiconductor *Phys. Rev. Lett.* **105** 136805
[9] Yao W, Xiao D and Niu Q 2008 Valley-dependent optoelectronics from inversion symmetry breaking *Phys. Rev. B* **77** 235406
[10] Mak K F, He K, Shan J and Heinz T F 2012 Control of valley polarization in monolayer $MoS_2$ by optical helicity *Nat. Nanotechnol.* **7** 494–498
[11] Zeng H, Dai J, Yao W, Xiao D and Cui X 2012 Valley polarization in $MoS_2$ monolayers by optical pumping *Nat. Nanotechnol.* **7** 490–493
[12] Cao T, Wang G, Han W, Ye H, Zhu C, Shi J, Niu Q, Tan P, Wang E, Liu B and Feng J 2012 Valley-selective circular dichroism of monolayer molybdenum disulphide *Nat. Commun.* **3** 887
[13] Jones A M, Yu H, Ghimire N J, Wu S, Aivazian G, Ross J S, Zhao B, Yan J, Mandrus D G, Xiao D, Yao W and Xu X 2013 Optical generation of excitonic valley coherence in monolayer $WSe_2$ *Nat. Nanotechnol.* **8** 634–638
[14] Mak K F, McGill K L, Park J and McEuen P L 2014 The valley Hall effect in $MoS_2$ transistors *Science* **344** 1489–1492
[15] Gorbachev R V, Song J C W, Yu G L, Kretinin A V, Withers F, Cao Y, Mishchenko A, Grigorieva I V, Novoselov K S, Levitov L S and Geim A K 2014 Detecting topological currents in graphene superlattices *Science* **346** 448–451
[16] Sui M, Chen G, Ma L, Shan W-Y, Tian D, Watanabe K, Taniguchi T, Jin X, Yao W, Xiao D and Zhang Y 2015 Gate-tunable topological valley transport in bilayer graphene *Nat. Phys.* **11** 1027–1031
[17] Shimazaki Y, Yamamoto M, Borzenets I V, Watanabe K, Taniguchi T and Tarucha S 2015 Generation and detection of pure valley current by electrically induced Berry curvature in bilayer graphene *Nat. Phys.* **11** 1032–1036
[18] Lee J, Mak K F and Shan J 2016 Electrical control of the valley Hall effect in bilayer $MoS_2$ transistors *Nat. Nanotechnol.* **11** 421–425
[19] Li Y, Ludwig J, Low T, Chernikov A, Cui X, Arefe G, Kim Y D, van der Zande A M, Rigosi A, Hill H M, Kim S H, Hone J, Li Z, Smirnov D and Heinz T F 2014 Valley splitting and polarization by the Zeeman effect in monolayer $MoSe_2$ *Phys. Rev. Lett.* **113** 266804
[20] Srivastava A, Sidler M, Allain A V, Lembke D S, Kis A and Imamoglu A 2015 Valley Zeeman effect in elementary optical excitations of monolayer $WSe_2$ *Nat. Phys.* **11** 141–147
[21] Aivazian G, Gong Z, Jones A M, Chu R-L, Yan J, Mandrus D G, Zhang C, Cobden D, Yao W and Xu X 2015 Magnetic control of valley pseudospin in monolayer $WSe_2$ *Nat. Phys.* **11** 148–152
[22] MacNeill D, Heikes C, Mak K F, Anderson Z, Kormanyos A, Zolyomi V, Park J and Ralph D C 2015 Breaking of valley degeneracy by magnetic field in monolayer $MoSe_2$ *Phys. Rev. Lett.* **114** 037401
[23] Kim J, Hong X, Jin C, Shi S-F, Chang C-Y S, Chiu M-H, Li L-J and Wang F 2014 Ultrafast generation of pseudo-magnetic field for valley excitons in $WSe_2$ monolayers *Science* **346** 1205–1208
[24] Sie E J, McIver J W, Lee Y-H, Fu L, Kong J and Gedik N 2015 Valley-selective optical Stark effect in monolayer $WS_2$ *Nat. Mater.* **14** 290–294
[25] Sie E J, Lui C H, Lee Y-H, Fu L, Kong J and Gedik N 2017 Large, valley-exclusive Bloch-Siegert shift in monolayer $WS_2$ *Science* **355** 1066–1069
[26] Rivera P, Seyler K L, Yu H, Schaibley J R, Yan J, Mandrus D G, Yao W and Xu X 2016 Valley-polarized exciton dynamics in a 2D semiconductor heterostructure *Science* **351** 688–691
[27] Trauzettel B, Bulaev D V, Loss D and Burkard G 2007 Spin qubits in graphene quantum dots *Nat. Phys.* **3** 192–196
[28] Schaibley J R, Yu H, Clark G, Rivera P, Ross J S, Seyler K L, Yao W and Xu X 2016 Valleytronics in 2D materials *Nat. Rev. Mater.* **1** 16055
[29] Vitale S A, Nezich D, Varghese J O, Kim P, Gedik N, Jarillo-Herrero P, Xiao D and Rothschild M 2018 Valleytronics: Opportunities, challenges, and paths forward *Small* **14** 1801483

[30] Mak K F, Xiao D and Shan J 2018 Light–valley interactions in 2D semiconductors *Nat. Photonics* **12** 451–460
[31] Rivera P, Yu H, Seyler K L, Wilson N P, Yao W and Xu X 2018 Interlayer valley excitons in heterobilayers of transition metal dichalcogenides *Nat. Nanotechnol.* **13** 1004–1015
[32] Zhao S, Li X, Dong B, Wang H, Wang H, Zhang Y, Han Z and Zhang H 2021 Valley manipulation in monolayer transition metal dichalcogenides and their hybrid systems: status and challenges *Rep. Prog. Phys.* **84** 026401

# 2. Valley Hall effects

**Sunit Das[1] and Amit Agarwal[1]**

[1]Department of Physics, Indian Institute of Technology, Kanpur 208016, India

E-mail: amitag@iitk.ac.in

### Fundamentals of Valley Physics

In crystalline solids, the periodic lattice potential shapes the electronic states into bands that often host multiple inequivalent extrema in momentum space. These extrema, known as valleys, provide an internal degree of freedom that distinguishes carriers by their crystal momentum rather than their charge or spin. When valleys are separated by a large momentum transfer in the Brillouin zone, intervalley scattering becomes inefficient [1]. This allows valley populations to persist over timescales relevant for transport and optical measurements. This elevates the valley index to a functional quantum degree of freedom.

Hexagonal lattices such as graphene allotropes and transition-metal dichalcogenides (TMDs) offer the clearest realization of valley physics. Their Brillouin zones contain two inequivalent corners, denoted by $\boldsymbol{\xi}$ and $\bar{\boldsymbol{\xi}}$ in Fig. 1(a), which are related by time-reversal symmetry. This symmetry allows for valley-contrasting band geometric properties as reflected in the Berry curvature $[\boldsymbol{\Omega}^{\xi}(\mathbf{k}) = -\boldsymbol{\Omega}^{\bar{\xi}}(\mathbf{k})]$, the momentum-space analogue of a magnetic field [2]. The most prominent transport manifestation of the opposite Berry curvature in the two valleys is the valley Hall effect (VHE) [2]. In the VHE, carriers at inequivalent valleys have opposite anomalous transverse velocities $\propto \mathrm{E} \times \Omega$ under an in-plane electric field. This results in a net valley current without a net charge Hall response [see Fig. 1]. Because the carriers in each valley possess contrasting orbital magnetic moments, analogous to the intrinsic Bohr magneton associated with electron spin, valley transport can also be viewed as the flow of carriers with opposite orbital magnetic moments, detectable through magneto-optical probes [3,4].

A central requirement for studying valley-dependent transport is the ability to initialize carriers in specific valleys. In atomically thin semiconductors, circularly polarized light couples selectively to one valley through helicity-dependent optical selection rules, producing a controlled valley imbalance [3]. This arises because the Bloch states at the two valleys carry opposite orbital angular momentum; therefore, right- and left-handed circularly polarized light couple to them through angular momentum-selective optical transitions. Furthermore, electrostatic and mechanical tuning via gating, strain, or layer asymmetry, provides additional control by reshaping the Berry curvature landscape and redistributing carriers between valleys.

Together, these mechanisms establish the foundation for generating, manipulating, and probing valley Hall transport. The next section surveys the present theoretical and experimental landscape of the VHE, encompassing advances across multiple materials as well as emerging variants of VHE.

**Status**

Since its theoretical proposal [2], the VHE has grown into an experimentally verified, tunable platform across TMDs, graphene heterostructures, and moiré systems. We summarize its key milestones in Fig. 1(d). The first experimental access to valley physics was achieved in monolayer $MoS_2$, where valley-specific optical selection rules enable helicity-resolved excitation and detection of valley polarization [3].

A key advance was the recognition that valley-contrasting transport does not require optical pumping. Gate-tunable VHE in bilayer $MoS_2$ demonstrated that inversion symmetry [5], and therefore Berry curvature, can be modified electrostatically, enabling all-electrical manipulation of valley currents. Spatially resolved Kerr microscopy directly imaged the resulting transverse valley-specific magnetic moment accumulation, validating the microscopic mechanism and highlighting the feasibility of device-compatible, all-electrical control of the valley degree of freedom.

Graphene-based systems provide a highly tunable platform for valley transport. In bilayer graphene, a perpendicular displacement field opens a controllable bandgap and induces valley-contrasting Berry curvature, yielding a gate-switchable VHE observable in nonlocal geometries [6], see Fig. 1(c). Moiré engineering in graphene/hBN and twisted multilayer graphene further amplifies these geometric effects due to multiple flat minibands with concentrated Berry curvature [7,8]. In ballistic graphene/hBN superlattices, these features manifest as giant nonlocal resistances capturing quantum valley Hall transport, and aligned graphene/hBN and electrically gapped bilayer graphene have shown similar signatures [9]. Complementing these graphene platforms, room-temperature valley Hall transport with micron-scale diffusion lengths has also been observed in atomically thin $MoS_2$ [10].

Beyond linear response, several new mechanisms have expanded the symmetry regimes in which VHE can occur. The nonlinear, second-order VHE, arising from the Berry connection polarizability, a field-induced geometric correction, allows valley currents even in inversion-symmetric crystals where the linear effect is forbidden [11]. Graphene/hBN moiré heterostructures indeed exhibit second-harmonic nonlocal signals consistent with this mechanism [12]. In parallel, altermagnets, which possess compensated magnetic order but break the combined spin–sublattice symmetry, have been predicted to host a `crystal valley Hall effect,' where crystalline symmetries generate valley-contrasting Berry curvature and the valley Hall response [13]. Recently, Euler curvature-based valley responses have been proposed, where real-valued Bloch wave functions support valley-selective transport even when the Berry curvature vanishes [14].

Beyond the single-particle picture, electron–electron interactions can generate additional topological structure in valley transport. In graphene-based moiré materials, interactions are predicted to stabilize quantum valley Hall (QVH) phases featuring quantized transverse response and valley-polarized edge modes [15]. Valley-dependent transverse transport is not limited to electrons: excitons in TMDs and valley-polarized phononic and acoustic modes in engineered metamaterials also exhibit Hall-like behavior [4]. These responses arise from distinct microscopic mechanisms and are discussed later in this article.

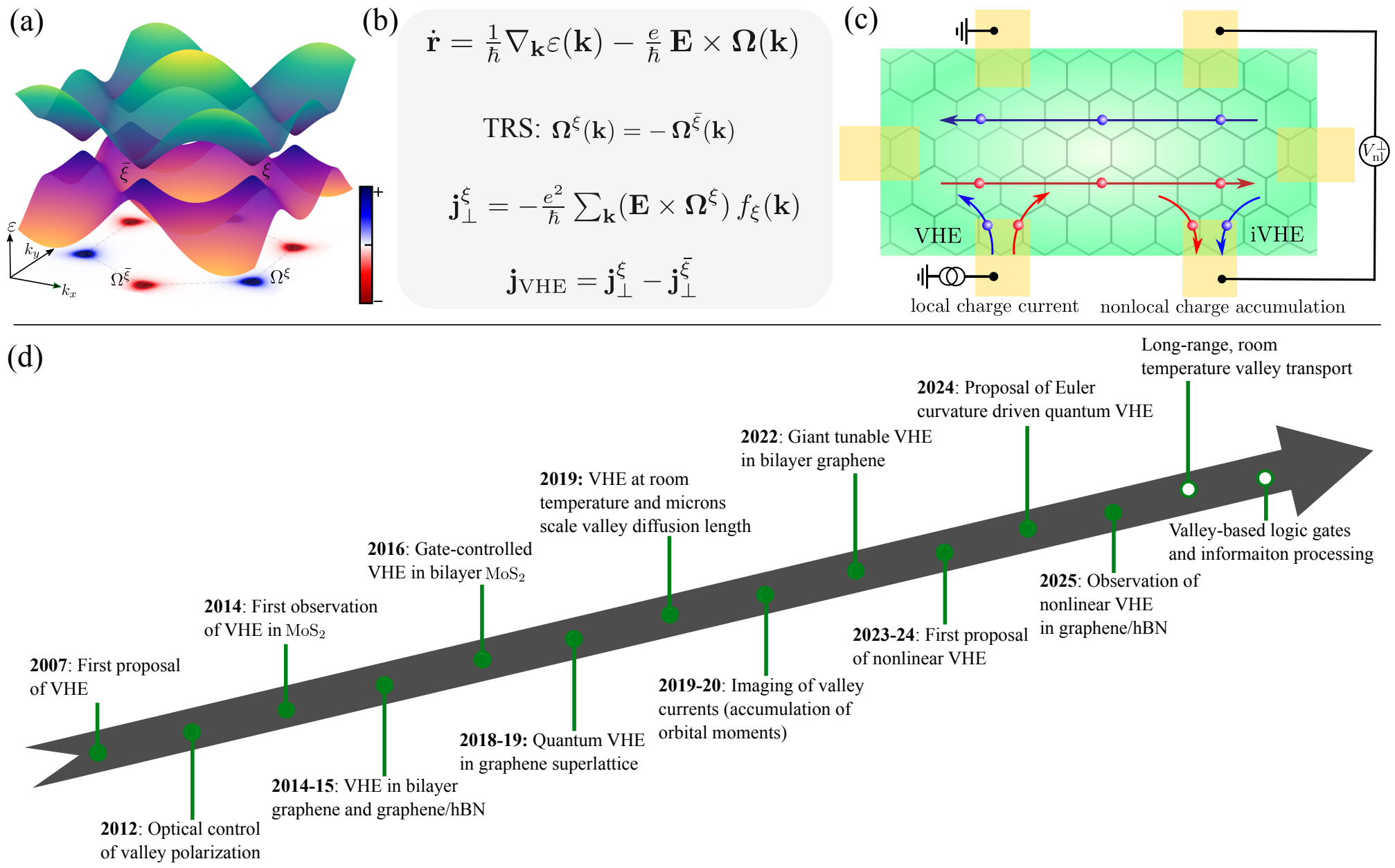

**Figure 1. Geometric origin, detection, and major milestones of the valley Hall effect.** (a) Low-energy electronic bands of an inversion-asymmetric hexagonal crystal (e.g., gapped graphene) exhibit two inequivalent valleys, $\xi$ and $\bar{\xi}$. The Berry curvature of the lower band is projected where the Berry curvatures $\mathbf{\Omega}^{\xi}$ and $\mathbf{\Omega}^{\bar{\xi}}$ are strongly localized in the valleys and have opposite sign. (b) In semiclassical dynamics, an electric field generates an anomalous velocity $(e/\hbar)\boldsymbol{E}\times\mathbf{\Omega}$. In a time-reversal symmetric (TRS) system, the opposite Berry curvatures deflect carriers from the two valleys in opposite directions. As a consequence, their charge Hall contributions cancel, but the difference produces a pure valley current $\mathbf{j}_{\mathrm{VHE}}$. (c) In a nonlocal measurement setup, the valley imbalance accumulates at opposite device edges and diffuses to distant contacts, producing an inverse VHE (iVHE) signal. The generated nonlocal transverse voltage $V_{\mathrm{nl}}^{\perp}$ serves as the characteristic electrical signature of the valley Hall effect. (d) Chronological milestones in the valley Hall effect. Solid markers denote established results, while the last two hollow markers denote outlook, such as long-range room-temperature valley transport and valley-based information technologies.

## Current and future challenges

The valley Hall effect faces several fundamental and practical challenges that must be addressed for reliable valleytronic applications. A primary limitation is the short valley lifetime and diffusion length. In most semiconductors, intervalley scattering arising from phonons, disorder, etc., rapidly diminishes the valley character of the carriers, keeping diffusion lengths below the micron scale [4]. Achieving long-range, room-temperature, and purely electronic valley transport remains a challenge.

Another problem is the unambiguous detection of valley Hall currents. Nonlocal resistance measurements, the most widely used all-electrical experimental signature, can be contaminated by parasitic effects including current spreading, trivial edge conduction, and anisotropic resistivity [16]. Scanning probe studies have shown that signals previously attributed to the VHE can potentially arise from trivial edge states. This underscores the need for standardized measurement geometries, local probes capable of analyzing valley-specific orbital magnetization, and quantitative frameworks to disentangle the contributions of valley effects from those of spin and thermoelectric responses.

A related challenge is the separation of intrinsic and extrinsic mechanisms. The intrinsic VHE originates from Berry curvature, but extrinsic mechanisms, such as skew scattering, side-jump effects, can obscure or even dominate the measured signal. Identifying systems in which the intrinsic contribution dominates the VHE is crucial for exploring valley topological effects and using them to ensure predictable device behavior.

From a materials perspective, scalable and reproducible control of inversion symmetry breaking remains difficult. In TMDs, inversion asymmetry is intrinsic but produces short valley lifetimes. In bilayer graphene, inversion breaking requires precise electrostatic control or moiré alignment. In twisted or stacked systems, small deviations in twist angle or strain can drastically alter the Berry curvature. Achieving reproducible, wafer-scale fabrication of valleytronic materials with consistent band geometric responses is a key challenge for the scalable integration of devices.

Looking forward, nonlinear and topological extensions of the VHE introduce both opportunities and complications. Nonlinear VHE expands valley transport to centrosymmetric crystals but often requires strong driving fields or engineered asymmetry [11]. Interaction-driven and Euler curvature-based valley Hall phases may offer routes to quantized or disorder-robust transport. However, their experimental realization remains in its early stages and requires improved theoretical understanding, material quality, and sensitive detection methods.

**Advances in science and technology to meet challenges**

Progress toward robust and scalable VHE has advanced across various materials platforms, symmetry engineering strategies, and detection techniques. These developments have significantly expanded the accessible regimes of valley transport, even though important challenges remain in translating proof-of-concept demonstrations into device-level performance [4,8].

Gate voltage control of inversion symmetry in bilayer graphene, bilayer TMDs, and their moiré heterostructures, remains one of the most efficient routes for tuning Berry curvature, offering quantitative control over the VHE. Materials such as Janus TMD monolayers, ferrovalley compounds, and altermagnetic systems offer built-in valley asymmetry that is often tunable by electrostatic or magnetic means, providing promising paths toward non-volatile valley configurations [17]. These symmetry-engineered materials directly address the challenge of valley initialization without the need for optical pumping.

Steady progress has been made in extending the lifetime of valley polarized carriers, a key requirement for observing VHE over device-relevant length scales. On the optical front, twisted TMD heterolayers host interlayer excitons with greatly enhanced valley lifetimes which facilitate improved initialization and readout of valley imbalance [18]. Hybrid photonic approaches, such as cavity or plasmonic coupling, further suppress depolarization [19]. These advances provide concrete strategies for increasing the VHE-relevant length scale, although integrating them with purely electronic devices remains an open challenge. Nonetheless, in electronic systems, cleaner interfaces, reduced disorder, and improved dielectric environments have enabled micron-scale valley diffusion lengths at room temperature, directly strengthening nonlocal VHE signals [10].

Real-space probes capable of imaging valley-contrasting orbital magnetization, such as Kerr rotation and nano-SQUID microscopy, have played a crucial role in clarifying ambiguous nonlocal signals, but they are not scalable for device integration. A growing set of device-level diagnostics now provides physically grounded criteria for identifying genuine valley Hall transport. In diffusive valley transport, the nonlocal voltage shows a characteristic spatial decay and a distinct scaling with the longitudinal resistivity [2,11]. These mechanisms enable reliable identification of valley Hall transport using purely electrical measurements [7], without the need for local imaging techniques.

Recent theoretical and experimental progress has broadened the materials landscape capable of producing valley currents. The giant second-order VHE observed in graphene/hBN moiré structures demonstrates that higher-order band geometry can generate valley transport even in inversion-symmetric systems [12]. Interaction-driven and Euler curvature-based valley phases further reveal that symmetry and topology alone can drive valley-selective responses, thereby extending VHE beyond the conventional Berry curvature paradigm.

Finally, advances in large-area 2D material growth, deterministic stacking, and twist control are enabling wafer-scale structures with reproducible symmetry tuning. Improved contacts, encapsulation, and strain engineering have reduced device variability, while low-power gating and integrated photonics support hybrid optoelectronic-valley functionality [4,18]. Demonstrations of room-temperature valley transport in $MoS_2$ [5,10] and promising high-temperature signals in bilayer graphene [20] suggest that technologically relevant VHE is increasingly within reach as materials and device platforms continue to improve.

### Concluding remarks

The valley Hall effect has emerged as a quantitatively understood and experimentally accessible manifestation of valley degrees of freedom in two-dimensional materials. Its continued development now hinges on overcoming several central obstacles: extending valley lifetimes, establishing unambiguous detection protocols, and achieving reproducible operation at technologically relevant temperatures and length scales. As valley physics intersects with spintronics, orbitronics, and quantum geometry, the VHE may evolve from a diagnostic tool into a design principle for multifunctional devices. Realizing this potential will require coordinated progress in materials synthesis, theoretical modelling, device engineering, and system-level integration.

### Acknowledgements

SD acknowledges the Ministry of Education, Government of India, for funding support through the Prime Minister's Research Fellowship. AA acknowledges funding from the Core Research Grant by Anusandhan National Research Foundation (ANRF, Sanction No. CRG/2023/007003), Department of Science and Technology, India.

### References

[1] Morozov S V, Novoselov K S, Katsnelson M I, Schedin F, Ponomarenko L A, Jiang D, and Geim A K, 2006 Strong suppression of weak localization in graphene *Phys. Rev. Lett.* **97**, 016801.

[2] Xiao D, Yao W and Niu Q 2007 Valley-contrasting physics in graphene: Magnetic moment and topological transport *Phys. Rev. Lett.* **99** 236809 .

[3] Mak K F, He K, Shan J and Heinz T F 2012 Control of valley polarization in monolayer $MoS_2$ by optical helicity *Nat. Nanotechnol.* **7** 494–498.

[4] Vitale S A, Nezich D, Varghese J O, Kim P, Gedik N, Jarillo-Herrero P, Xiao D and Rothschild M 2018 Valleytronics: Opportunities, challenges, and paths forward *Small* **14** 1801483.

[5] Hung T Y T, Camsari K Y, Zhang S, Upadhyaya P and Chen Z 2019 Direct observation of valley-coupled topological current in $MoS_2$ *Sci. Adv.* **5** eaau6478.

[6] Sui M, Chen G, Ma L, Shan W-Y, Tian D, Watanabe K, Taniguchi T, Jin X, Yao W, Xiao D and Zhang Y 2015 Gate-tunable topological valley transport in bilayer graphene *Nat. Phys.* **11** 1027–1031.

[7] Gorbachev R V, Song J C W, Yu G L, Kretinin A V, Withers F, Cao Y, Mishchenko A, Grigorieva I V, Novoselov K S, Levitov L S and Geim A K 2014 Detecting topological currents in graphene superlattices *Science* **346** 448–451.

[8] Adak P C, Sinha S, Agarwal A and Deshmukh M M 2024 Tunable moiré materials for probing Berry physics and topology *Nat. Rev. Mater.* **9** 481–498.

[9] Komatsu K, Morita Y, Watanabe E, Tsuya D, Watanabe K, Taniguchi T and Moriyama S 2018 Observation of the quantum valley Hall state in ballistic graphene superlattices *Sci. Adv.* **4** eaaq0194.

[10] Wu Z, Zhou B T, Cai X, Cheung P, Liu G-B, Huang M, Lin J, Han T, An L, Wang Y, Xu S, Long G, Cheng C, Law K T, Zhang F and Wang N 2019 Intrinsic valley Hall transport in atomically thin $MoS_2$ *Nat. Commun.* **10** 611.
[11] Das K, Ghorai K, Culcer D and Agarwal A 2024 Nonlinear valley Hall effect *Phys. Rev. Lett.* **132** 096302.
[12] He P, Zhang M, Cao J, Li J, Liu H, Zhai J, Wang R, Xiao C, Yang S A and Shen J 2025 Observation of giant nonlinear valley Hall effect arXiv:2503.03147.
[13] Tan C-Y, Gao Z-F, Yang H-C, Liu Z-X, Liu K, Guo P-J and Lu Z-Y 2025 Crystal valley Hall effect Phys. Rev. B **111** 094411.
[14] Ghadimi R, Mondal C, Kim S and Yang B-J 2024 Quantum valley Hall effect without Berry curvature Phys. Rev. Lett. **133** 196603.
[15] Marino E C, Nascimento L O, Alves V S and Morais Smith C 2015 Interaction induced quantum valley Hall effect in graphene Phys. Rev. X **5** 011040.
[16] Aharon-Steinberg A, Marguerite A, Perello D J, Bagani K, Holder T, Myasoedov Y, Levitov L S, Geim A K and Zeldov E 2021 Long-range nontopological edge currents in charge-neutral graphene Nature **593** 528–534.
[17] Luo C, Huang Z, Qiao H, Qi X and Peng X 2024 Valleytronics in two-dimensional magnetic materials J. Phys.: Mater. **7** 022006.
[18] Zhao S, Li X, Dong B, Wang H, Wang H, Zhang Y, Han Z and Zhang H 2021 Valley manipulation in monolayer transition metal dichalcogenides and their hybrid systems: status and challenges Rep. Prog. Phys. **84** 026401.
[19] Lee C-J, Pan H-C, HadavandMirzaee F, Lu L-S, Cheng F, Her T-H, Shih C-K and Chang W-H 2025 Exciton–polariton valley Hall effect in monolayer semiconductors on plasmonic metasurface ACS Photonics **12** 1351–1358.
[20] Huang K, Fu H, Watanabe K, Taniguchi T and Zhu J 2024 High-temperature quantum valley Hall effect with quantized resistance and a topological switch Science **385** 657–661.

# 3. Exciton valley physics

**Ioannis Paradisanos[1,2] and Mikhail M. Glazov[3]**

[1] Department of Materials Science and Engineering, University of Crete, Heraklion, 70013, Greece
[2] Institute of Electronic Structure and Laser, Foundation for Research and Technology-Hellas, Heraklion, 70013, Greece
[3] orcid.org/0000-0003-4462-0749

E-mail: iparad@iesl.forth.gr

### Status

Exciton valley physics has become a leading direction in two-dimensional semiconductors, particularly transition-metal dichalcogenides (TMDs). The valley degree of freedom arises from inequivalent extrema at the $K$ and $K'$ points. In monolayer TMDs, broken inversion symmetry, time-reversal symmetry, and strong spin-orbit coupling lock spin and valley quantum numbers, allowing selective optical addressing with circularly polarized light - forming the basis of valleytronics.

Excitons, Coulomb-bound electron–hole pairs, dominate the optical properties of TMD monolayers. Their binding energies of several hundred meV, a consequence of reduced dielectric screening, the relatively heavy carrier effective masses, and spatial confinement, keep them stable well above room temperature [1].

A central feature of exciton valley physics is the valley-contrasting optical selection rules. At normal incidence, right-handed ($\sigma^+$) circularly polarized light couples exclusively to interband transitions in the $K$ valley, whereas left-handed ($\sigma^-$) light couples to the $K'$ valley. This property allows the optical generation of valley-polarized excitons with pseudospin of $|+\mathbf{1}\rangle$ or $|-\mathbf{1}\rangle$ (figure 1(a)). The emitted photoluminescence can retain the helicity of the excitation, providing a direct probe of valley polarization (also known as optical orientation), as shown in early studies [2, 3]. Beyond neutral excitons, trions, biexcitons, momentum-dark excitons, and interlayer excitons in heterobilayers inherit valley-specific optical properties, further enriching the field (figure 1(b)). In addition to valley polarization, excitons can also be prepared in coherent superpositions of the $K$ and $K'$ valleys using linearly polarized excitation, opening the way to manipulate exciton valley coherence (also known as exciton alignment) and highlighting the potential of the valley degree of freedom as a qubit basis beyond population imbalance (figure 1(a)) [4]. The latter property is particularly attractive because the valley coherent - linearly polarized - exciton is an entangled state in the electron and hole valley degrees of freedom.

While the valley index of free carriers is expected to be stable mainly due to the large momentum transfer required for intervalley scattering and the need for simultaneous spin flip, excitons display much faster valley depolarization. This behavior arises from the long-range electron–hole exchange interaction [5]. The exchange interaction couples excitonic states from opposite valleys, acting like an effective, exciton-momentum-dependent, magnetic field on the exciton pseudospin (figure 1(a)), akin to artificial Dresselhaus and Rashba fields produced by the spin-orbit interaction in electron and hole gases [6]. Because the strength of this effective field grows with exciton center-of-mass momentum, the associated pseudospin precession frequency increases with temperature, leading to progressively faster polarization decay. Exciton valley polarization typically decays on a picosecond timescale even at low temperatures, as confirmed by time-resolved photoluminescence and Kerr rotation experiments [7].

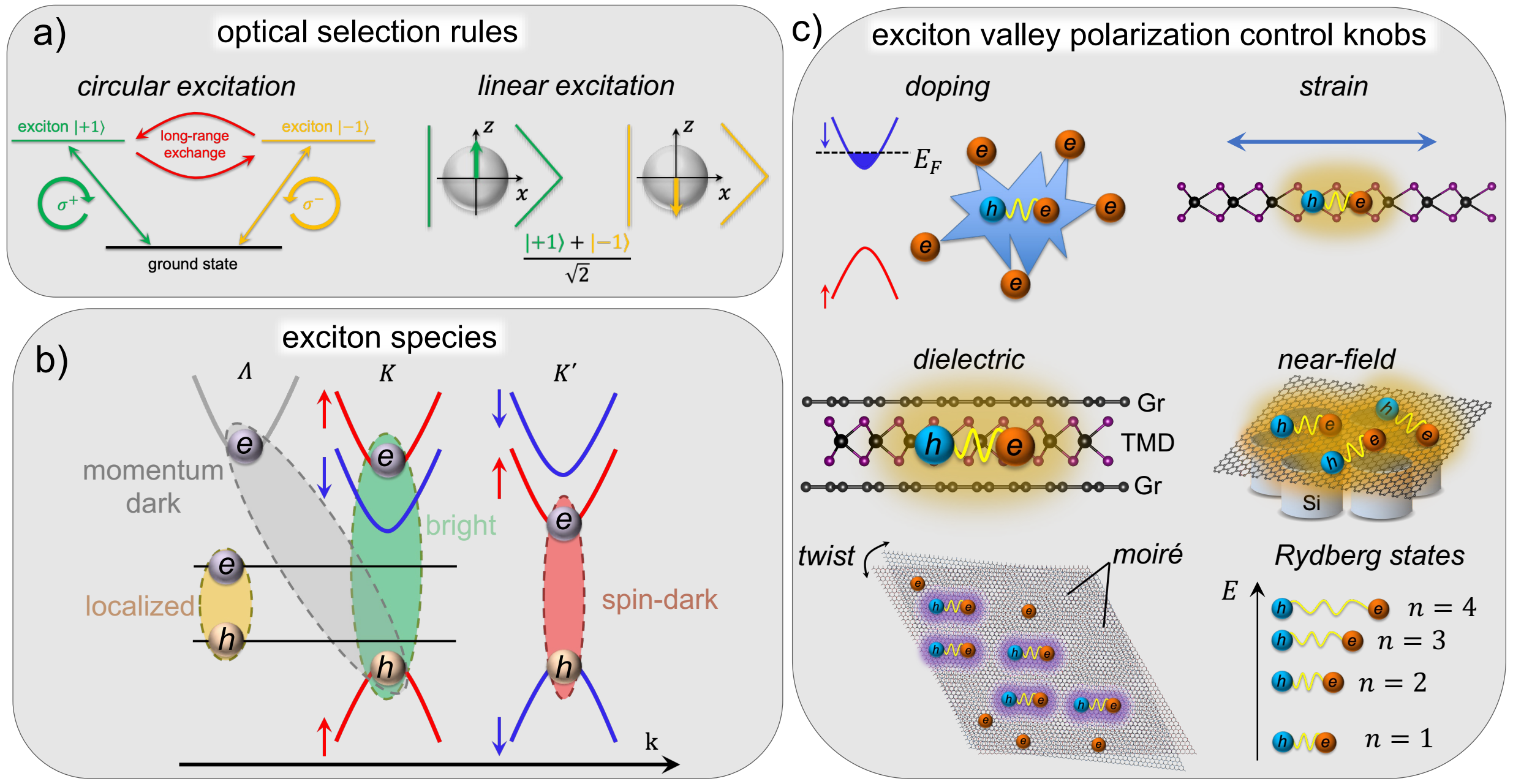

**Figure 1.** (a) Valley-selective optical selection rules in TMD monolayers: $\sigma^+$ or $\sigma^-$ circularly polarized light addresses excitons with pseudospin of $|+1\rangle$ or $|-1\rangle$, respectively, while linear polarization prepares valley-coherent excitons. (b) Excitonic landscape in TMDs, including spin- and momentum-bright excitons (green), spin-dark excitons (red), momentum-dark excitons (grey), localized excitons (yellow). (c) Representative control knobs for engineering exciton valley polarization, including carrier doping, strain, dielectric environment, near-field coupling (nanoantennas/cavities), twist-induced moiré superlattices, and Rydberg exciton states.

Beyond intrinsic effects, several external tuning knobs have emerged in the field. Electrostatic [8] and photochemical [9] doping, strain [10], and dielectric engineering [11] can modify valley polarization and exciton-scattering times, while coupling to optical cavities promotes the formation of exciton–polaritons [12], enriching the landscape of exciton valley physics (figure 1(c)). The field has expanded further with the introduction of van der Waals heterostructures and moiré superlattices, which support new regimes of exciton localization for singlet and triplet interlayer excitons associated with local symmetry [13] and a plethora of many body effects sensitive to the valley degrees of freedom (figure 1(c)).

## Current and future challenges

Despite major advances, specific obstacles limit the development of exciton valley physics. Following optical excitation, excitons typically do not recombine directly but undergo a cascade of relaxation processes involving phonons, impurities, defects, and interactions with other excitons. These processes introduce dephasing before radiative emission and potentially shorten the lifetime of valley polarization. A further complication arises from the high oscillator strength of bright excitons: while it enables strong light–matter interactions, it also enhances the long-range exchange interaction. Since the exchange field scales with oscillator strength, pseudospin precession accelerates, leading to rapid polarization decay. Valley coherence is even more fragile and typically dephases within a picosecond under exchange- and phonon-driven mechanisms. A central challenge involves the extension of both polarization and coherence lifetimes, while balancing strong optical activity with long-lived quantum states which is impossible without detailed investigation of the underlying mechanisms.

Experimental reports of valley polarization reveal additional subtleties. Early measurements under cryogenic and resonant excitation showed polarization degrees approaching 100% [2]. However, in today's cleaner and encapsulated samples, the observed values are significantly reduced. This counterintuitive trend arises from two related effects. First, suppressing defect-assisted recombination prolongs the exciton lifetime, providing more time for depolarization. Second, the reduction of exciton-defect scattering in high-quality samples increases the exciton scattering time, which in turn shortens the valley lifetime and reduces the degree of valley polarization and coherence [14].

Beyond these intrinsic processes, the band structure specifics impose constraints on valley preservation. Dark excitons act as long-lived reservoirs with suppressed exchange interaction and depolarization. Spin-dark excitons, common in tungsten-based TMDs, couple to $z$-polarized light and thus remain hidden in circular photoluminescence, while momentum-dark excitons such as $K$–$K'$, $K$–$\Lambda$ or $K$–$\Gamma$ states require phonon or defect assistance for recombination (figure 1(b)). Their presence, however, can impact bright-exciton polarization, making their control a critical challenge. In compounds such as $MoSe_2$, the different effective masses of spin-split conduction bands cause crossings between bright and dark states, accelerating valley depolarization through state mixing [15].

Finally, van der Waals moiré heterostructures highlight another open challenge. Moiré patterns induce local optical selection rules tied to atomic registry and local symmetry, yet these occur on length scales of a few to tens of nanometers, well below the diffraction limit of conventional optics. Probing and exploiting such local rules will require techniques capable of matching the moiré wavelength. Additionally, presence of free charge carriers strongly affects exciton valley physics both on a few particle level related to the trion formation and via many body, Fermi-polaron-like, effects [16]. Beyond these microscopic hurdles, the integration of valley physics into practical devices remains difficult. Most demonstrations rely on cryogenic temperatures and resonant excitation. Extending valley control to room temperature and electrical operation will require substantial advances in materials growth, defect management, and interface quality, as well as reducing dielectric disorder to achieve narrow lines and long coherence times. These remain critical obstacles on the path to room-temperature valleytronic devices based on excitons.

In parallel, several fundamental problems remain unresolved. These include clarifying the role of exciton-exciton and exciton-electron interactions in depolarization and coherence loss, developing a comprehensive description of dark exciton spin and valley relaxation mechanisms, and probing valley-dependent exciton transport in both monolayers and heterostructures. Additional directions involve understanding the valley physics of excited Rydberg excitons (figure 1(c)), establishing links between valley degrees of freedom and topological properties, and realizing valley–photon interfaces capable of transferring both polarization and coherence.

### Advances in science and technology to meet challenges

Progress in exciton valley physics will likely come from steady improvements in materials, experimental and theoretical methods, and device concepts rather than a single breakthrough. To address these challenges, a range of strategies is being developed to stabilize and control valley excitons. A central theme is engineering the exciton lifetime. Graphene/TMD/graphene heterostructures [17] provide ultrafast extraction pathways that shorten lifetimes and reduce depolarization. Controlled doping can introduce additional scattering channels that counteract and randomize exchange-driven precession (figure 1(c)). Conversely, coupling TMDs to optical cavities or placing them in the near field of tailored nanoantennas offers a route to accelerate radiative decay through the Purcell effect, providing flexible control over valley polarization dynamics (figure 1(c)). Simultaneously, dielectric environment is predicted to affect the long-range exchange interaction via screening providing a pathway to control exciton valley depolarization and decoherence times [11]. At the same time, progress has been made in addressing the role of dark excitons: coupling to ferromagnetic substrates, plasmonic/dielectric nanostructures, or optical cavities can activate spin-dark states, allowing them to serve as long-lived valley reservoirs.

Band-structure engineering provides another avenue. Material selection critically determines achievable polarization degrees, as the magnitude of spin-orbit splitting and the effective mass of conduction-band states influence bright-dark state mixing. Carefully chosen compounds or alloys can control these effects, while strain offers additional means to reshape the band structure, exciton fine structure, oscillator strengths, and relaxation pathways.

Moiré superlattices open new directions. Interlayer excitons, with reduced wavefunction overlap, naturally exhibit weaker oscillator strengths and thus longer valley lifetimes [18]. The moiré potential creates local optical selection rules defined by atomic registry, effectively generating ordered arrays of valley-specific emitters (figure 1(c)) [19]. A key technological need is the development of near-field optical techniques or other nanoscale probes capable of resolving and manipulating valley selection rules on the moiré wavelength scale. Such approaches would allow direct access to the spatially varying excitonic landscape and unlock new modes of valley control. In this respect, the role of doping may become particularly pronounced as excitons interact with spin/valley ordered electronic phases.

Finally, progress toward device integration is accelerating. Deterministic encapsulation, defect engineering, and dielectric control are improving sample homogeneity and reducing disorder. Electrical injection of spin-polarized carriers and gating schemes are being explored to move beyond purely optical pumping. Integration with photonic platforms such as waveguides and silicon photonics is under way, aiming to exploit valley-polarized excitons for optoelectronic and quantum technologies [20]. Combining cavity coupling, moiré engineering, and interlayer excitons may ultimately support scalable arrays of long-lived valley qubits and entangled photon sources, bridging fundamental science and applications.

### Concluding remarks

Exciton valley physics in two-dimensional semiconductors has become a central area of research, linking fundamental studies of quantum degrees of freedom with prospects for new optoelectronic functionality. Significant progress has been made in understanding valley-selective optical selection rules, strong excitonic effects, and the mechanisms that govern valley polarization and coherence. Yet short valley lifetimes, the oscillator strength of bright excitons, the role of dark excitons, and sensitivity to disorder or band-structure details remain persistent challenges.

Advances in materials synthesis, photonic structures, straintronic devices and heterostructure engineering are providing tools to address these issues. Improved encapsulation and growth reduce disorder, while cavity coupling and near-field environments in proximity with nanoantennas offer lifetime control. Moiré superlattices introduce new regimes of exciton localization with local valley selection rules, and near-field methods are emerging to probe them directly. In parallel, progress toward deterministic integration and electrical operation is moving valley physics closer to device concepts.

Overall, the field continues to mature while retaining significant open questions. Improved understanding and control of valley degrees of freedom will deepen knowledge of excitonic processes in two dimensions and open new avenues for optoelectronic and photonic applications.

## Acknowledgements

I.P. acknowledges financial support by the Hellenic Foundation for Research and Innovation (H.F.R.I.) under the “3rd Call for H.F.R.I. Research Projects to support Post-Doctoral Researchers” (Project Number:7898).

## References

[1] Wang G, Chernikov A, Glazov M M, Heinz T F, Marie X, Amand T & Urbaszek B 2018 Colloquium: Excitons in atomically thin transition metal dichalcogenides *Rev. Mod. Phys.* **90** 021001

[2] Mak K F, He K, Shan J and Heinz T F 2012 Control of valley polarization in monolayer $MoS_2$ by optical helicity *Nature nanotechnology* **7** 494

[3] Cao T, Wang G, Han W, Ye H, Zhu C, Shi J, Niu Q, Tan P, Wang E, Liu B and Feng J, 2012 Valley-selective circular dichroism of monolayer molybdenum disulphide *Nature communications* **3** 887

[4] Wang G, Marie X, Liu B L, Amand T, Robert C, Cadiz F, Renucci P and Urbaszek B, 2016 Control of exciton valley coherence in transition metal dichalcogenide monolayers *Physical review letters* **117** 187401.

[5] Glazov M M, Ivchenko, E L, Wang G, Amand T, Marie X, Urbaszek B and Liu B L, 2015 Spin and valley dynamics of excitons in transition metal dichalcogenide monolayers *physica status solidi (b)*, **252** 2349

[6] Žutić I, Fabian J and Sarma S D 2004 Spintronics: Fundamentals and applications *Reviews of modern physics* **76** 323

[7] Zhu C R, Zhang K, Glazov M, Urbaszek B, Amand T, Ji Z W, Liu B L and Marie X 2014 Exciton valley dynamics probed by Kerr rotation in $WSe_2$ monolayers *Physical Review B* **90** 161302.

[8] Wu Y C, Taniguchi T, Watanabe K and Yan J 2021 Enhancement of exciton valley polarization in monolayer $MoS_2$ induced by scattering *Physical Review B* **104** L121408.

[9] Katsipoulaki E, Mourzidis K, Jindal V, Lagarde D, Taniguchi T, Watanabe K, Kopidakis G, Marie X, Glazov M M, Stratakis E, Kioseoglou G and Paradisanos I 2025 Spin-Valley Polarization Control in $WSe_2$ Monolayers using Photochemical Doping *Advanced Optical Materials* **13** e00575.

[10] Kourmoulakis G, Michail A, Paradisanos I, Marie X, Glazov M M, Jorissen B, Covaci L, Stratakis E, Papagelis K, Parthenios J and Kioseoglou G 2023 Biaxial strain tuning of exciton energy and polarization in monolayer $WS_2$ *Applied Physics Letters* **123** 223103

[11] Prazdnichnykh A I, Glazov M M, Ren L, Robert C, Urbaszek B and Marie X 2021 Control of the exciton valley dynamics in atomically thin semiconductors by tailoring the environment *Physical Review B* **103** p.085302

[12] Chen Y J, Cain J D, Stanev T K, Dravid V P and Stern N P 2017 Valley-polarized exciton–polaritons in a monolayer semiconductor *Nature Photonics* **11** 431

[13] Tran K, Choi J and Singh A 2020 Moiré and beyond in transition metal dichalcogenide twisted bilayers *2D Materials* **8** 022002

[14] D’yakonov M I and Perel V I 1971 *JETP* **33** 1053

[15] Yang M, Robert C, Lu Z, Van Tuan D, Smirnov D, Marie X and Dery H 2020 Exciton valley depolarization in monolayer transition-metal dichalcogenides *Physical Review B* **101** 115307.

[16] Sidler M, Back P, Cotlet O, Srivastava A, Fink T, Kroner M, Demler E and Imamoglu A 2017 Fermi polaron-polaritons in charge-tunable atomically thin semiconductors *Nature Physics* **13** 255

[17] Lorchat E, Azzini S, Chervy T, Taniguchi T, Watanabe K, Ebbesen T W, Genet C and Berciaud S, 2018 Room-temperature valley polarization and coherence in transition metal dichalcogenide–graphene van der Waals heterostructures *ACS photonics* **5** 5047

[18] Shree S, Paradisanos I, Marie X, Robert C and Urbaszek B 2021 Guide to optical spectroscopy of layered semiconductors Nature Reviews Physics 3 39.

[19] Seyler K L, Rivera P, Yu H, Wilson N P, Ray E L, Mandrus D G, Yan J, Yao W and Xu X, 2019 Signatures of moiré-trapped valley excitons in $MoSe_2/WSe_2$ heterobilayers *Nature* **567** 66

[20] Zotev P G, Bouteyre P, Wang Y, Randerson S A, Hu X, Sortino L, Wang Y, Shegai T, Gong S H, Tittl A, Aharonovich I and Tartakovskii A I 2025 Nanophotonics with multilayer van der Waals materials *Nature Photonics*, **19** 788

# 4. Ultrafast valleytronics

**Oleg Dogadov[1,2†], Francesco Gucci[1†], Giulio Cerullo[1,3*] and Stefano Dal Conte[1]**

[1] Dipartimento di Fisica, Politecnico di Milano, Piazza L. da Vinci 32, 20133 Milano, Italy
[2] Department of Physical Chemistry, Fritz Haber Institute of the Max Planck Society, 14195 Berlin, Germany
[3] IFN-CNR, Department of Physics, Politecnico di Milano, Piazza L. da Vinci 32, Milano 20133, Italy
[†] These authors contributed equally to this work

E-mail: giulio.cerullo@polimi.it

### Status

Monolayers (1L) of transition metal dichalcogenides (TMDs) have sparked renewed interest in valleytronics, owing to their unique band structure characterized by strongly bound excitons in two energy-degenerate valleys at the $K$ and $K'$ points, with opposite Berry curvatures and orbital magnetic moments, which can be selectively excited by circularly polarized light. The practical exploitation of the valley degree of freedom in TMDs has however been impeded by its intrinsically short lifetime, which is of the order of a few picoseconds at cryogenic temperatures and becomes progressively shorter with increasing temperature [1, 2]. This rapid valley polarization decay, measured by either time-resolved circular dichroism (TR-CD) or time-resolved Faraday/Kerr rotation (TRFR/TRKR) (see Fig. 1a,b), is predominantly due to the intervalley exchange interaction, leading to resonant coupling between the $K$ and $K'$ valleys. (Fig. 1c). The physical landscape is further enriched by the presence of additional spin-dark and momentum-indirect excitonic states, which are nearly degenerate in energy. These states can be efficiently populated via phonon-assisted electron scattering and other Coulomb-mediated processes, such as Dexter-like coupling mechanisms that hybridize A and B excitons residing in opposite valleys [3].

In recent years, research on valleytronics in 1L-TMDs has progressed along two main directions. The first aims to extend the valley lifetime. One of the proposed ways to increase it is the use of artificially stacked TMD heterostructures (HSs), in which the exchange interaction is suppressed (Fig. 1d). It has been shown that at low temperature electrons and holes preserve their spin upon the interlayer charge transfer and that the holes exhibit microsecond-long lifetime [4, 5, 6].

An alternative way to manipulate the valley polarization dynamics relies on the ability to control interexcitonic interactions, which is another important challenge for ultrafast valleytronic applications. One of the explored strategies is the use of mechanical strain, whereby different electronic bands undergo different energy shifts, allowing therefore to externally modify valley depolarization channels [7] (Fig. 1e). By careful strain engineering, one could make use of the created hybridized states for valley manipulation and spin-valley transport, as well as utilize the in-plane pseudo-magnetic and out-of-plane electric fields, depending on the strain profile.

The second important research direction of ultrafast valleytronics focuses on developing protocols based on the use of few-cycle laser pulses to manipulate the valley degree of freedom on timescales shorter than its intrinsic lifetime. Although some theoretical works have proposed different schemes for valley polarization control [8, 9], to date, this area remains largely unexplored.

## Current and future challenges

Although a deep understanding of fundamental processes governing valley (de)polarization dynamics in TMDs has been achieved during several years of dedicated studies, the majority of experimental results have been obtained by either TR-CD or TRFR/TRKR techniques [1–3], which only probe optically bright states. Therefore, experiments directly accessing dark excitonic states would be beneficial.

The main challenge for potential valleytronic applications is a relatively short valley polarization time in 1L-TMDs. Although some improvement has been achieved in TMD HSs [4–6], the effect is significantly reduced at room temperature, limiting prospective applications. The search for conditions that could potentially preserve spin-valley polarization in TMD HSs is still an ongoing challenge.

Artificial stacking of TMD layers provides further opportunities for the design of excitonic states and for manipulating valleytronic effects by controlling the stacking twist angle. Moiré patterns, producing periodic potential wells for excitons, allow to engineer new states in TMD HSs. Although recent studies have demonstrated the possibility to generate localized excitons with distinct spin-layer-valley configurations [10], the effects of the moiré potential on spin-valley dynamics still need to be explored [11].

Despite the successful demonstration of tuning of the spin-valley dynamics in 1L-TMDs by applying strain [7], such strain devices require a careful design. One of the main challenges lies in controlling the applied strain precisely and achieving uniform strain across the sample. Also, the theoretical understanding of the strain effect on spin-valley dynamics in 1L-TMDs and their HSs requires further studies.

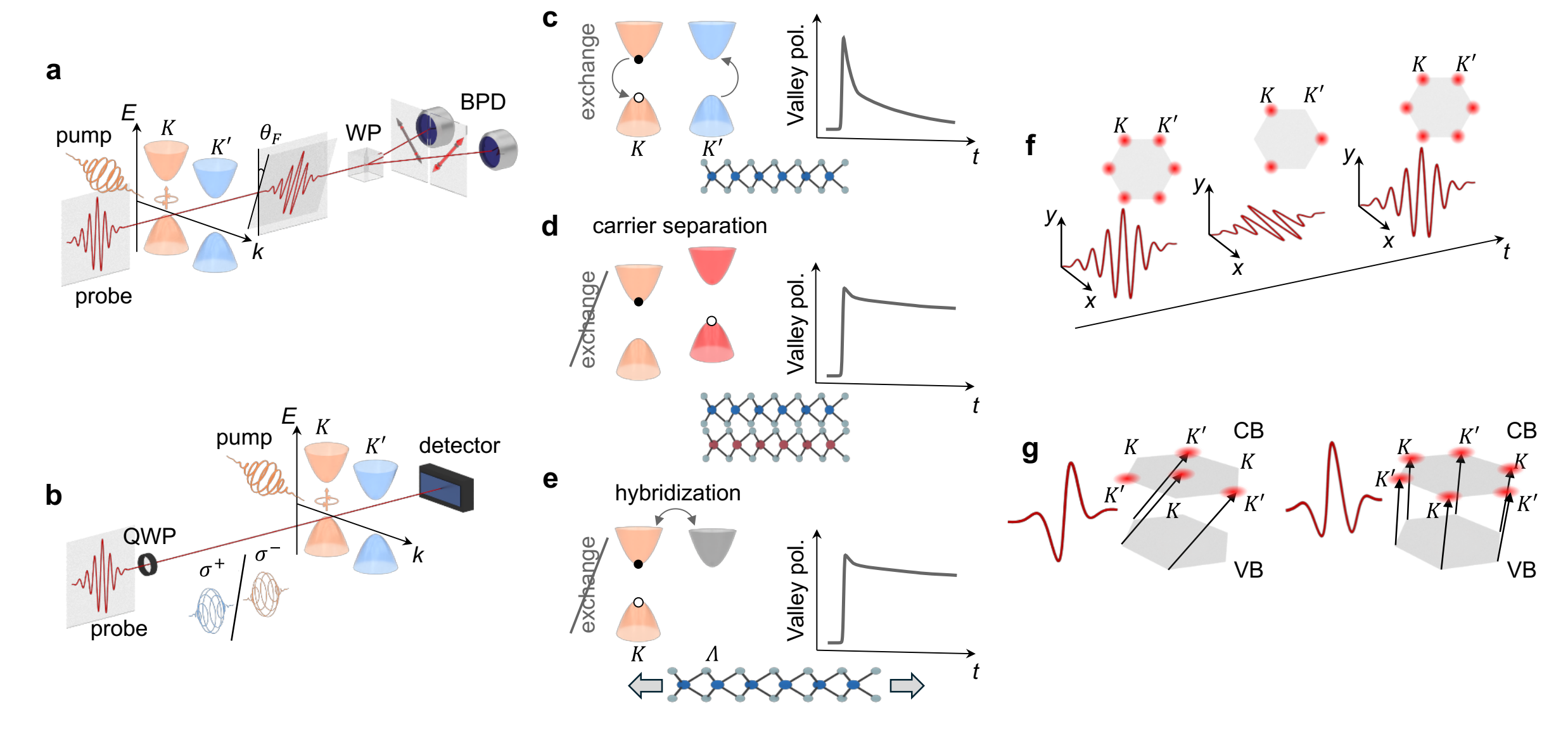

**Figure 1.** Time resolved spectroscopic tools to study spin-valley dynamics: time-resolved Faraday rotation (a) and time-resolved circular dichroism (b). (c–e) Valley depolarization time in TMDs can be increased compared to pristine 1L-TMD (c) when stacked into HSs (d) or when strain is applied (e). (f) Manipulation of the valley degree of freedom through multiple pulse excitation. (g) Strong field, single-cycle, non-resonant excitation through carrier tunnelling. By controlling the driver's CEP, one could obtain a valley-selective excitation (left: asymmetric electric field) or a homogeneous population (right: symmetric electric field).

Another possible way to affect valley dynamics, which remains largely unexplored, is by exploiting proximity effects, for instance, by placing TMDs on a magnetic substrate or by stacking them with layered magnetic materials [12].

A distinct paradigm in ultrafast valleytronics focuses on manipulating valley polarization before it decays, rather than attempting to extend its lifetime (see Fig. 1f,g). Strong terahertz fields can drive efficient valley population transfer, although the resulting carriers are delocalized in momentum space [13]. Alternative approaches seek to actively steer the valley pseudospin within its coherence time using resonant pulses. Building on seminal works, such as [14], recent theoretical studies have proposed several schemes to encode information and perform logic operations in the time domain through controlled valley superpositions [9, 15]. However, these protocols, based on the excitation with multiple, few-cycle, phase-locked pulses, tuned to the excitonic resonances, are technically challenging. Furthermore, since the concept of valleytronics is inherently general, developing techniques that are material independent and do not rely on broken centrosymmetry is highly desirable.

### Advances in science and technology to meet challenges

The information about the dynamics of different momentum and spin states in TMDs can be obtained in such potential experiments, as spin-, time-, and angle-resolved photoemission spectroscopy, or solid-state high-harmonic generation. While the former allows a direct visualization of excited electron states, the latter would allow studying valley dynamics under ambient conditions. On the other hand, techniques offering both sub-picosecond temporal and atomic spatial resolution could unveil the spin-valley dynamics of confined moiré excitons and would be beneficial to spatio-temporally record the valley Hall effect. In this direction, Siday et al. have recently demonstrated an all-optical tip-based microscope combining picometric precision with femtosecond temporal resolution [16]. Additionally, future optical studies would benefit from a technique, combining broadband detection, commonly applied in TR-CD, with the high sensitivity of TRFR/TRKR. Such an approach could be realized, for instance, through interferometric Fourier-transform balanced detection [17].

Among the schemes proposed for ultrafast valley manipulation, multiple-pulse excitation with phase-locked, linearly polarized, resonant pulses enables coherent control of valley states by driving the pseudospin around the Bloch sphere, where the relative phase and timing between pulses define the trajectory in pseudospin space [15]. Within this protocol, the achievable valley switching rate is ultimately limited by the pulse duration: For materials with large bandgaps, such as hexagonal boron nitride (hBN), the single-cycle limit corresponds to attosecond optical periods, enabling in principle valley switching on a few-femtosecond timescale [9, 18]. Recent experimental demonstrations of room-temperature coherent valley manipulation on sub-picosecond timescales highlight the feasibility of controlling valley polarization within its intrinsic lifetime in ambient conditions [19]. Within this excitation protocol based on a series of femtosecond pulses (see Fig. 1f), a linearly polarized pulse creates an excitonic coherence between the valleys. A subsequent, orthogonally polarized pulse drives a valley-selective population either in $K$ or $K'$ valley, depending on the relative phase delay between the two pulses. A third pulse can then restore the balance, resetting the valley polarization. Extending beyond the resonant, linear regime [9], sub-cycle strong-field control exploits carrier-envelope-phase (CEP) stable, non-resonant driving pulses with asymmetric vector potentials to induce carriers tunnelling towards specific valleys [20]. Here, the valley selectivity arises directly from the single-cycle field asymmetry, offering a material-independent, symmetry-agnostic route to all-optical valley control on ultrafast timescales [8]. By controlling the CEP and the field strength of the driving pulse, one can selectively excite $K'$ (left in Fig. 1g), $K$, or a homogeneous combination of the two valleys (right in Fig. 1g).

## Concluding remarks

The ultrafast valley dynamics in TMDs and their HSs has attracted much attention in recent years. Some open fundamental questions will likely be addressed in the near future with the advances of experimental techniques. At the same time, we expect that future studies on ultrafast valleytronics in TMDs will mainly be focused on potential applications, rather than fundamental investigations. Given the significant results demonstrating successful manipulation of the valley degree of freedom, we envision future technological developments targeted at prolonging valley polarization time, as well as at controlling it on ultrafast timescales. The recent advances in sample fabrication further ensure increasing possibilities for the use of TMDs and their combinations in HSs with other layered materials, opening new ways for control of the spin-valley dynamics in these systems.

## Acknowledgements

G.C. and S.D.C. acknowledge support from the European Union's Next Generation EU Programme with the I-PHOQS Infrastructure 423 [IR0000016, ID D2B8D520, CUP B53C22001750006] "Integrated infrastructure initiative in Photonic and Quantum Sciences." G.C. acknowledges support by the Horizon Europe European Innovation Council (101130384, HORIZONEIC-2023-PATHFINDEROPEN-01, QUONDENSATE). S.D.C. acknowledges support from the European Union's NextGenerationEU—Investment 1.1, M4C2—Project No. 2022LA3TJ8—CUP D53D23002280006.

## References

[1] Dal Conte S, Bottegoni F, Pogna E A A, De Fazio D, Ambrogio S, Bargigia I, D'Andrea C, Lombardo A, Bruna M, Ciccacci F, Ferrari A C, Cerullo G and Finazzi M 2015 Ultrafast valley relaxation dynamics in monolayer $MoS_2$ probed by nonequilibrium optical techniques *Physical Review B* **92** 235425

[2] Zhu C R, Zhang K, Glazov M, Urbaszek B, Amand T, Ji Z W, Liu B L and Marie X 2014 Exciton valley dynamics probed by Kerr rotation in $WSe_2$ monolayers *Physical Review B* **90** 161302

[3] Dogadov O, Mittenzwey H, Bertolotti M, Olsen N, Thomas D, Trovatello C, Zhu X, Daniele B, Cerullo G, Knorr A and Dal Conte S 2025 Dissecting intervalley coupling mechanisms in monolayer transition metal dichalcogenides *arXiv preprint* arXiv:2507.16665

[4] Kim J, Jin C, Chen B, Cai H, Zhao T, Lee P, Kahn S, Watanabe K, Taniguchi T, Tongay S, Crommie M F and Wang F 2017 Observation of ultralong valley lifetime in $WSe_2/MoS_2$ heterostructures *Science Advances* **3**(7) e1700518

[5] Kumar A, Yagodkin D, Stetzuhn N, Kovalchuk S, Melnikov A, Elliott P, Sharma S, Gahl C and Bolotin K I 2021 Spin/Valley Coupled Dynamics of Electrons and Holes at the $MoS_2$–$MoSe_2$ Interface *Nano Letters* **21**(17) 7123

[6] Wagner J, Bernhardt R, Rieland L, Abdul-Aziz O, Li Q, Zhu X, Dal Conte S, Cerullo G, van Loosdrecht P H M and Hedayat H 2025 Unveiling Ultrafast Spin-Valley Dynamics and Phonon-Mediated Charge Transfer in $MoSe_2/WSe_2$ Heterostructures *Advanced Optical Materials* **13**(8) 2402703

[7] Kumar A M, Bock D J, Yagodkin D, Wietek E, Höfer B, Sinner M, Dewambrechies A, Hernández López P, Kovalchuk S, Dhingra R, Heeg S, Gahl C, Libisch F, Chernikov A, Malic E, Rosati R and Bolotin K I 2025 Strain control of valley polarization dynamics in a 2D semiconductor via exciton hybridization *Nano Letters* **25**(42) 15164

[8] Rana N and Dixit G 2023 All-Optical Ultrafast Valley Switching in Two-Dimensional Materials *Physical Review Applied* **19**(3) 034056

[9] Silva R E F, Ivanov M, and Jiménez-Galán Á 2022 All-optical valley switch and clock of electronic dephasing *Optics Express* **30**(17) 30347

[10] Brotons-Gisbert M, Baek H, Molina-Sánchez A, Campbell A, Scerri E, White D, Watanabe K, Taniguchi T, Bonato C and
Gerardot B D 2020 Spin–layer locking of interlayer excitons trapped in moiré potentials *Nature Materials* **19**(6) 630

[11] Shinokita K, Watanabe K, Taniguchi T and Matsuda K 2022 Valley Relaxation of the Moiré Excitons in a $WSe_2/MoSe_2$ Heterobilayer *ACS Nano* **16**(10) 16862
[12] Castro E C, Brandão D S, Bragança H, Martins A S, Riche F, Dias A C, Zhao J H, Fonseca A L A and Qu F 2023 Mechanisms of interlayer exciton emission and giant valley polarization in van der Waals heterostructures *Physical Review B* **107** 035439
[13] Langer F, Schmid C P, Schlauderer S, Gmitra M, Fabian J, Nagler P, Schüller C, Korn T, Hawkins P G, Steiner J T, Huttner U, Koch S W, Kira M and Huber R 2018 Lightwave valleytronics in a monolayer of tungsten diselenide *Nature* **557**(7703) 76
[14] Ye Z, Sun D and Heinz T F 2017 Optical manipulation of valley pseudospin *Nature Physics* **13**(1) 26
[15] Sharma S, Elliott P and Shallcross S 2022 Valley control by linearly polarized laser pulses: example of $WSe_2$ *Optica* **9**(8) 947
[16] Siday T, Hayes J, Schiegl F, Sandner F, Menden P, Bergbauer V, Zizlsperger M, Nerreter S, Lingl S, Repp J, Wilhelm J, Huber M A, Gerasimenko Y A and Huber R 2024 All-optical subcycle microscopy on atomic length scales *Nature* **629**(8011) 329
[17] Gucci F, Iudica A, Tacchi A V Y, Schirato A, Crotti G, Kim R M, Lee S M, Han J H, Villa A, Ali D, Rizzo A, Maiuri M, Nam K T, Della Valle G and Cerullo G 2025 Broadband ultrafast self-heterodyned chiro-optical spectroscopy *arXiv preprint* arXiv:2511.10574
[18] Li W, Zhu X, Lan P, Li L, Wang K, He W, Hübener H, De Giovannini U and Lu P 2025 Attosecond All-Optical Retrieval of Valley Polarization via Circular Dichroism in Transient Absorption *Physical Review Letters* **135**(8) 086404
[19] Gucci F, Molinero E B, Russo M, San-Jose P, Camargo F V A, Maiuri M, Ivanov M, Jiménez-Galán Á, Silva R E F, Dal Conte S and Cerullo G 2026 Encoding and Manipulating Ultrafast Coherent Valleytronic Information with Lightwaves *Nature Photonics*
[20] Jiménez-Galán Á, Silva R E F, Smirnova O and Ivanov M 2021 Sub-cycle valleytronics: control of valley polarization using few-cycle linearly polarized pulses *Optica* **8**(3) 277

## 5. Lightwave valleytronics

**Shubhadeep Biswas**[1]

[1] Department of Physics, Indian Institute of Science, Bangalore-560012, India

E-mail:shubha@iisc.ac.in

### Status

Valleytronics is an emerging branch of condensed matter physics that seeks to exploit the valley degree of freedom of electrons in crystalline solids as a new carrier of information, analogous to charge in electronics and spin in spintronics. In a crystal's electronic band structure, valleys refer to the inequivalent local extrema in the conduction or valence bands where electrons or holes can reside. The valley index thus acts as a binary pseudospin — commonly labelled as K and K' — that can potentially encode information [1].

The foundation of valleytronics lies in materials that possess multiple, energetically degenerate but momentum-distinguishable valleys. In most conventional semiconductors, valleys cannot be individually addressed because they lack valley-dependent physical quantities. However, the isolation of two-dimensional (2D) materials such as graphene, transition metal dichalcogenides (TMDs) like $MoS_2$, $WS_2$, $MoSe_2$, and $WSe_2$, and hexagonal boron nitride (hBN), which all possess hexagonal symmetry, has revolutionized the field. These 2D crystals, in either their monolayer limit or under external perturbation, exhibit broken inversion symmetry, which gives rise to valley-contrasting properties such as Berry curvature and orbital magnetic moments, enabling the selective manipulation of valley states with optical, electrical, and magnetic means [1].

Due to the valley-specific orbital magnetic moments, circularly polarized light with different helicities couples selectively to different valleys, enabling optical initialization of valley polarization [2,3], see Fig. 1 (a-b). Such control paves the way for valley-based logic, optoelectronic, and quantum information devices, offering low-dissipation and non-volatile functionalities.

Since the initiation of this field around 2012, most of the works have used resonant circularly polarized single-photon excitation to address a specific valley. However, in recent years, resonant or non-resonant light in the strong field regime has been used to initiate and manipulate the valley polarization as well as the production of valley current (e.g. see Fig. 1(c-f)). This approach, namely 'lightwave valleytronics', primarily utilizes the symmetry of the light field to break certain symmetries of the system, giving rise to valley polarization and current. As one can design the symmetry of the light field with nearly infinite possibilities, it has given a new handle to explore the valley physics. Additionally, this also provides the possibility of ultrafast (hundreds of attoseconds to a few femtoseconds) initiation and manipulation of the valley degree of freedom, which enables the possibility of multiple classical or quantum logic gate operations well within the decoherence time.

Several schemes have been proposed under lightwave valleytronics, some of which have been experimentally verified in recent years. However, many unique frontiers remain to be explored. The development of new light technologies and experimental techniques provides the feasibility to explore this unique new regime of valleytronics, which can revolutionize next-generation information technologies that integrate quantum-electronic and photonic degrees of freedom on an ultrafast time scale.

**Current and future challenges**

The field of lightwave valleytronics has been propelled by advances in the precise control of light fields on sub-cycle timescales—a capability that also underpins the emergence of attosecond physics [4–7]. In this context, lightwave valleytronics leverages ultrafast optical control in two key ways: (i) by employing tailored light fields with well-defined symmetries that resonate with the intrinsic structural symmetry of two-dimensional materials, and (ii) by utilizing ultrashort, near–single-cycle light waveforms to drive and probe valley-selective electron dynamics with extreme temporal precision.

By interferometrically combining two counter-rotating circularly polarized pulses of fundamental frequency (ω) and its second harmonic (2ω), one can generate a trefoil-shaped light waveform that exhibits $C_3$ rotational symmetry. Jiménez-Galán et al. first proposed employing such an off-resonant tailored trefoil waveform to perform valleytronic operations in inversion-symmetry–broken hexagonal systems [8]. This concept was experimentally realized only recently—by Mitra et al. in monolayer hBN [9] (see Fig. 2) and by Tyulnev et al. in bulk $MoS_2$ [10]—where the trefoil light field was shown to induce complex next-nearest-neighbour hopping interactions within the two-dimensional lattice (see Fig. 2(a-c)). According to the Haldane model [11], these induced hopping terms transiently break time-reversal (TR) symmetry, thereby lifting the valley degeneracy. The orientation of the trefoil waveform with respect to the lattice can be actively rotated, allowing dynamic control of TR-symmetry breaking on sub-cycle timescales (see Fig. 2(d-h)). The resulting bandgap asymmetry between the K and K′ valleys modifies the light-induced electron tunnelling probabilities, leading to a net valley polarization (see Fig. 2(i-l)). With precise sub-cycle rotation, the character of this polarization can be reversibly flipped, enabling ultrafast valley switching. This approach opens a pathway to selective valley excitation and manipulation using strong-field interactions, thereby relaxing the conventional constraint of matching the photon energy with the bandgap of the material. The valley polarization was experimentally probed using time-delayed harmonic polarimetry, establishing a fully optical valleytronic scheme that integrates all three essential steps: polarization initiation, switching, and readout. Earlier, Langer et al. demonstrated valley polarization switching in $WSe_2$ via a two-color strong-field scheme that combined resonant linearly polarized excitation with a terahertz streaking field [12], inspiring subsequent theoretical studies exploring valley polarization and current control in graphene and gapped graphene systems [13, 14]. More recently, Weitz et al. extended this concept by examining tailored waveforms with varying symmetries in graphene [15], while Mrudul et al. proposed a trefoil-waveform–based strategy to break inversion symmetry in monolayer graphene [16], rendering it an attractive platform for lightwave-driven valleytronics. Although these pioneering demonstrations mark important milestones, significant challenges remain. A key limitation lies in the relatively weak strong-field–induced valley asymmetry signals, which are often masked by symmetric population backgrounds. Enhancing the sensitivity and selectivity of valley-resolved detection remains an immediate experimental challenge.

In the single-cycle limit, the controllable asymmetry of a light waveform—governed by the carrier–envelope phase (CEP)—can be exploited to generate and manipulate valley polarization. Jiménez-Galán et al. predicted such CEP-controlled valley manipulation in gapped graphene, where a single-cycle linearly polarized pulse enables sub-cycle precision in valley-selective excitation [17]. Similar control has been proposed using single-cycle circularly polarized pulses, which are also predicted to induce a measurable valley current, contingent on $C_3$-symmetry breaking of the valley dipole moment [18]. However, such single-cycle-limit schemes are yet to be realized experimentally. More recently, an orthogonally polarized, time-delayed resonant pulse pair has been employed to control valley polarization in TMDs. This experimentally realized scheme [19]—originally predicted theoretically [20,21]—utilizes sub-cycle tuning of the resultant ellipticity of circularly polarized light, thereby modulating circular dichroism–driven absorption and achieving ultrafast optical control of valley polarization.

Looking forward, many fundamental and technological frontiers remain open. The rapid development of new 2D materials, heterostructures, and moiré systems offers rich opportunities to apply symmetry-tailored strong fields for realizing novel optoelectronic functionalities. Extending these concepts toward quantum information processing, where coherent superpositions of valley states can act as quantum bits [22], remains an exciting but formidable goal. Moreover, strong-field valleytronics predicts light-induced topological phase transitions [8], potentially enabling ultrafast, dissipationless topological optoelectronics.

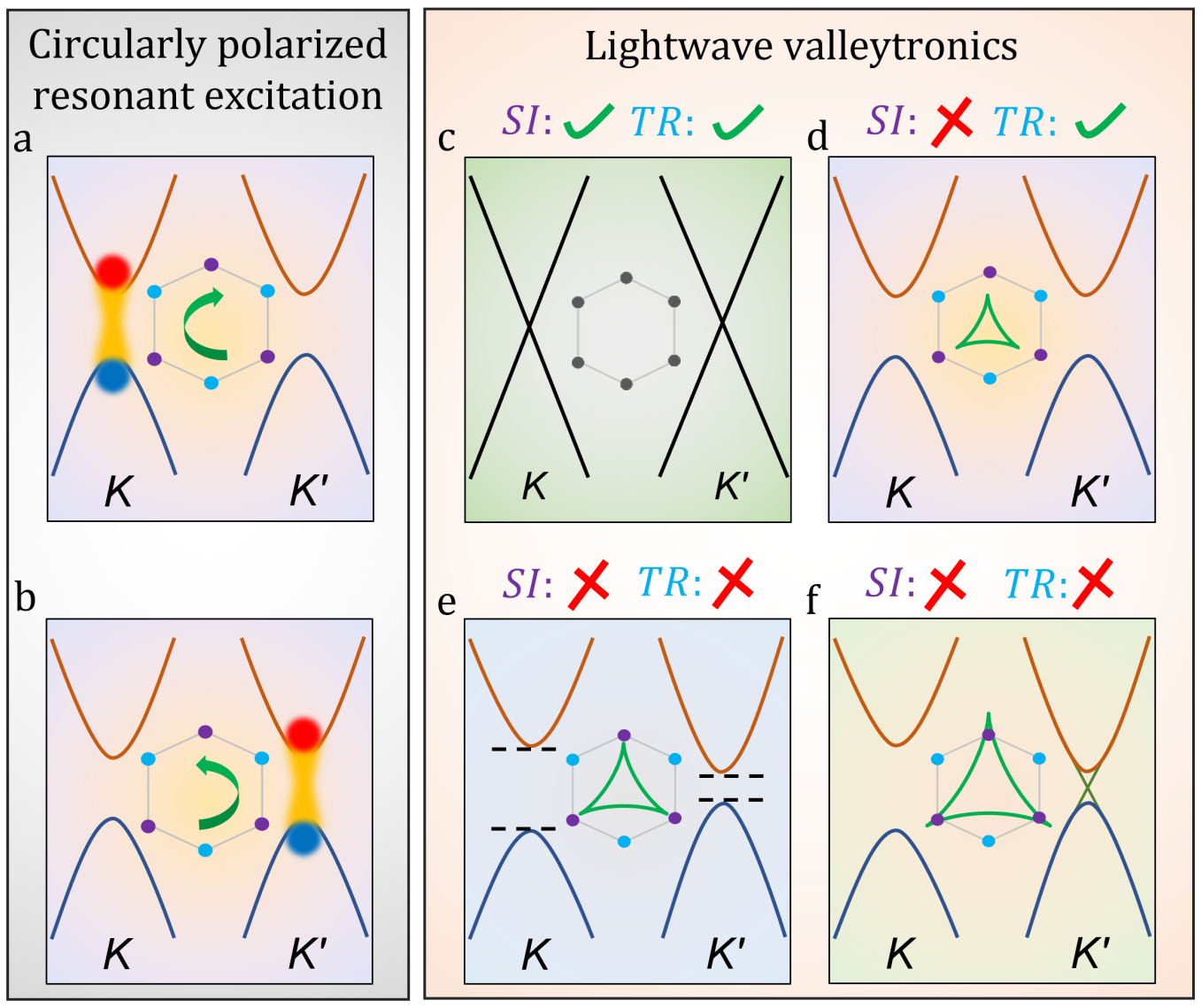

**Figure 1.** (a-b) Resonant circularly polarized light excitation. Following circular dichroism, the opposite helicity of light excites opposite (K or K') valleys. (c) Valley configuration for a spatial inversion (SI) and time reversal (TR) symmetry preserved system like monolayer graphene. (d) A low amplitude tailored trefoil waveform (green) breaks the SI symmetry opening a band gap. (e) Higher amplitude of waveform breaks the TR symmetry of the SI symmetry broken system making an asymmetric band gap. (f) Further increase of amplitude can induce topological phase transition.

## Advances in science and technology to meet challenges

The advancement of lightwave valleytronics is closely tied to progress in laser and optical technologies. The emergence of long-wavelength (>2 μm) coherent light sources has significantly advanced the realization of off-resonant valleytronic schemes. However, the high field strengths required for such experiments often risk material damage, particularly in the monolayer limit. This challenge can be mitigated by confining pulse energy to the single- or few-cycle regime, where strong fields can be applied without damaging the sample. Achieving such short pulses at longer wavelengths remains difficult, though recent pulse-compression techniques have shown encouraging progress [23]. Similarly, polarization shaping in this spectral range is technically demanding but has recently been addressed through the use of spatial light modulators [24].

Together, these advances are poised to expand the experimental reach of ultrafast lightwave valleytronics into previously unexplored regimes.

**Figure 2.** (a) A tailored trefoil light waveform coherently manipulates the band structure of monolayer hBN. (b) For a two-dimensional system with hexagonal symmetry, the band structure exhibits minima at the K and K′ valleys. In monolayer hBN, the bandgap is $\Delta \approx 5.9$ eV. (c) Upon interaction with a strong trefoil waveform, atomic couplings are modified, and the laser-dressed system is described by a Haldane-type Hamiltonian with complex next-nearest-neighbour hoppings that differ in phase by π between sublattices having on-site energies $\pm\Delta/2$. (d-g) As the field (and its vector potential) rotates, these hoppings evolve from real to imaginary, corresponding to directions toward M or K/K′. This lifts the degeneracy between K and K′, reducing the bandgap in one valley while enlarging it in the other. (h) Consequently, the effective bandgap oscillates with the trefoil field's rotation, producing valley-dependent excitation dynamics. (i-l) The normalized conduction-band populations show maximum electron density in the valley with the reduced bandgap. The first Brillouin zone (green hexagon) and vector potential orientation (purple triangles) are shown in the insets. Figure is adapted from Ref. [13].

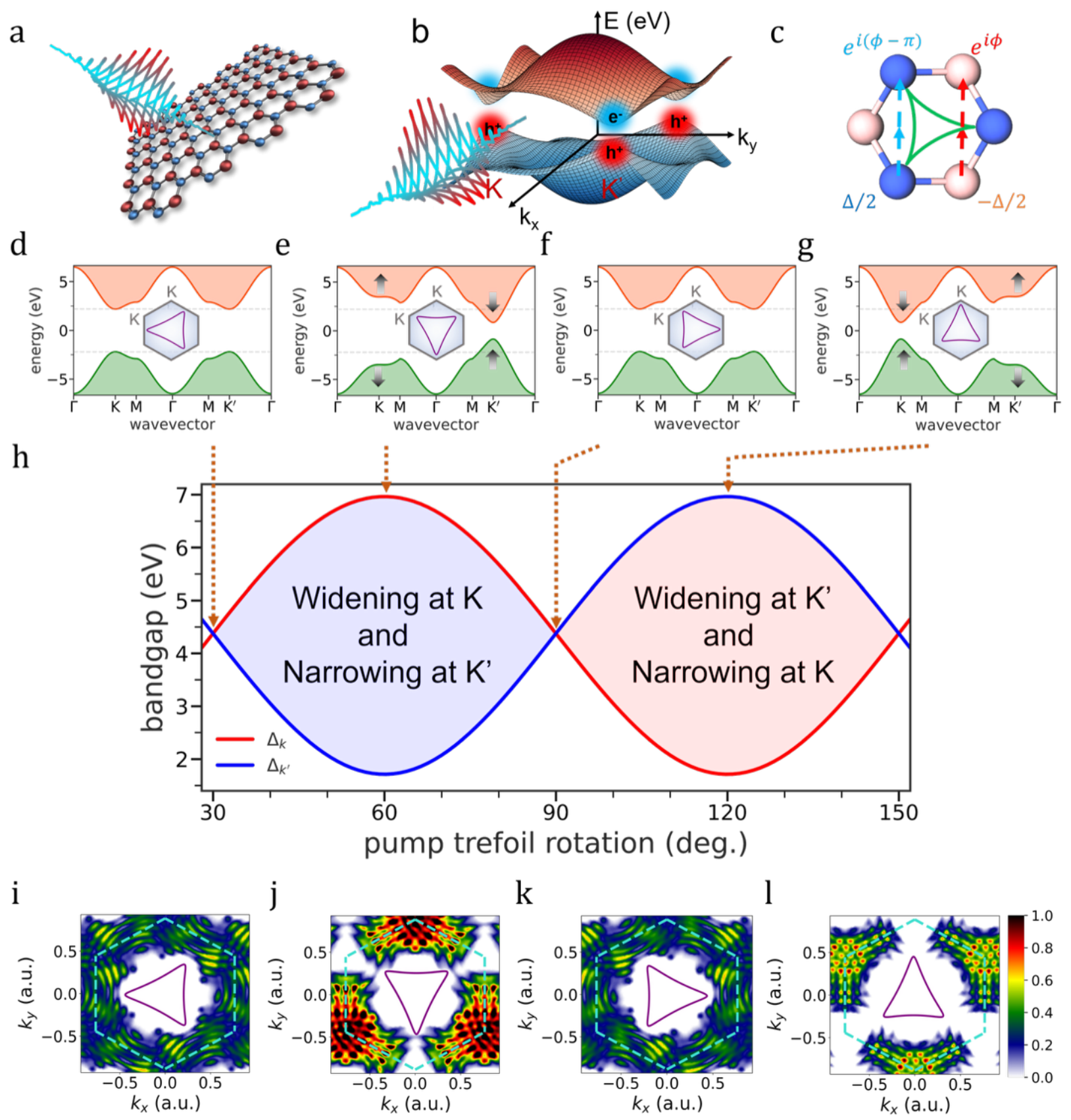

Traditionally, valleytronic phenomena have been probed through optical measurements. However, the advent of ultrashort light pulses in the ultraviolet to hard X-ray range offers unprecedented opportunities to observe valley dynamics in momentum space with elemental specificity and attosecond-to-femtosecond temporal resolution. Recent progress in high-harmonic-generation [25] and free-electron laser [26] sources has opened the door to such investigations, promising a deeper, time-resolved understanding of valleytronic processes in quantum materials.

## Concluding remarks

Lightwave valleytronics represents a rapidly evolving frontier at the intersection of ultrafast optics, quantum materials, and topological science. By exploiting sub-cycle control of light fields, it enables direct manipulation of the valley degree of freedom on attosecond-to-femtosecond timescales—far beyond the reach of conventional optical methods. Recent advances in tailored lightwave engineering, including trefoil have demonstrated the feasibility of all-optical valley initialization, switching, and readout, establishing the foundation for future valley-based logic and quantum information architectures.

Looking ahead, the confluence of long-wavelength few-cycle light sources, polarization shaping, and ultrashort high-energy probes promises unprecedented insight into symmetry breaking, electron dynamics, and light–matter coupling in two-dimensional materials and their heterostructures. Equally exciting is the prospect of strong-field–induced topological phase transitions, which may enable ultrafast, dissipationless optoelectronic devices or stepping into the possibility of valleytronic qubit technology. Continued progress in laser science, material design, and time-resolved spectroscopy will be pivotal in translating these concepts into functional technologies.

Ultimately, lightwave valleytronics not only deepens our understanding of nonequilibrium quantum phenomena but also opens transformative routes toward integrated quantum–electronic–photonic platforms operating on the fastest timescales permitted by nature.

## Acknowledgements

S B acknowledges the financial support from Indian Institute of Science, Bangalore, India.

## References

[1] Schaibley J R, Yu H, Clark G, Rivera P, Ross J S, Seyler K L, Yao W and Xu X 2016 Valleytronics in 2D materials *Nat. Rev. Mater.* **1** 16055

[2] Mak K F, He K, Shan J and Heinz T F 2012 Control of valley polarization in monolayer $MoS_2$ by optical helicity *Nat. Nanotechnol.* **7** 494–8

[3] Zeng H, Dai J, Yao W, Xiao D and Cui X 2012 Valley polarization in $MoS_2$ monolayers by optical pumping *Nat. Nanotechnol.* **7** 490–3

[4] Mitra S, Biswas S, Schötz J, Schötz J, Pisanty E, Förg B, Förg B, Kavuri G A, Burger C, Burger C, Okell W, Okell W, Högner M, Högner M, Pupeza I, Pervak V, Pervak V, Pervak V, Lewenstein M, Wnuk P, Wnuk P, Kling M F and Kling M F 2020 Suppression of individual peaks in two-colour high harmonic generation *Journal of Physics B: Atomic, Molecular and Optical Physics* **53** 134004

[5] Schötz J, Förg B, Schweinberger W, Liontos I, Masood H A, Kamal A M, Jakubeit C, Kling N G, Paasch-Colberg T, Biswas S, Högner M, Pupeza I, Alharbi M, Azzeer A M and Kling M F 2020 Phase-Matching for Generation of Isolated Attosecond XUV and Soft-X-Ray Pulses with Few-Cycle Drivers *Phys. Rev. X* **10** 041011

[6] Biswas S, Förg B, Ortmann L, Schötz J, Schweinberger W, Zimmermann T, Pi L, Baykusheva D, Masood H A, Liontos I, Kamal A M, Kling N G, Alharbi A F, Alharbi M, Azzeer A M, Hartmann G, Wörner H J, Landsman A S and Kling M F 2020 Probing molecular environment through photoemission delays *Nat. Phys.* **16** 778–83

[7] Biswas S, Trabattoni A, Rupp P, Magrakvelidze M, Madjet M E A, De Giovannini U, Castrovilli M C, Galli M, Liu Q, Månsson E P, Schötz J, Wanie V, Wnuk P, Colaizzi L, Mocci D, Reduzzi M, Lucchini M, Nisoli M, Rubio A, Chakraborty H S, Kling M F and Calegari F 2025 Correlation-driven attosecond photoemission delay in the plasmonic excitation of C60 fullerene *Sci. Adv.* **11** eads0494

[8] Jiménez-Galán, Silva R E F, Smirnova O and Ivanov M 2020 Lightwave control of topological properties in 2D materials for sub-cycle and non-resonant valley manipulation *Nat. Photon.* **14** 728–32

[9] Mitra S, Jiménez-Galán Á, Aulich M, Neuhaus M, Silva R E F, Pervak V, Kling M F and Biswas S 2024 Light-wave-controlled Haldane model in monolayer hexagonal boron nitride *Nature* **628** 752–7

[10] Tyulnev I, Jiménez-Galán Á, Poborska J, Vamos L, Russell P S J, Tani F, Smirnova O, Ivanov M, Silva R E F and Biegert J 2024 Valleytronics in bulk MoS2 with a topologic optical field *Nature* **628** 746–51

[11] Haldane F D M 1988 Model for a quantum Hall effect without Landau levels: condensed-matter realization of the parity anomaly *Phys. Rev. Lett.* **61** 2015–8

[12] Langer F, Schmid C P, Schlauderer S, Gmitra M, Fabian J, Nagler P, Schüller C, Korn T, Hawkins P G, Steiner J T, Huttner U, Koch S W, Kira M and Huber R 2018 Lightwave valleytronics in a monolayer of tungsten diselenide *Nature* **557** 76–80
[13] Sharma S, Gill D, Krishna J, Dewhurst J K, Elliott P and Shallcross S 2025 Combining THz and Infrared Light to Control Valley Charge and Current in Gapless Graphene *Nano Lett.* **25** 3791–8
[14] Sharma S, Elliott P and Shallcross S 2023 THz induced giant spin and valley currents *Sci Adv* 9 eadf3673
[15] Lesko D M B, Weitz T, Wittigschlager S, Li W, Heide C, Neufeld O and Hommelhoff P 2025 Optical control of electrons in a Floquet topological insulator *arXiv*:2407.17917
[16] Mrudul M S, Jiménez-Galán Á, Ivanov M and Dixit G 2021 Light-induced valleytronics in pristine graphene *Optica* **8** 422–7
[17] Jiménez-Galán Á, Silva R E F, Smirnova O, Smirnova O, Ivanov M, Ivanov M and Ivanov M 2021 Sub-cycle valleytronics: control of valley polarization using few-cycle linearly polarized pulses *Optica* **8** 277–80
[18] Sharma S, Gill D, Krishna J, Dewhurst J K and Shallcross S 2024 Direct coupling of light to valley current *Nat Commun* **15** 1–9
[19] Gucci F, Molinero E B, Russo M, San-Jose P, Camargo F V A, Maiuri M, Ivanov M, Jiménez-Galán Á, Silva R E F, Conte S D and Cerullo G 2026 Encoding and manipulating ultrafast coherent valleytronic information with lightwaves *Nat Photon.* https://doi.org/10.1038/s41566-025-01823-w
[20] Silva R E F, Silva R E F, Ivanov M, Ivanov M, Ivanov M, Jiménez-Galán Á and Jiménez-Galán Á 2022 All-optical valley switch and clock of electronic dephasing *Opt. Express* **30** 30347–55
[21] Elliott P, Sharma S and Shallcross S 2022 Valley control by linearly polarized laser pulses: example of WSe2 *Optica* **9** 947–52
[22] Ye Z, Sun D and Heinz T F 2016 Optical manipulation of valley pseudospin *Nat Phys* **13** 26–9
[23] Safaei R, Fan G, Kwon O, Légaré K, Lassonde P, Schmidt B E, Ibrahim H and Légaré F 2020 High-energy multidimensional solitary states in hollow-core fibres *Nat. Photon.* **14** 733–9
[24] Ogawa K, Kanda N, Murotani Y and Matsunaga R 2024 Programmable generation of counterrotating bicircular light pulses in the multi-terahertz frequency range *Nat. Commun.* **15** 1–8
[25] Sie E J, Rohwer T, Lee C and Gedik N 2019 Time-resolved XUV ARPES with tunable 24–33 eV laser pulses at 30 meV resolution *Nat. Commun.* **10** 1–11
[26] Shokeen V, Heber M, Kutnyakhov D, Wang X, Yaroslavtsev A, Maldonado P, Berritta M, Wind N, Wenthaus L, Pressacco F, Min C H, Nissen M, Mahatha S K, Dziarzhytski S, Oppeneer P M, Rossnagel K, Elmers H J, Schönhense G and Dürr H A 2024 Real-time observation of non-equilibrium phonon-electron energy and angular momentum flow in laser-heated nickel *Sci. Adv.* **10** eadj2407

# 6. Nonlinear optics

**Jan Wilhelm**[1]

[1] Institute of Theoretical Physics and Regensburg Center for Ultrafast Nanoscopy (RUN), University of Regensburg, 93059 Regensburg, Germany

E-mail: jan.wilhelm@ur.de

## Status

Valleytronic operations in monolayer transition metal dichalcogenides (TMDs) have attracted continued interest due to their ability to encode and manipulate information in the valley degree of freedom. Pioneering work focused on linear optical processes, where circularly polarized light enables valley-selective photoexcitation, exciting electrons at either the +K or −K valley [1, 2]. Valley polarization can be read out via helicity-resolved photoluminescence. A related effect is the valley-exclusive optical Stark and Bloch-Siegert (OSBS) shift [3], which modulates the bandgap of the Floquet band structure selectively at one valley (Fig. 1a).

More recently, nonlinear optical valleytronics [4–11] has opened new pathways for manipulating and detecting valley-dependent properties in a non-invasive and ultrafast manner. These methods rely on virtual carrier dynamics, avoiding real carrier excitation and enabling sub-picosecond control. A representative example is the detection of a valley-contrasting bandgap using third-order Kerr rotation [10] (Fig. 1b). In this scheme, an elliptical laser beam induces a valley-selective OSBS shift at +K via its circular component. When tuned to generate resonant third-harmonic generation (THG) at +K, the beam produces THG with helicity +1. At −K, THG is off-resonant and carries helicity −1. The imbalance in THG phase and intensity between the two valleys leads to a rotation of the THG polarization ellipse, revealing the underlying bandgap asymmetry.

Theoretical analysis of nonlinear valleytronic operations is essential not only for explaining present results but also for engineering new schemes. Two main approaches are emerging. The phenomenological route uses the wave-vector group at the valleys [12] to constrain the structure of the nonlinear susceptibility tensor $\boldsymbol{\chi}^{(n)}$ [13, 9, 10]. For effects originating from valley imbalance, magnetic point groups are needed to account for broken time-reversal symmetry. Polarization properties of the nonlinear optical signal are then described via Stokes parameters.

The microscopic approach derives $\boldsymbol{\chi}^{(n)}$ from the k-resolved band structure and dipole matrix elements [14, 15, 9, 10]. Recent work [10] used the Floquet band structure to describe the OSBS and its impact on $\boldsymbol{\chi}^{(3)}$, predicting a quadratic scaling of THG Kerr rotation with the driving field (Fig. 1b). The model also captures key observables such as spectral line shape and dependence on the transition dipole matrix elements and on band structure parameters like trigonal warping.

Continued progress in theoretical modelling (as outlined below) and experiment (e.g., Ref. [4]) are expected to enable various nonlinear optical schemes to extract valley-contrasting information.

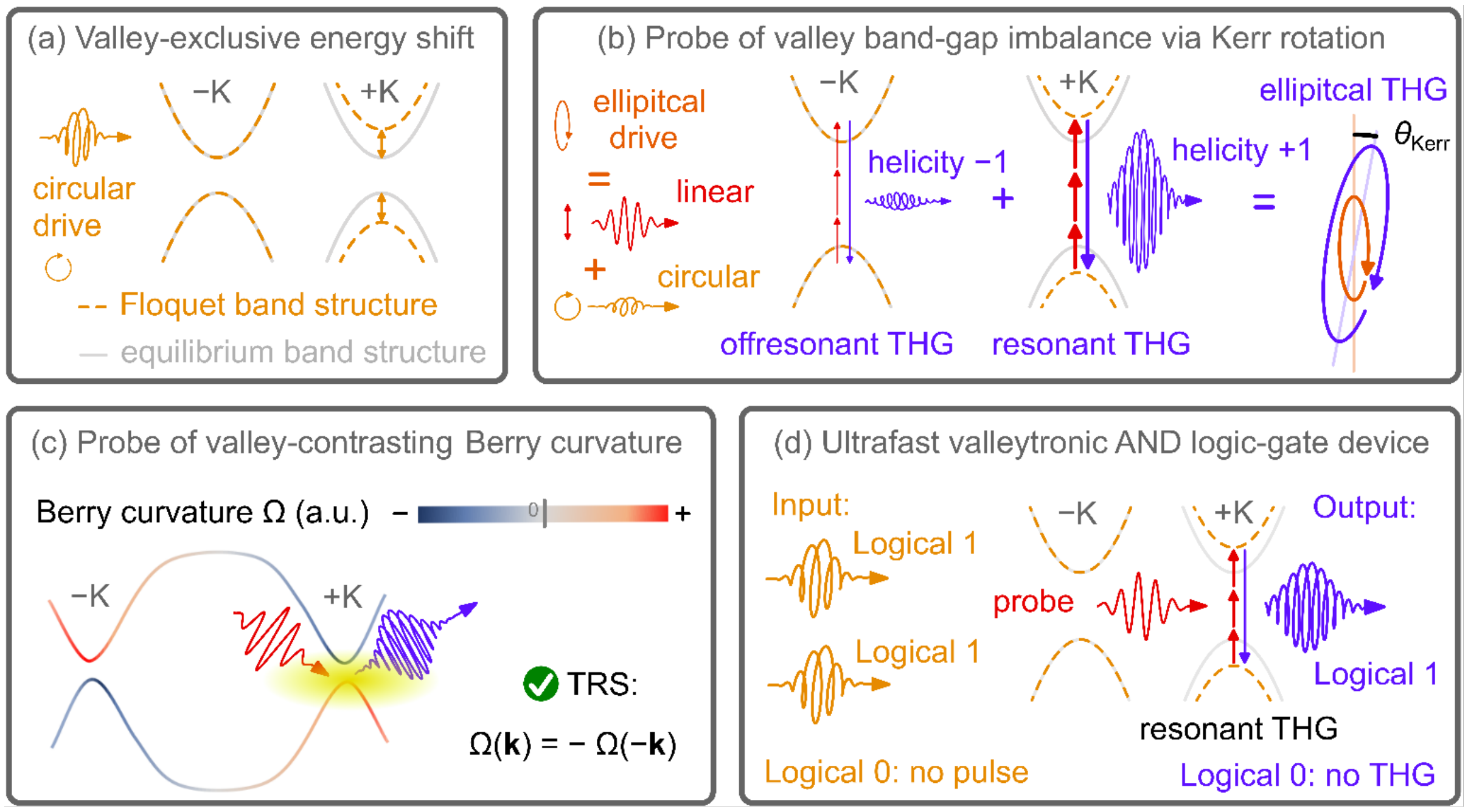

**Figure 1**. (a) Floquet band structure (dashed orange) under circular driving shows a valley-selective energy shift of the bandgap, originating from OSBS [3]. (b) Nonlinear-optical detection of a valley bandgap imbalance via THG Kerr rotation [10]: an elliptical drive induces an OSBS bandgap shift predominantly at +K. The resulting THG is resonant at +K (THG helicity +1) and off-resonant at −K (THG helicity −1), causing a rotation of the emitted THG polarization ellipse. (c) Valley-contrasting Berry curvature Ω(k), where time-reversal symmetry enforces Ω(k) = −Ω(−k). This antisymmetry offers a route to valley-resolved nonlinear optical probes of quantum-geometric properties. (d) Concept for an ultrafast, valleytronic AND logic gate based on nonlinear optics. Two input pulses (logical "1") are required to induce a sufficient bandgap shift for resonant THG emission (logical "1" output). In the absence of one input pulse (logical "0"), no THG is generated (logical "0").

## Current and future challenges

Nonlinear optical valleytronics builds on initial demonstrations of valley-selective nonlinear optical processes [4–11]. To further advance the field, several theoretical and conceptual challenges remain.

(i) *Extracting quantum geometric information from nonlinear optical signals.* A central challenge is to determine the extent to which valley-contrasting quantum-geometric properties such as Berry curvature contribute to nonlinear responses. In monolayer TMDs, time-reversal symmetry imposes $\Omega(\mathbf{k}) = -\Omega(-\mathbf{k})$, making Berry curvature a valley-contrasting quantity, as sketched in Fig. 1c. Low-order nonlinear processes, such as second- and third-harmonic generation, offer analytically tractable regimes where the influence of quantum geometry could be isolated. A first work in this direction has identified k-derivatives of the Berry curvature in second-order optical responses [15]. Establishing signatures of Berry curvature in valley-selective nonlinear signals would represent a major conceptual advance.

(ii) *Role of excitonic effects in nonlinear valleytronics*. Excitons dominate the optical response in TMDs. Their impact on nonlinear valley-selective signals remains to be investigated. The interplay between excitonic states and Berry curvature could modify selection rules or obscure purely geometric contributions. While analytical and numerical methods for including excitonic effects in nonlinear response exist [16-18], they have yet to be systematically applied to the context of nonlinear-optical valleytronics.

(iii) *Role of dephasing from electron-phonon and electron-electron interactions*. In models of light–matter interactions, dephasing is typically introduced through a simplified dephasing rate. To justify and refine this approximation, a more realistic treatment of electron-phonon and electron-electron interactions is needed.

(iv) *Generalizing beyond TMDs using Floquet band engineering*. Valley-selective nonlinear responses in TMDs often rely on OSBS shifts. However, Floquet band structure engineering is a general consequence of periodic driving and can be applied to other materials. Identifying materials with valley degrees of freedom that support controlled Floquet bandgap shifts would broaden the applicability of nonlinear valleytronics beyond TMDs.

(v) *Formal treatment of high-order nonlinear susceptibility*. Using a Floquet Hamiltonian inserted into the expression for the nonlinear susceptibility tensor [9, 10] captures an important contribution to valley-selective nonlinear responses. However, this approach may omit higher-order correction terms that affect the optical signal. A complete treatment requires evaluation of susceptibility tensors such as $\boldsymbol{\chi}^{(3)}(3\omega;\omega,\omega,\omega)$ and $\boldsymbol{\chi}^{(5)}(3\omega;\omega, \omega,\omega,\omega,-\omega)$ (evaluation solely with the equilibrium band structure) relevant for THG Kerr rotation [10]. Comparison between the Floquet-derived response and these terms could offer deeper mechanistic insight.

(vi) *Optical logic using nonlinear valleytronic effects*. All-optical logic based on valley-selective harmonic generation might offer a new computing paradigm. For example, logical operations can be encoded through the presence or absence of input pulses that modulate the bandgap via OSBS, enabling conditional activation of resonant harmonic generation, as illustrated in Fig. 1d.

### Advances in science and technology to meet challenges

Advancing nonlinear optical valleytronics relies on further progress in theoretical modeling, both to clarify the microscopic mechanisms underlying valley-selective nonlinear responses and to inspire new valley-sensitive probe designs. Early theoretical approaches based on Floquet Hamiltonians [9, 10] have already provided mechanistic insights, but these must be expanded to account for all higher-order corrections and more realistic material parameters.

A key need is the evaluation of nonlinear susceptibility tensors starting from first-principles band structures and dipole matrix elements. Analytical approaches, particularly those involving symbolic computation tools, will enable the identification of dominant contributions and correction terms beyond leading-order Floquet approximations. This approach also offers a way to separate the influence of quantum geometry from other mechanisms driving valley selectivity.

Capturing quantum geometric properties, such as Berry curvature and quantum metric, within nonlinear optical theory remains a central goal. Systematic perturbative frameworks are needed to isolate valley-contrasting geometric contributions and connect them to observable signatures in nonlinear optical processes. These developments are essential for interpreting nonlinear valley-selective signals in terms of their geometric origin.

A major frontier is the integration of excitonic effects into analytical theories of nonlinear-optical valleytronics [9, 10]. While numerical methods are established to describe excitonic effects in nonlinear optics, analytical treatments are less established in describing nonlinear optical phenomena. Bridging this gap requires advancing analytical treatments of many-body and Green's function techniques to describe how excitons modify both selection rules and, importantly, the influence of quantum geometry in various nonlinear responses.

More broadly, it may prove useful to develop a theoretical framework that links Floquet theory, perturbative nonlinear optics, and many-body physics. Such a framework should consistently account for realistic band structures and valley-specific interactions. It must also incorporate dephasing due to electron-phonon and electron-electron interactions.

While experimental progress is essential for validation and application, theoretical progress can help design new valleytronic probes based on nonlinear optics and define their fundamental limits. Both an improved mechanistic and improved quantitative understanding of nonlinear valley-selective phenomena will clarify existing observations and will guide the design of new nonlinear valleytronic protocols and application to new materials.

### Concluding remarks

Nonlinear optical valleytronics offers new opportunities for probing and controlling valley-specific phenomena using ultrafast and non-invasive light–matter interactions. Early microscopic models, such as the analytical evaluation of nonlinear susceptibilities based on Floquet band structures, have provided mechanistic insight into valley-selective responses. Key open challenges include incorporating excitonic effects into these frameworks and identifying signatures of valley-contrasting quantum geometry like Berry curvature in nonlinear optical signals. Continued theoretical development, analytical and numerical, in close connection with experimental progress, will be essential for guiding the design of future valley-selective optical probes and functionalities.

### Acknowledgements

J.W. gratefully acknowledges continued valuable discussions and collaborations with F. Evers, F. Friedrich, M. Graml, P. Herrmann, S. Klimmer, T. Lettau, S. Piehler, J. Schmitz, A. Seith, S. S. Shanbhag, and G. Soavi. Funding from the German Research Foundation (Deutsche Forschungsgemeinschaft, DFG) is acknowledged through the Emmy Noether Programme (project number 503985532), CRC 1277 (project number 314695032, sub-project A03), and RTG 2905 (project number 502572516).

### References

[1] Xiao D, Liu G, Feng W, Xu X, Yao W 2012 Coupled spin and valley physics in monolayers of $MoS_2$ and other group-VI dichalcogenides *Phys Rev. Lett.* **108** 196802

[2] Mak K F, He K, Shan J and Heinz T F 2012 Control of valley polarization in monolayer $MoS_2$ by optical helicity *Nat. Nanotechnol.* **7** 494

[3] Sie E J, Lui C H, Lee Y-H, Fu L, Kong J and Gedik N 2017 Large, valley-exclusive Bloch-Siegert shift in monolayer $WS_2$ *Science* **355** 1066

[4] García de Abajo F J, Basov D N, Koppens F H L, Orsini L, Ceccanti M, Castilla S, Cavicchi L, Polini M, Gonçalves P A D, Costa A T, Peres N M R, Mortensen N A, Bharadwaj S, Jacob Z, Schuck P J, Pasupathy A N, Delor M, Liu M K, Mugarza A, Merino P, Cuxart M G, Chávez-Angel E, Svec M, Tizei L H G, Dirnberger F, Deng H, Schneider C, Menon V, Deilmann T, Chernikov A, Thygesen K S, Abate Y, Terrones M, Sangwan V K, Hersam M C, Yu L, Chen X, Heinz T F, Murthy P, Kroner M, Smolenski T, Thureja D, Chervy T, Genco A, Trovatello C, Cerullo G, Dal Conte S, Timmer D, De Sio A, Lienau C, Shang N, Hong H, Liu K, Sun Z, Rozema L A, Walther P, Alù A, Cotrufo M, Queiroz R, Zhu X Y, Cox J D, Dias E J C, Rodríguez Echarri Á, Iyikanat F, Marini A, Herrmann P, Tornow N, Klimmer S, Wilhelm J, Soavi G, Sun Z, Wu S, Xiong Y, Matsyshyn O, Kumar R K, Song J C W, Bucher T, Gorlach A, Tsesses S, Kaminer I, Schwab J, Mangold F, Giessen H, Sánchez Sánchez M, Efetov D K, Low T, Gómez-Santos G, Stauber T, Álvarez-Pérez G, Duan J, Martín-Moreno L, Paarmann A, Caldwell J D, Nikitin A Y, Alonso-González P, Mueller N S, Volkov V, Jariwala D, Shegai T, van de Groep J, Boltasseva A, Bondarev I V, Shalaev V M, Simon J, Fruhling C, Shen G, Novko D, Tan S, Wang B, Petek H, Mkhitaryan V, Yu R, Manjavacas A, Ortega J E, Cheng X, Tian R, Mao D, Van Thourhout D, Gan X, Dai Q, Sternbach A, Zhou Y, Hafezi M, Litvinov D, Grzeszczyk M, Novoselov K S, Koperski M, Papadopoulos S, Novotny L, Viti L, Vitiello M S, Cottam N D, Dewes B T, Makarovsky O, Patanè A, Song Y, Cai M, Chen J, Naveh D, Jang H, Park S, Xia F, Jenke P K, Bajo J, Braun B, Burch K S, Zhao L and Xu X 2025 Roadmap for Photonics with 2D Materials *ACS Photonics* **12** 3961

[5] Ho, Y W, Rosa H G, Verzhbitskiy I, Rodrigues M J L F, Taniguchi T, Watanabe K, Eda G, Pereira V M and Viana-Gomes J C 2020 Measuring Valley Polarization in Two-Dimensional Materials with Second-Harmonic Spectroscopy *ACS Photonics* **7** 925
[6] Mouchliadis L, Psilodimitrakopoulos S, Maragkakis G M, Demeridou I, Kourmoulakis G, Lemonis A, Kioseoglou G and Stratakis E 2021 Probing valley population imbalance in transition metal dichalcogenides via temperature-dependent second harmonic generation imaging *npj 2D Mater. Appl.* **5** 6
[7] Dogadov O, Trovatello C, Yao B, Soavi G, Cerullo G 2022 Parametric Nonlinear Optics with Layered Materials and Related Heterostructures *Laser Photonics Rev.* **16** 2100726
[8] Herrmann P, Klimmer S, Lettau T, Monfared M, Staude I, Paradisanos I, Peschel U and Soavi G 2023 Nonlinear All-Optical Coherent Generation and Read-Out of Valleys in Atomically Thin Semiconductors *Small* **13** 2301126
[9] Herrmann P, Klimmer S, Lettau T, Weickhardt T, Papavasileiou A, Mosina K, Sofer Z, Paradisanos I, Kartashov D, Wilhelm J and Soavi G 2025 Nonlinear valley selection rules and all-optical probe of broken time-reversal symmetry in monolayer $WSe_2$ *Nat. Photonics* **19** 300
[10] Friedrich F, Herrmann P, Shanbhag S S, Klimmer S, Wilhelm J and Soavi G 2025 Direct measurement of broken time-reversal symmetry in centrosymmetric and non-centrosymmetric atomically thin crystals with nonlinear Kerr rotation *Nat. Photonics*
[11] Liu X, Liu D, Sun Y, Li Y, Cui Z 2025 Signatures of intervalley interferences and valley asymmetries in high-order harmonic generation in pristine graphene *Phys. Rev. A* **111**, 063112
[12] Dresselhaus, M, Dresselhaus G and Jorio A 2007, Springer Group Theory: Application to the Physics of Condensed Matter
[13] Gallego S V, Etxebarria J, Elcoro L, Tasci E S and Perez-Mato J M 2019 Automatic calculation of symmetry-adapted tensors in magnetic and non-magnetic materials: a new tool of the Bilbao Crystallographic Server *Acta Crystallogr. A* **75** 438
[14] Aversa C, Sipe J E 1995 Nonlinear optical susceptibilities of semiconductors: Results with a length-gauge analysis *Phys. Rev. B* **52**, 14636
[15] Soavi G, Wilhelm J 2025 The role of Berry curvature derivatives in the optical activity of time-invariant crystals arXiv preprints arXiv:2501.03684
[16] Taghizadeh A, Pedersen T G 2019 Nonlinear optical selection rules of excitons in monolayer transition metal dichalcogenides *Phys. Rev. B* **99** 235433
[17] Ruan J, Chan Y-H, Louie S-G 2024 Exciton Enhanced Nonlinear Optical Responses in Monolayer h-BN and $MoS_2$: Insight from First-Principles Exciton-State Coupling Formalism and Calculations *Nano Lett.* **24** 15533
[18] Kira M, Koch S W 2006 Many-body correlations and excitonic effects in semiconductor spectroscopy *Prog. Quantum Electron.* **30** 155

# 7. Proximity control of valleys

**Igor Žutić[1], Konstantin S. Denisov[1] and Tong Zhou[2]**

[1] Department of Physics, University at Buffalo, State University of New York, Buffalo, New York 14260, USA
[2] Eastern Institute for Advanced Study, Eastern Institute of Technology, Ningbo, Zhejiang 315200, China

E-mail: zigor@buffalo.edu, tzhou@eitech.edu.cn

### Status

For almost a century materials have been transformed through proximity effects, acquiring properties from the neighbours, to become superconducting, magnetic, topologically nontrivial, or with an enhanced spin–orbit coupling (SOC) [1]. With most of the proximity effects limited to only several or tens of nm, two-dimensional (2D) materials and their van der Waals (vdW) heterostructures [1, 2] are particularly suitable to display emergent valley-dependent phenomena of proximitized materials. Similar to lifting the spin degeneracy in spintronics, removing the valley degeneracy is often crucial in manipulating valley degrees of freedom. Instead of a small Zeeman splitting of 0.1–0.2 meV/T in transition metal dichalcogenides (TMDs) [2], which requires large applied magnetic fields, removing the valley degeneracy is more effective using magnetic substrates, where the proximity-induced spin splitting in TMDs can reach tens of meV [1]. The resulting magnetic proximity effect is not simply an induced uniform effective magnetic field but can become valley selective, as observed through excitons [3], a common fingerprint of various proximity effects [1], shown in Figs.1(a),(b).

Valley-selective proximity effects can alter the band topology and reverse the helicity of the emitted light in Fig.1(c) [4]. In 2D materials with broken inversion symmetry, such as gapped graphene and TMDs, the opposite sign of the momentum-space Berry curvature $\Omega(k)$ [5] in different valleys is responsible for a valley Hall effect [2, 5]. A sign reversal of $\Omega(k)$ along an internal boundary in a 2D material is realized in quantum valley Hall kink (QVHK) states in Fig.1(d) [6, 7]. The resulting topological defect supports counterpropagating 1D chiral electrons, protected by the valley-inversion symmetry to realize waveguides, valley valves, and beam splitters, with valley-momentum locking that could enable coherent quantum transport networks [7]. However, just as quantum spin Hall (QSH) states with the spin-momentum locking, such kink states are fragile under short-range disorder [8]. This limitation can be overcome in spin-valley-momentum double locking of the quantum spin-valley Hall kink (QSVHK) states, illustrated in Fig.1(e) [8]. The underlying QVH and QSH states differ by the ratio of the SOC and staggered potential strength, a large effective electric field, implemented through a ferroelectric proximity effect [9, 10], controls this ratio to realize robust QSVHK states in a single material, by incorporating valley-dependent properties in a monolayer derived from bismuthene on SiC with a record topological gap [11]. 2D ferroelectric materials themselves or, through ferroelectric proximity effects, can host and manipulate topological kink states and realize movable Dirac cones, shown in Fig.1(f), that lead to the nonlinear optical and transport properties [12].

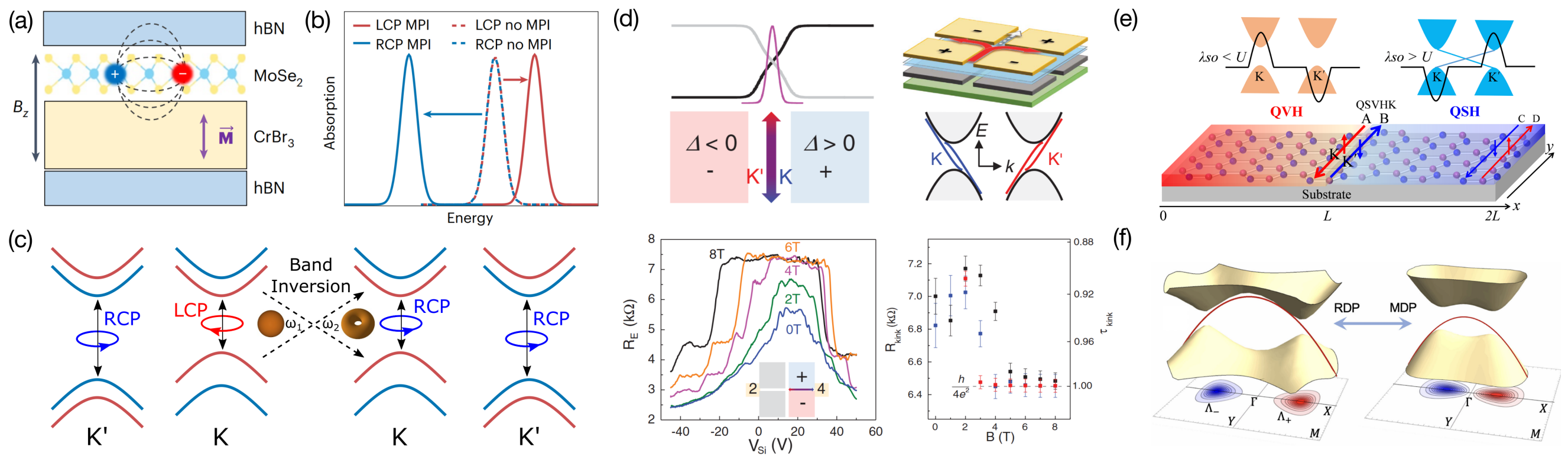

**Figure 1.** (a) Schematic of the $MoSe_2/CrBr_3$ heterostructure. An exciton is a bound state of the + (-) charged hole (electron) with marked electric field lines. $CrBr_3$ has an out-of-plane magnetization, M. (b) Valley-dependent exciton absorption shift induced by the magnetic proximity interactions (MPIs) [3]. The solid (dashed) lines: Magnetic (nonmagnetic) $CrBr_3$. (c) The exciton helicity reversal when the trivial topology (ball) at the K valley is transformed into a nontrivial topology (doughnut) through the band inversion. (d) The valley router device, the potential profile and wave function distribution of the valley-momentum locked kink states and the left- and right-turn configurations of the waveguide controlled by the top gates voltage signs and quantized conductance due to ballistic channels [7]. (e) The quantum spin-valley Hall kink states (QSVHK) in a junction formed by QVH and QSH insulators, distinguished by the relative strength of the spin-orbit coupling, lSO [8]. (f) The energy dispersions of movable Dirac electrons in 2D bulk around the two valleys, L+, L-, and of the kink states (red line) pinned by a domain wall are transformed by moving the Dirac points [12].

## Current and future challenges

Even for extensively studied graphene or TMDs, the resulting proximity-induced phenomena challenge their accurate description, including the sensitivity to the stacking twist angle of the resulting superconducting and strongly-correlated states [1], novel forms of emergent SOC [13], or exciton formation [14]. While proximity effects are known to influence equilibrium valley-dependent properties, how they transform nonequilibrium phenomena and dynamical response is less understood. The valley Zeeman SOC proximity-induced in graphene from TMD $PdSe_2$ is anisotropic leading to the tenfold modulation of spin lifetimes [15]. Graphene on a magnetic substrate acquires both proximity-induced SOC and exchange splitting. An oscillatory electric field induces spin-flip transitions in the electric dipole spin resonance (EDSR) with a THz response in Fig.2a, the tenfold-enhanced optical absorption, compared to that for the pristine graphene [16]. The resulting enhancement arises from the resonant behaviour of the coupled spin-pseudospin torque in Fig.2b [16]. Similar coupled spin-pseudospin dynamics is inherent to other manifestations of the spin-charge conversion, beyond EDSR, whose dynamical properties remain to be understood. Since the spin-orbit torque (SOT) is studied for various energy-efficient spintronics [1, 17], could valley-dependent spin-pseudospin dynamics enable ultrafast SOT?

While SOC proximity appears central for valley-dependent properties and emergent phenomena [1], it is important to explore possible alternatives. A growing class of unconventional magnets with nonrelativistic spin splitting, including their subset altermagnets [18], have complex magnetic textures. The resulting proximity-imprinted information to 2D materials is not limited to spin splitting, but also includes effective SOC. A similar scenario was considered from magnetic arrays to transform normal and superconducting states, including for creating elusive Majorana states for topological fault-tolerant quantum computing [1]. An outstanding challenge is demonstrating that the tunability of altermagnets enables also tunable valley-dependent proximity effects, illustrated in Fig.2(c)-(e), both in the normal and the superconducting state [19]. Since proximity effects in 2D materials were monitored through the transformation of excitons [1], their valley-dependent counterparts in altermagnets [20], shown in Fig.2(e), offer an important platform to describe the resulting many-body effects and dynamical phenomena.

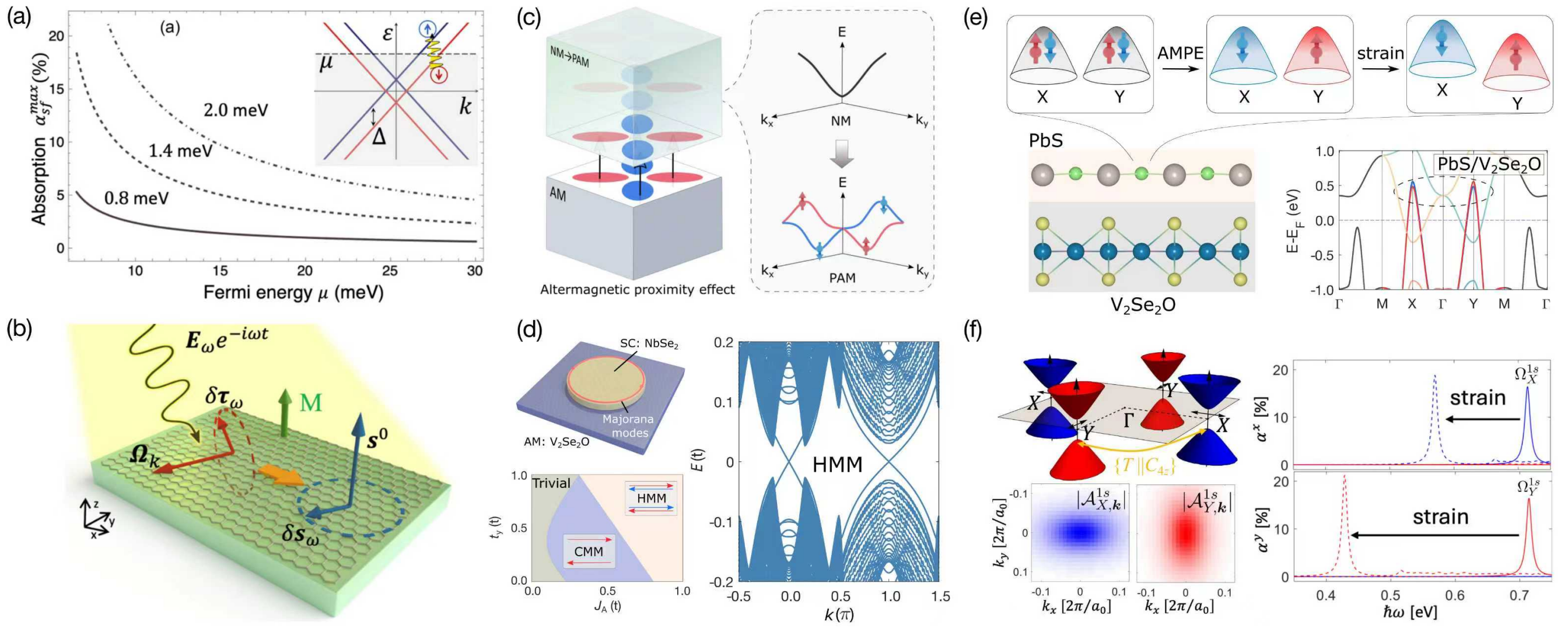

**Figure 2.** (a) THz EDSR absorption with the Femi energy, SOC 0.8, 1.4, 2.0 meV, and the Dirac spectrum [16]. (b) THz radiation (yellow) with electric field $\boldsymbol{E}_\omega$ is absorbed creating a pseudospin component, $\delta\boldsymbol{\tau}_\omega$, which precesses in the pseudospin field, $\mathbf{W}_k$, exerting a torque on a spin due to SOC (orange arrow) [16]. (c) The altermagnetic proximity effect (AMPE): The AM order penetrates into a nonmagnetic (NM) layer, inducing band evolution as the NM transforms into a proximitized altermagnet (PAM) [19]. (d) Transformation of an *s*-wave superconductor $NbSe_2$ into a topological superconductor hosting edge Majorana modes (red circle) via AMPE from $V_2Se_2O$, together with the calculated topological phase diagram and ribbon spectra of the PAM/$NbSe_2$ showing helical Majorana (HMM) modes, which can be converted into chiral Majorana modes (CMM) [19]. (e) The band structure of the PbS/$V_2Se_2O$ heterostructure, demonstrating AMPE- and strain-induced spin and valley splitting in monolayer PbS [19]. (f) Left: The band structure of a *d*-wave 2D AM with spin-polarized X (blue) and Y (red) valleys, connected by the symmetry {T II $C_{4z}$}, shown with the 1*s* excitonic momentum-space wave functions. Right: Absorption spectra of X/Y polarized light under strain, where $\Omega_{X,Y}$ are the strain-tunable excitonic energies in the valleys X/Y [20].

There is a continued challenge to better quantify proximity effects and their characteristic length scales. Even in TMDs, density-functional theory may overestimate the proximity-induced spin splitting by an order of magnitude [1], so it is unclear which are the optimal materials candidates. Similar uncertainties are also accompanying SOC proximity. Many 2D materials have inherent limitations for potential application by being air sensitive, available only as small flakes, or described by the ordering temperature below the room temperature, as typical for 2D ferromagnets [1, 17].

## Advances in science and technology to meet challenges

Some guidance what to expect from the advances in proximity control of valleys comes from the superconducting proximity effects and the insights derived from their microscopic mechanism described by Andreev reflection [1]. Quantifying the underlying proximity effect is offered by comparing the magnitude between the proximity-induced and the native superconducting gap. The resulting transformation of materials by proximity effects is exemplified in Josephson junctions. 2025 Nobel Prize in Physics shows that the fundamental manifestations of superconducting proximity effects and macroscopic quantum tunnelling are also closely related to the advances in quantum computing. An ambitious vision calls for realizing elusive Majorana states and their non-Abelian statistics for fault-tolerant quantum computing, where the topological protection is not native, but derived from proximity effects [1]. Here recent research on altermagnets could provide a platform to study such Majorana states, where their tunability is derived from the tunability of the valley-dependent spin splitting inherent to altermagnets, illustrated in Figs.2(c)-(e) [19]. Exploring altermagnets may provide room-temperature valley- and spin-dependent proximity effects, even when 2D ferromagnets remain scarce at room temperature [1].

2D materials and proximity effects could offer exciting opportunities for phenomena already realized in more conventional systems. For example, spintronics, photonic, and electronics were integrated at room-temperature by using SOT to reverse the magnetization and injected spin in the neighbouring semiconductor and, through SOC and the conservation of angular momentum transfer, reverse the helicity of the emitted light [21]. The resulting transfer of spin information from carriers, limited to nanoseconds and microns in nonmagnetic materials, to the circularly polarized light, allows spin information to travel much faster and farther, opening a path for high-performance interconnects and even interplanetary communication [21]. A 2D counterpart of this scheme may expand the capabilities of such SOT-light-emitting diodes and provide a platform for coherent light and 2D spin-lasers where the valley-dependent effects enable ultrafast modulation of the emitted polarized light [1].

Until 2017 [1] magnetic proximity effects were constrained to the single-particle description. With optical response in 2D materials often dominated by excitons this necessitates a different approach and an accurate inclusion of the Coulomb interaction. A similar generalization is needed in the predictive first-principles description, beyond density functional theory, where the self-consistent Green-function description [22, 23] offers not only an accurate description of bandgaps, but also excitations, strong correlations, and disorder. This accuracy is essential for correctly understanding proximity and valley-dependent effects in multilayer heterostructures which closely depend on the band alignment and interfacial properties.

### Concluding remarks

Tailoring valley degrees of freedom enables novel phenomena and emerging applications. While valley effects have been studied for decades in materials such as silicon, diamond, or AlAs, a related success in emulating the better-known manipulation of spin and spintronic applications has been modest. However, this situation is changing with a rapidly growing number of 2D materials and their vdW heterostructures where proximity effects can transform their properties and control valley-dependent phenomena. Possible outcomes of these proximitized materials can be seen from enhancing topological protection against different types of disorder to dynamical manifestations where the interplay between spin and valley degrees of freedom could enable ultrafast phenomena and enhance desirable THz response. These proximity effects can go both ways, just as in superconductor/ferromagnet junction a leaking superconductivity in a ferromagnet is also accompanied by the net magnetization in the superconductor. Even seemingly well-understood magnetic proximity effects can be valley-dependent and, as known from their influence on excitons, go beyond the single-particle picture. This both poses challenges for the underlying description where the accurate first-principles methods could offer predictive tools to determine the magnitude of proximity effects, as well as offers important opportunities to advance the valley control.

## Acknowledgements

This work was supported by the U.S. DOE, Office of Science BES, Award No. DE-SC0004890 (I. Ž., K.D.), by the National Science Foundation Grants No. ECCS-2512491 (I. Ž.), by the Air Force Office of Scientific Research under Award No. FA9550-22-1-0349 (I. Ž., K.D.), and by the National Natural Science Foundation of China (12474155), and the Zhejiang Provincial Natural Science Foundation of China (LR25A040001) (T.Z.).

## References

[1] Žutić I, Matos-Abiague A,Scharf B, Dery H and Belashchenko K 2019 Proximitized materials *Mater Today* **22** 85
[2] Schaibley J R, Yu H, Clark G, Rivera P, Ross J S, Seyler K L, Yao W and Xu X 2016 Valleytronics in 2D materials *Nat Rev Mater* **1** 1
[3] Choi J, Lane C, Zhu J-X and Crooker S A 2023 Asymmetric magnetic proximity interactions in $MoSe_2$/$CrBr_3$ van der Waals heterostructures *Nat Mater* **22** 305
[4] Xu G, Zhou T, Scharf B and Žutić I 2020 Optically probing tunable band topology in atomic monolayers *Phys Rev Lett* **125** 157402
[5] Xiao D, Chang M-C and N. Q 2010 Berry phase effects on electronic properties *Rev Mod Phys* **82** 1959
[6] Ju L, Shi Z, Nair N, Lv Y, Jin C, Velasco Jr J, Ojeda-Aristizabal C, Bechtel H A, Martin M C, Zettl A, Analytis J and Wang F 2015 Topological valley transport at bilayer graphene domain walls *Nature* **520** 650
[7] Li J, Zhang R-X, Yin Z, Zhang J, Watanabe K, Taniguchi T, Liu C, Zhu J 2018 A valley valve and electron beam splitter *Science* **362** 1149
[8] Zhou T, Cheng S, Schleenvoigt M, Schüffelgen P, Jiang H, Yang Z and Žutić I 2021 Quantum spin-valley Hall kink States: From concept to materials design *Phys Rev Lett* **127** 116402
[9] Chen J, Cui P and Zhang Z 2024 Ferroelectricity-tuned band topology and superconductivity in 2D materials and related *Adv Funct Mater* **34** 2408625
[10] Niranjan M K, Wang Y, Jaswal S S and Tsymbal E Y 2009 Prediction of a switchable two-dimensional electron gas at ferroelectric oxide interfaces *Phys Rev Lett* **103** 016804
[11] Reis F, Li G, Dudy L, Bauernfeind M, Glass S, Hanke W, Thomale R, Schäfer J and Claessen R 2017 Bismuthene on a SiC substrate: A candidate for a high-temperature quantum spin Hall *Science* **257** 287
[12] Denisov K S, Liu Y and Žutić I 2025 Movable Dirac points with ferroelectrics: Kink states and Berry curvature dipole *Phys Rev Lett* **134** 246602
[13] Li Y and Koshino M 2019 Twist-angle dependence of the proximity spin-orbit coupling in graphene on transition-metal dichalcogenides *Phys Rev B* **99** 075438
[14] Faria Junior P E, Naimer T, McCrear K M, Jonker B T, Finley J J, Crooker S A, Fabian J and Stier A V 2023 Proximity-enhanced valley Zeeman splitting at the $WS_2$/graphene interface *2D Mater* **10** 034002
[15] Sierra J F, Světlík J, Savero Torres W, Camosi L, Herling F, Guillet T, Xu K, Sebastián Reparaz J, Marinova V, Dimitrov D and Valenzuela S O 2025 Room-temperature anisotropic in-plane spin dynamics in graphene induced by $PdSe_2$ proximity *Nat Mater* **24** 876
[16] Denisov K S, Rozhansky I V, Valenzuela S O and Žutić I 2024 Terahertz spin-light coupling in proximitized Dirac materials *Phys Rev B* **109** L201406
[17] Sierra J F, Fabian J, Kawakami R K, Roche S and Valenzuela S O 2021 van der Waals heterostructures for spintronics and opto-spintronics *Nat Nanotechnol* **16** 856
[18] Liu Q, Dai X and Blügel S 2025 Different facets of unconventional magnetism *Nat Phys* **21** 329
[19] Zhu Z, Huang R, Chen X, Duan X, Zhang J, Žutić I and Zhou T 2025 Altermagnetic proximity effect arXiv:2509.06790 preprint
[20] Cao J D, Denisov K S, Liu Y and Žutić I 2025 Symmetry classification for alternating excitons in two-dimensional altermagnets *Phys Rev Lett* **135** 266703
[21] Dainone P A, Prestes N F, Renucci P, Bouche A, Morassi M, Devaux X, Lindemann M, George J-M, Jaffres H, Lemaitre A, Xu B, Stoffel M, Chen T, Lombez L, Lagarde D, Cong G, Ma T, Pigeat P, Vergnat M, Rinnert H, Marie X, Han X, Mangin S, Rojas- Sanchez J C, Wang J P, Beard M C, Žutić I and Lu Y 2024 Controlling the helicity of light by electrical magnetization switching *Nature* **627** 783
[22] Pashov D, Acharya S, Lambrecht W R L, Jackson J, Belashchenko K D, Chantis A Jamet F and van Schilfgaarde M 2020 Questaal: A package of electronic structure methods based on the linear muffin-tin orbital technique *Comp Phys Commun* **249** 107065

[23] Acharya S, Pashov D, Rudenko A N, Rösner M, van Schilfgaarde M and Katsnelson M I 2021 Importance of charge self-consistency in description of strongly correlated systems *npj Comput Mater* **7** 208

# 8. Quantum geometry in moiré valleytronics

**Huiyuan Zheng[1], Wang Yao[2,3] and Hongyi Yu[4,5,*]**

[1] Department of Materials Science and Engineering, University of Washington, Seattle, Washington 98195, USA
[2] New Cornerstone Science Lab, Department of Physics, The University of Hong Kong, Hong Kong Special Administrative Region of China, People's Republic of China
[3] HK Institute of Quantum Science & Technology, The University of Hong Kong, Hong Kong Special Administrative Region of China, People's Republic of China
[4] Guangdong Provincial Key Laboratory of Quantum Metrology and Sensing & School of Physics and Astronomy, Sun Yat-Sen University (Zhuhai Campus), Zhuhai 519082, People's Republic of China
[5] State Key Laboratory of Optoelectronic Materials and Technologies, Sun Yat-Sen University (Guangzhou Campus), Guangzhou 510275, People's Republic of China

*E-mail: yuhy33@mail.sysu.edu.cn

## Status

The quantum geometry plays important roles in condensed matter physics and material sciences, which can result in exotic quantum phenomena including the various Hall effects and magnetoelectric responses. The quantum geometry measures how the particle's internal quantum structure varies with parameters like position and momentum. In long-wavelength moiré patterns of the layered transition metal dichalcogenides (TMDs), electrons (as well as holes) and excitons exhibit varieties of quantum degrees-of-freedom (DoF) including the spin, valley, layer-pseudospin, moiré potential sublattice pseudospin, and exciton Rydberg states, see Table I. The periodic modulation of the local stacking registry in moiré patterns (Fig. 1a) can lead to variations of these DoF with position or momentum. These underlie the quantum geometric quantities associated with electrons and excitons, such as real- and momentum-space Berry curvatures, which can affect the transport, topological and optical properties of the system.

In monolayer TMDs, the variation of the Bloch state orbital composition with the momentum gives rise to finite Berry curvatures $\pm\Omega_{\mathbf{K}}$ for electrons in $\pm\mathbf{K}$ valleys. An in-plane driving force applied on the electron can introduce an anomalous transverse velocity proportional to $\pm\Omega_{\mathbf{K}}$, resulting in the valley Hall effect (Fig. 1b). The exciton can inherit quantum geometries from its electron and hole constituents, giving rise to excitonic valley Hall effects under a thermal or density gradient [1, 2]. Besides these transport phenomena, quantum geometries also manifest as energy corrections to exciton Rydberg states, resulting in a ~15 meV splitting between 2p± excitons [3].

The layer DoF has attracted great interest recently. In R-type homobilayer $MoTe_2$ moiré patterns, the competition between the interlayer hopping and layer energy offset due to interfacial electric polarizations, both varying with local stacking registries, leads to spatially modulating layer-hybridizations [4, 5]. For instance, the hole layer-pseudospin exhibits a skyrmion lattice structure with the moiré periodicity, and a periodic real-space Berry curvature emerges which plays the role of an out-of-plane magnetic field with opposite signs in $\pm\mathbf{K}$ valleys (Fig. 1c). Combined with the moiré potential, the low-energy minibands can be well-described by a three-orbital lattice model (Fig. 1d), with the basis being wavepackets localized at AB, BA and AA regions [5]. The nontrivial topology of low-energy minibands [4, 5] can be attributed to the Berry flux threading the lattice, with valley-contrasted Chern numbers tunable through varying the moiré periodicity and interlayer bias. In this system, quantum anomalous Hall effects have been observed by a series of transport/optical experiments [6-9], whereas layer-pseudospin skyrmion patterns has been visualized by STM/S tomography [10, 11].

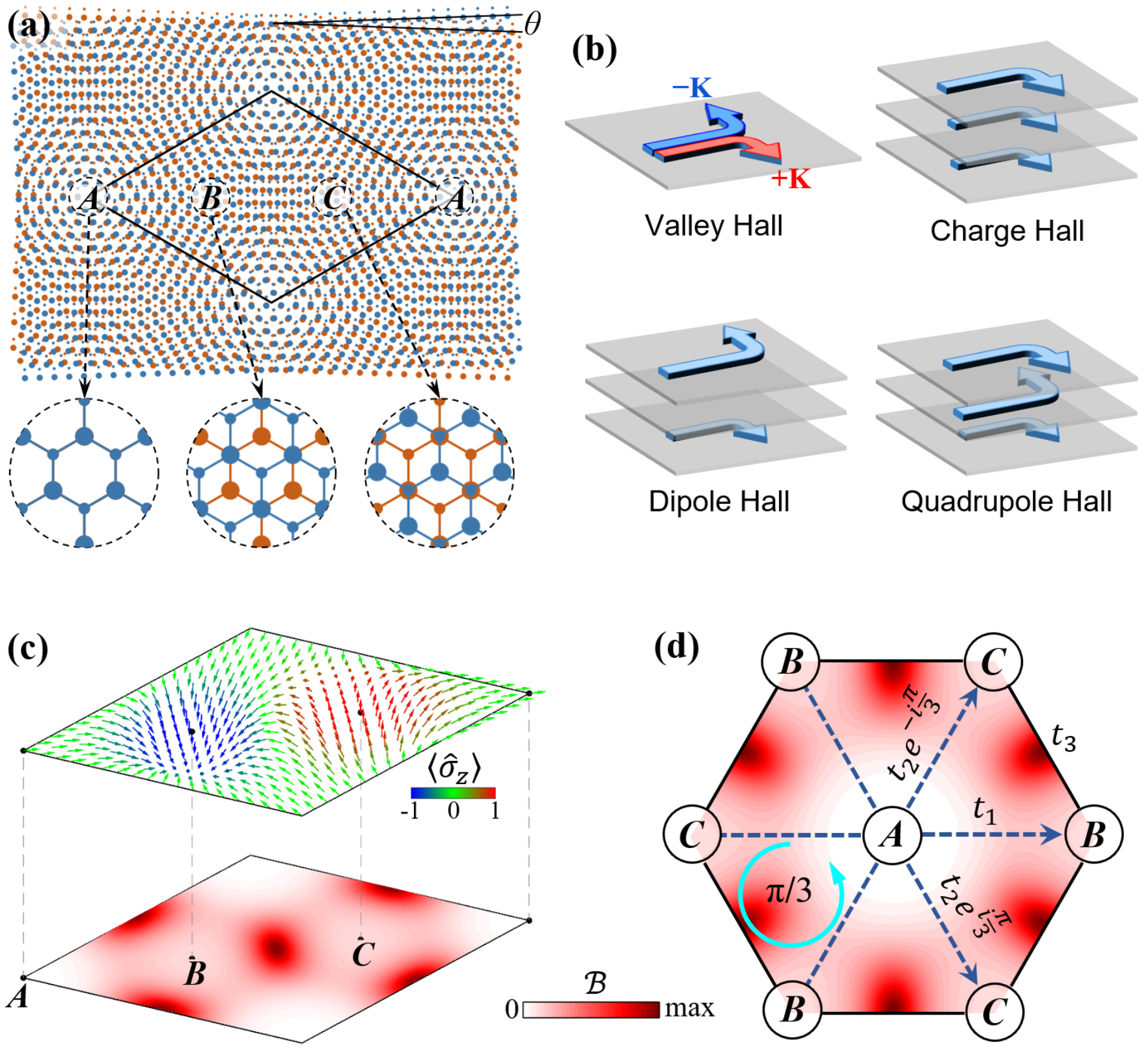

**Figure 1.** (a) The upper-panel is an illustration of the twisted bilayer TMDs moiré pattern, with the diamond shape being a moiré supercell. $\theta$ is the interlayer twist angle. The lower-panels show the atomic stacking registries of the three $2\pi/3$-rotationally symmetric local regions $\boldsymbol{A}$, $\boldsymbol{B}$ and $\boldsymbol{C}$, which correspond to AA, AB and BA stackings, respectively. (b) Illustrations of the valley, charge, dipole and quadrupole Hall effects in monolayer and multilayer TMDs systems. Reproduced from [14] by permission of ACS publications. (c) The upper-panel shows the position-dependent layer-pseudospin texture for the $-\mathbf{K}$ valley hole in twisted homobilayer $MoTe_2$ moiré patterns, with the colour indicating the out-of-plane layer-pseudospin component $\langle\hat{\sigma}_z\rangle$. The lower-panel shows the resulting real-space Berry curvature $\mathcal{B}$, which acts as an out-of-plane magnetic field. (d) The three-orbital flux lattice model for the low-energy hole in twisted homobilayer $MoTe_2$ moiré patterns, with the basis being wavepackets localized at local potential extrema $\boldsymbol{A}$, $\boldsymbol{B}$ and $\boldsymbol{C}$. The phases of nearest-neighbour hoppings are determined by the flux $\pi/3$ of each equilateral triangle. (c) and (d) are reproduced from [5] (CC BY 4.0).

**Table 1.** A brief summary for various DoF in TMDs moiré patterns.

| Quantum DoF | Description | Ways for manipulation | Variations with parameters |
|---|---|---|---|
| Spin/valley | Spin and valley are locked for electrons and bright excitons near ±**K** valleys due to the strong spin-orbit-coupling. | Magnetic and optical fields can break the degeneracy between ±**K**. The electron-hole exchange interaction couples ±**K** valleys of bright excitons. | The exchange-induced intervalley coupling of bright excitons varies with the centre-of-mass momentum. |
| Atomic orbital | Conduction/valence band Bloch states near ±**K** mainly consist of $d_{z^2}$ and $d_{x^2-y^2} \pm id_{xy}$ orbitals, which can be well described by a massive Dirac model. | ———— | The atomic orbital compositions of Bloch states vary with momentum. |
| Layer-pseudospin | The layer-pseudospin state of the electron/exciton is determined by the layer energy offset and interlayer hopping. | An out-of-plane electric field introduces layer energy differences. | Both the layer energy offset and interlayer hopping vary with local stacking registries at different positions. |
| Sublattice sites of the moiré potential | The moiré potential can exhibit multiple inequivalent local minima with distinct layer stacking registries. | An out-of-plane electric field, as different sublattice sites can exhibit different layer-pseudospins. | The sublattice-pseudospins of minibands vary with the momentum due to the in-plane inter-site hopping. |
| Exciton Rydberg states | Electron-hole relative motion of the exciton, forming discrete states usually denoted as 1s, $2p_\pm$, 2s, … similar to the 2D hydrogen model. | An in-plane electric field couples two Rydberg states with angular momentum difference ±1. | A periodic electrostatic potential introduces momentum-dependent hybridizations between Rydberg states. |

## Current and future challenges

The numerous DoF in moiré patterns provide additional opportunities to manipulate quantum geometries. Different DoF can couple with each other, which not only enriches the quantum geometry but also offers ways to manipulation one DoF through another. This research field faces challenges from both experimental and theoretical sides. Experimentally, clean samples with improved qualities as well as moiré patterns with high spatial homogeneities are needed. Many theoretical ideas have been proposed but further explorations are still needed. A critical challenge involves the experimental verification of theoretical proposals and detection of the related quantum geometries. For example, as an intrinsic property of monolayer valley electrons, what novel geometric effects can the Berry curvature $\Omega_{\mathbf{K}}$ introduce in the context of moiré? Theories have suggested that the combination of $\Omega_{\mathbf{K}}$ and a superlattice potential can lead to topologically nontrivial minibands, even without the layer-pseudospin structure [12-14]. The controllability of certain DoF also poses a challenge. The ±**K** valleys of the bright exciton are coupled by the electron-hole exchange interaction, but its strength is hard to control since both the magnitude and phase depend sensitively on the centre-of-mass momentum. The Rydberg states correspond to an internal DoF for the exciton, whose hybridization can be manipulated by an in-plane electric field [15]. However, a strong field is required to overcome the large splitting between the ground and excited Rydberg states.

For the quantum geometric properties associated with layer DoF, finding their manifestations in transport phenomena would be an interesting direction. In van der Waals multilayer systems, the charge current is expected to be layer-resolved and electrons can exhibit layer Hall counterflows. The interlayer hopping creates a layer-hybridized state with an out-of-plane electric dipole in the bilayer or quadrupole in the trilayer, enabling the multipole Hall effect [16], see Fig. 1b. The multipole Hall effect generates an in-plane magnetic moment which couples to an in-plane magnetic field, which can give rise to an intrinsic planar Hall effect [16]. The dipole Hall effect underlies novel forms of magnetoelectric responses in correlated magnetic insulating states in twisted TMDs homobilayer moiré. Such effects can serve as experimental signatures for the underlying quantum geometric quantities encoded by the layer DoF.

Besides the experimentally verified topological hole states, excitons with nontrivial topologies are also appealing, as it offers a platform for realizing the bosonic fractional Chern insulator. Critical challenges are the realization and detection of topological exciton bands. In bilayer TMDs, the large tunability of the interlayer exciton combined with its long lifetime and strong dipolar interaction makes it an excellent candidate, with its out-of-plane electric dipole facilitating the detection through transport measurements. Various proposals have been suggested to introduce nontrivial topologies to the exciton, for example, by attaching an electron to the topological hole in twisted bilayer $MoTe_2$ moiré [17-19]. Meanwhile, momentum-dependent hybridizations between exciton Rydberg states can be introduced by a periodic electrostatic potential. The resulting quantum geometry can lead to topologically nontrivial exciton minibands [20]. Moreover, the electron-hole exchange interaction enables layer-hybrid moiré excitons to hop coherently between moiré potential minima separated by $\sim$10 nm and/or across layers, with sizable valley-flipping channels [21]. In the context of twisted bilayer $MoTe_2$, such layer-hybrid moiré excitons feature opposite electric dipoles on the two sublattices. Exciton molecules can form through the dipolar attraction on the nearest-neighbour bonds of the honeycomb moiré lattice, and their valley-flipping hopping through the electron-hole exchange gives rise to non-Abelian lattice gauge field [22].

### Advances in science and technology to meet challenges

The presence of multiple DoF in moiré systems introduces substantial complexities in understanding the underlying electric, magnetic, and optical properties. For analysing novel quantum geometric effects in these systems, effective models need to be constructed which should incorporate the various intertwining DoF. Conventional continuum models of the moiré pattern with low computational cost can offer insights, as well as semi-quantitative descriptions with parameters fitted from band structure calculations. Meanwhile, the recently developed large-scale DFT calculations combined with machine learning simulations can offer quantitative results with improved accuracy, as they can well reproduce the scalar and vector potentials induced by the lattice relaxation effect in moiré patterns [23]. Experimentally, advances in fabrication technologies for the van der Waals bilayer structure have greatly improved the sample qualities of the fabricated moiré devices. However, further improvements are still needed to reduce the defect densities and unintentional twist-angle/strain variations across the sample, which can affect the spatial homogeneity of the moiré pattern. For probing quantum geometries associated with the layer DoF, it is desirable to have the capability for resolving the layer-dependent transport.

### Concluding remarks

Moiré patterns of TMDs and other layered materials have proved to be a versatile platform for exploring the quantum geometry and topology in condensed matter physics. There exists varieties of quantum DoF in this system including the spin, valley, layer, sublattice and exciton Rydberg states, whose complex interplay can result in new concepts for the quantum geometry. Combined with the high tunability through modulating the twist angle, substrate/strain, doping density, etc., the low-energy electrons and excitons in moiré patterns can be exploited to explore a broad range of phenomena, as well as practical device applications.

### Acknowledgements

H.Y. acknowledges support by NSFC under Grant No. 12274477. W.Y. acknowledges support by the National Natural Science Foundation of China (No. 12425406) and the Research Grant Council of Hong Kong (HKU SRFS2122-7S05).

### References

[1] Onga M, Zhang Y, Ideue T, Iwasa Y 2017 Exciton Hall effect in monolayer MoS2 *Nat. Mater.* **16** 1193-1197
[2] Ubrig N, Jo S, Philippi M, Costanzo D, Berger H, Kuzmenko A B, Morpurgo A F 2017 Microscopic Origin of the Valley Hall Effect in Transition Metal Dichalcogenides Revealed by Wavelength-Dependent Mapping *Nano Lett.* **17** 5719-5725
[3] Yong C-K, Utama M I B, Ong C S, Cao T, Regan E C, Horng J, Shen Y, Cai H, Watanabe K, Taniguchi T, Tongay S, Deng H, Zettl A, Louie S G, Wang F 2019 Valley-dependent exciton fine structure and Autler-Townes doublets from Berry phases in monolayer MoSe2 *Nat. Mater.* **18** 1065-1070
[4] Wu F, Lovorn T, Tutuc E, Martin I, Macdonald A H 2019 Topological Insulators in Twisted Transition Metal Dichalcogenide Homobilayers *Phys. Rev. Lett.* **122** 086402
[5] Yu H, Chen M, Yao W 2020 Giant magnetic field from moiré induced Berry phase in homobilayer semiconductors *Natl. Sci. Rev.* **7** 12-20
[6] Cai J, Anderson E, Wang C, Zhang X, Liu X, Holtzmann W, Zhang Y, Fan F, Taniguchi T, Watanabe K, Ran Y, Cao T, Fu L, Xiao D, Yao W, Xu X 2023 Signatures of Fractional Quantum Anomalous Hall States in Twisted MoTe2 *Nature* **622** 63-68
[7] Zeng Y, Xia Z, Kang K, Zhu J, Knüppel P, Vaswani C, Watanabe K, Taniguchi T, Mak K F, Shan J 2023 Thermodynamic evidence of fractional Chern insulator in moiré MoTe2 *Nature* **622** 69-73
[8] Park H, Cai J, Anderson E, Zhang Y, Zhu J, Liu X, Wang C, Holtzmann W, Hu C, Liu Z, Taniguchi T, Watanabe K, Chu J-H, Cao T, Fu L, Yao W, Chang C-Z, Cobden D, Xiao D, Xu X 2023 Observation of Fractionally Quantized Anomalous Hall Effect *Nature* **622** 74-79
[9] Xu F, Sun Z, Jia T, Liu C, Xu C, Li C, Gu Y, Watanabe K, Taniguchi T, Tong B, Jia J, Shi Z, Jiang S, Zhang Y, Liu X, Li T 2023 Observation of Integer and Fractional Quantum Anomalous Hall Effects in Twisted Bilayer MoTe2 *Phys. Rev. X* **13** 031037
[10] Zhang F, Morales-Durán N, Li Y, Yao W, Su J-J, Lin Y-C, Dong C, Liu X, Chen F-X R, Kim H, Watanabe K, Taniguchi T, Li X, Robinson J A, Macdonald A H, Shih C-K 2025 Experimental signature of layer skyrmions and implications for band topology in twisted WSe2 bilayers *Nat. Phys.* **21** 1217-1223
[11] Thompson E, Chu K T, Mesple F, Zhang X-W, Hu C, Zhao Y, Park H, Cai J, Anderson E, Watanabe K, Taniguchi T, Yang J, Chu J-H, Xu X, Cao T, Xiao D, Yankowitz M 2025 Microscopic signatures of topology in twisted MoTe2 *Nat. Phys.* **21** 1224-1230
[12] Su Y, Li H, Zhang C, Sun K, Lin S-Z 2022 Massive Dirac fermions in moiré superlattices: A route towards topological flat minibands and correlated topological insulators *Phys. Rev. Research* **4** L032024
[13] Suri N, Wang C, Hunt B M, Xiao D 2023 Superlattice engineering of topology in massive Dirac fermions *Phys. Rev. B* **108** 155409
[14] Tan T, Reddy A P, Fu L, Devakul T 2024 Designing Topology and Fractionalization in Narrow Gap Semiconductor Films via Electrostatic Engineering *Phys. Rev. Lett.* **133** 206601
[15] Zhu B, Xiao K, Yang S, Watanabe K, Taniguchi T, Cui X 2023 In-Plane Electric-Field-Induced Orbital Hybridization of Excitonic States in Monolayer WSe2 *Phys. Rev. Lett.* **131** 036901

[16] Zheng H, Zhai D, Xiao C, Yao W 2025 Layer Coherence Origin of Planar Hall Effect: From Charge to Multipole and Valley *Nano Lett.* **25** 10096-10101
[17] Xie M, Hafezi M, Sarma S D 2024 Long-Lived Topological Flatband Excitons in Semiconductor Moiré Heterostructures: A Bosonic Kane-Mele Model Platform *Phys. Rev. Lett.* **133** 136403
[18] Perea-Causin R, Liu H, Bergholtz E J 2025 Exciton fractional Chern insulators in moiré heterostructures *Phys. Rev. Research* **7** L042033
[19] Guo Z-H, Yan T, Zhao J-Z, Jin Y-J, Zhu Q Topological exciton bands and many-body exciton phases in transition metal dichalcogenide trilayer heterostructures *arXiv:2504.10189*
[20] Zhang N, Yao W, Yu H 2025 Engineering topological exciton structures in two-dimensional semiconductors by a periodic electrostatic potential *Phys. Rev. B* **112** 165406
[21] Zheng H, Li C, Yu H, Yao W 2025 Förster valley-orbit coupling and topological lattice of hybrid moiré excitons *Commun. Phys.* **8** 193
[22] Wang H, Yao W 2025 Emergent Kagome lattice and non-Abelian lattice gauge field of biexcitons in t-MoTe2 *Natl. Sci. Rev.*
[23] Zhang X-W, Wang C, Liu X, Fan Y, Cao T, Xiao D 2024 Polarization-driven band topology evolution in twisted MoTe2 and WSe2 *Nat. Commun.* **15** 4223

# 9. Flat band valleytronics

**Ting Cao[1], Dacen Waters[2] and Matthew Yankowitz[1,3]**

[1] Department of Materials Science and Engineering, University of Washington, Seattle, Washington, USA
[2] Department of Physics and Astronomy, University of Denver, Denver, CO, USA
[3] Department of Physics, University of Washington, Seattle, Washington, USA

E-mail: tingcao@uw.edu; dacen.waters@du.edu; myank@uw.edu

## Status

Flat electronic bands in van der Waals (vdW) materials provide a setting where Coulomb interactions dominate over kinetic energy, enabling a wide range of correlated and topological phenomena [1]. These bands can be realized in two ways: through intrinsic crystalline stacking, such as rhombohedral (ABC) graphene (Fig. 1a); or through moiré superlattices, where a small twist angle or lattice mismatch creates a long-wavelength interference pattern that reconstructs the band structure (Fig. 1b), as in twisted graphene and twisted transition metal dichalcogenides (TMDs).

A defining feature of many such systems is that the relevant electronic states lie near the corners of the Brillouin zone. In lattices with hexagonal Brillouin zones, low-energy states reside at the K and K' points, establishing valley as a two-level degree of freedom. Crucially, these valleys host Berry curvature, which acts as a lattice analogue of a magnetic field (Fig. 2a). In the single-particle limit, opposite valleys are degenerate and related by time-reversal symmetry; their Berry curvatures are equal and opposite and cancel in total. Strong Coulomb interactions in the flat-band regime can spontaneously lift this degeneracy, polarizing the system into a single valley. The resulting time-reversal symmetry breaking produces spin and orbital magnetizations (Fig. 2b). This can yield a nonzero Chern number (a topological invariant that corresponds to the number of chiral edge modes), realizing a zero-field analogue of the quantum Hall effect. Interaction-driven valley polarization underpins the emergence of a variety of phases, including integer and fractional quantum anomalous Hall states in both moiré and crystalline flat-band platforms.

Several material platforms now realize these ingredients in different combinations. Graphene-based systems—including twisted multilayers and rhombohedral graphene, either naturally stacked or aligned to hexagonal boron nitride (hBN)—provide highly tunable flat bands with nontrivial Berry curvature. Twisted TMD homo- and heterobilayers add intrinsic spin–orbit coupling (SOC), which locks spin and valley degrees of freedom. Appreciable SOC can also be imparted to graphene via proximity effects. Together, these platforms support a wide range of interaction-driven phases that leverage valley physics, including integer and fractional Chern insulators [2-6], quantum spin Hall states [7], multiferroics [8], intervalley-coherent phases [9, 10], states with broken translational symmetry [11, 12], and chiral superconductors [13, 14].

## Current and future challenges

The rapid progress of flat band valleytronics brings sharp challenges for both experiment and theory. Foremost are sample quality and reproducibility. Several forms of disorder still impede truly pristine devices, including atomic-scale defects in source crystals of TMDs and hBN dielectrics, and structural inhomogeneities in moiré superlattices introduced by local strain during sample assembly. The highest-quality devices are currently built from mechanically exfoliated flakes using polymer-free transfer techniques, but throughput remains low, lateral sample dimensions are typically tens of micrometers or smaller, and—crucially for moiré systems—precise control of the target superlattice wavelength is not yet reliable.

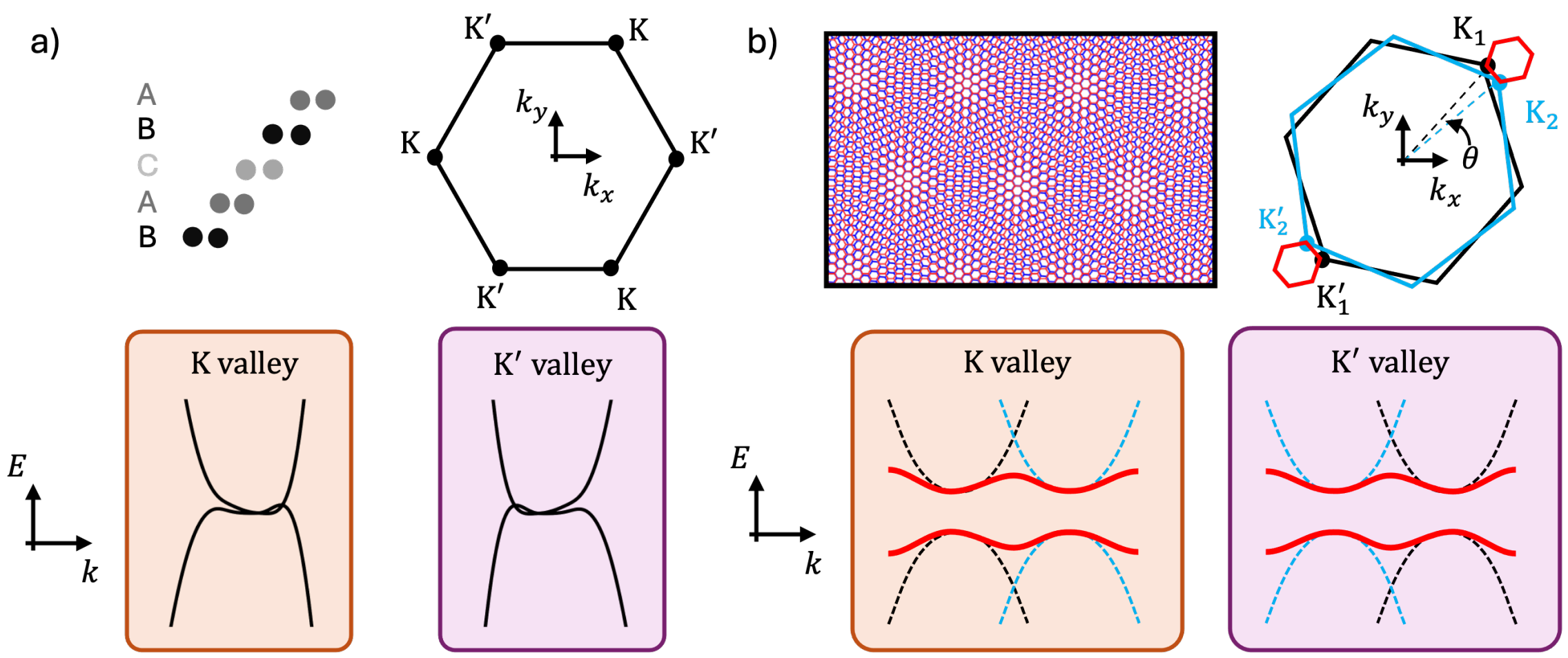

**Figure 1. Flat bands in vdW materials.** (a) Atomic structure of pentalayer rhombohedral (ABCAB) graphene (top left), Brillouin zone with K and K' points (top right), and the low energy bands of each valley (bottom). The bands become flat and host a large density of states near their touching points. (b) Moiré pattern formed by misaligning two triangular lattices with a twist angle $\theta$ (top left). Brillouin zones of each layer (black and blue) and the moiré Brillouin zone (red) formed in each valley due to the relative twist (top right). Schematic formation of flat bands (red) in a twisted moiré material.

Evidence for superconductivity coexisting with orbital ferromagnetism has emerged in rhombohedral graphene and twisted $MoTe_2$ [13, 14], most naturally attributed to finite-momentum pairing in a valley-polarized background. Superconductivity in twisted trilayer graphene has likewise been reported to show a diode effect [15], argued to arise from spontaneous breaking of time-reversal and inversion symmetries via partial valley polarization. Determining the pairing symmetry, topology, and precise valley order in these superconductors remains a central open problem.

In parallel, key questions surround the growing family of integer and fractional quantum anomalous Hall (QAH) states. While most appear consistent with complete valley polarization, moiré $MoTe_2/WSe_2$ heterobilayers show signatures compatible with more elaborate intervalley-coherent ordering [9]. For fractional Chern insulators (FCIs), the observed sequence largely mirrors the Jain hierarchy of the lowest Landau level [4-6], but the gap hierarchy, quasiparticle exchange statistics, and low-lying neutral modes remain only sparsely charted. Theory further indicates that the moiré superlattice can give composite fermions a dispersive band structure by breaking translational symmetry, opening a pathway to anyonic superconductivity within fully valley-polarized phases. The empirical adjacency of chiral superconductivity and FCIs in rhombohedral graphene and twisted $MoTe_2$ is suggestive of such a link, but direct evidence is still lacking.

Despite dramatic experimental progress, predictive understanding of flat bands in multilayer moiré and super-moiré systems is hindered by complex interplays of lattice reconstruction, long-range Coulomb interactions, and symmetry breaking that span multiple length scales [1]. Although quantum geometric properties such as Berry curvature and the Fubini-Study metric are now relatively straightforward to compute given a model [16, 17], optimizing them for topological many-body states remains elusive due to difficulties in constructing effective Hamiltonians and identifying parameters for system sizes beyond the reach of density functional theory (DFT) calculations. Competing ground states, such as FCIs versus charge density waves, have energetics depending sensitively on modeling details. This issue is even more pronounced for non-Abelian candidates. Moreover, predictions for valley physics at the M points of the Brillouin zone are beginning to emerge in other materials with distinct structural and electronic motifs [18, 19], requiring new modeling tools. Another frontier lies in characterizing collective excitations in flat bands—particularly terahertz-frequency valley-polarized excitons between moiré minibands. Unlike visible-light excitons in monolayer TMDs or magnetorotons in the fractional quantum Hall regime, these excitations couple valley order and topology under a tunable moiré potential, offering a new window into both quasiparticle dynamics and nonequilibrium valleytronic control.

### Advances in science and technology to meet challenges

Direct probes of the valley degree of freedom are essential. Circularly polarized light already addresses valleys in semiconducting TMDs with visible-wavelength gaps, but graphene and related platforms demand terahertz probes with either on-chip coupling or near-field enhancement from scanning probe tips. Momentum-resolved tools that distinguish valleys are likewise needed. Advanced nano-ARPES could provide this in principle, yet millikelvin quantum twisting microscopy (QTM) offers a compelling path to momentum-selective readout with device compatibility and sub-meV energy resolution. A parallel priority is joint measurement of valley polarization and orbital magnetism, the latter of which receives contributions from both bulk filled states and topological edges. Experimental probes that combine terahertz spectroscopy, magnetometry, and transport would clarify their nontrivial interplay.

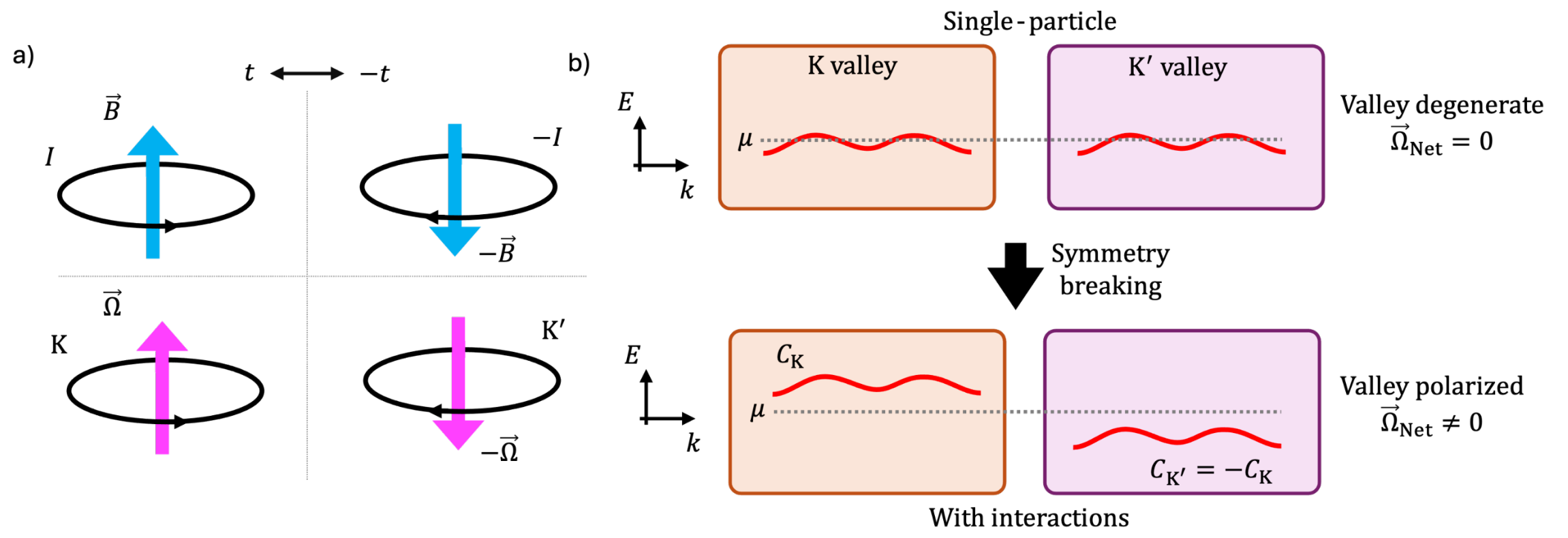

**Figure 2. Berry curvature and spontaneous valley polarization.** (a) Schematic showing the analogy between ordinary magnetic fields, ***B***, in real space and Berry curvature, ***Ω***, in momentum space. The K and K' valleys are related by time-reversal symmetry, just like loops of current, $I$, with opposite orientations. (b) Single-particle bands of an example flat band system (top), where time-reversal symmetry is present and each valley carries equal and opposite Berry curvature. When the chemical potential, $\mu$, is tuned into the flat bands, Coulomb interactions can spontaneously break time-reversal symmetry by lifting the valley degeneracy, leading to net non-zero Berry curvature and Chern numbers, $C$, (bottom).

Real-space visualization and engineered control of valley-ordered phases will also be crucial. For topological valley-polarized phases, scanning probes such as STM, scanning SQUID, and microwave impedance microscopy (MIM) can map edge conduction, local thermodynamic gaps, and topological domain textures to test bulk–boundary correspondence. Probes with nanometer-scale resolution are also needed to image valley-polarized states with simultaneous translational symmetry breaking. At the device level, advances are required to create clean lateral junctions between distinct valley orders, including superconductors interfaced with Chern insulators, or Josephson junctions with valley-polarized weak links. Such junctions enable phase-sensitive tests of pairing symmetry and provide platforms to engineer and detect non-Abelian zero modes, including Majoranas or parafermions. If successful, future work will also be needed to develop measurement protocols for anyon braiding.

New experimental strategies should also target subtle collective phenomena in flat-band valley systems. Sensitive spectroscopies and interferometric geometries that couple to intervalley coherence, often time-reversal symmetric and therefore invisible to conventional magnetometry, will be important. Low-loss cavities, on-chip resonators, and momentum-selective tunneling can access neutral modes and valley magnons over broad frequency and field ranges. Materials synthesis must also keep pace. Higher-purity source crystals for TMDs and hBN with reduced point-defect densities and control of strain landscapes are immediate goals. In the longer term, reproducible valley physics will benefit from increased throughput offered by wafer-scale crystal growth and vdW stack assembly.

A new generation of theory tools is poised to transform flat-band valleytronics into a predictive and designable platform. Multiscale machine-learning (ML) models are expanding beyond atomistic force fields toward learning full electronic structure across arbitrary moiré combinations, substrate environments, and lattice deformations [20]. This could enable fast, tunable predictions of flat bands and interaction form factors, paving the way for systematic design of topological phases. In parallel, modeling optical excitations remains a grand challenge: GW-BSE calculations are prohibitively expensive for large moiré supercells, while simplified models struggle to capture interband and interlayer coupling across multiple minibands. ML-driven approximations to excitonic response offer an emerging path forward. Looking further ahead, progress in quantum computing introduces powerful new motivations: superconducting circuits now reach hundreds of qubits, but error correction remains a barrier to scalability. Topologically ordered phases such as FCIs and non-Abelian anyonic states offer a tantalizing alternative for fault-tolerant architectures, if materials advances can be translated into quantum control schemes. Meeting this goal requires bridging condensed matter theory with device engineering, readout protocols, and qubit control—all areas where a new class of theoretical models, trained on experimental observables and quantum hardware constraints—will be essential.

### Concluding remarks

As flat-band valleytronics moves from discovery to design, progress will hinge on further developments of valley-resolved and momentum-selective probes, advanced device architectures that stitch together distinct valley orders, and new synthesis approaches that improve sample yield and uniformity. On the theory side, multiscale models calibrated to experiment should connect quantum geometry to many-body energetics and predict when chiral superconductivity, FCIs, or other exotic valley-ordered states prevail. Near-term milestones include decisive tests of superconducting pairing symmetry and valley order, gap and excitation measurements for FCIs, and reproducible junctions that enable anyon interferometry and braiding protocols. Achieving these will turn valley degrees of freedom into controllable resources for metrology, low-dissipation electronics, and ultimately fault-tolerant quantum information.

## Acknowledgements

M.Y. acknowledges support from the Department of Energy, Basic Energy Science Programs under award DE-SC0023062. T.C. acknowledges support from the Department of Energy, Basic Energy Science Programs under award DE-SC0025327.

## References

[1] Andrei E Y, Efetov D K, Jarillo-Herrero P, MacDonald A H, Mak K F, Senthil T, Tutuc E, Yazdani A and Young A F 2021 The marvels of moiré materials *Nat. Rev. Mater.* **6** 201

[2] Adak P C, Sinha S, Agarwal A and Deshmukh M M 2024 Tunable moiré materials for probing Berry physics and topology *Nat. Rev. Mater.* **9** 481

[3] Han T, Lu Z, Yao Y, Yang J, Seo J, Yoon C, Watanabe K, Taniguchi T, Fu L, Zhang F and Ju L 2024 Large quantum anomalous Hall effect in spin-orbit proximitized rhombohedral graphene *Science* **384** 647

[4] Park H, Cai J, Anderson E, Zhang Y, Zhu J, Liu X, Wang C, Holtzmann W, Hu C, Liu Z, Taniguchi T, Watanabe K, Chu J H, Cao T, Fu L, Yao W, Chang C Z, Cobden D, Xiao D and Xu X 2023 Observation of fractionally quantized anomalous Hall effect *Nature* **622** 74

[5] Xu F, Sun Z, Jia T, Liu C, Xu C, Li C, Gu Y, Watanabe K, Taniguchi T, Tong B, Jia J, Shi Z, Jiang S, Zhang Y, Liu X and Li T 2023 Observation of integer and fractional quantum anomalous Hall effects in twisted bilayer MoTe2 *Phys. Rev. X* **13** 031037

[6] Lu Z, Han T, Yao Y, Reddy A P, Yang J, Seo J, Watanabe K, Taniguchi T, Fu L and Ju L 2024 Fractional quantum anomalous Hall effect in multilayer graphene *Nature* **626** 759

[7] Kang K, Qiu Y, Watanabe K, Taniguchi T, Shan J and Mak K F 2024 Double Quantum Spin Hall Phase in Moiré WSe2 *Nano Lett.* **24** 14901

[8] Han T, Lu Z, Scuri G, Sung J, Wang J, Han T, Watanabe K, Taniguchi T, Fu L, Park H and Ju L 2023 Orbital multiferroicity in pentalayer rhombohedral graphene *Nature* **623** 41

[9] Li T, Jiang S, Shen B, Zhang Y, Li L, Tao Z, Devakul T, Watanabe K, Taniguchi T, Fu L, Shan J and Mak K F 2021 Quantum anomalous Hall effect from intertwined moiré bands *Nature* **600** 641

[10] Arp T, Sheekey O, Zhou H, Tschirhart C L, Patterson C L, Yoo H M, Holleis L, Redekop E, Babikyan G, Xie T, Xiao J, Vituri Y, Holder T, Taniguchi T, Watanabe K, Huber M E, Berg E and Young A F 2024 Intervalley coherence and intrinsic spin–orbit coupling in rhombohedral trilayer graphene *Nat. Phys.* **20** 1413

[11] Polshyn H, Zhang Y, Kumar M A, Soejima T, Ledwith P, Watanabe K, Taniguchi T, Vishwanath A, Zaletel M P and Young A F 2022 Topological charge density waves at half-integer filling of a moiré superlattice *Nature Physics* 18 42

[12] Su R, Waters D, Zhou B, Watanabe K, Taniguchi T, Zhang Y-H, Yankowitz M and Folk J 2025 Moiré-driven topological electronic crystals in twisted graphene *Nature* **637** 1084

[13] Han T, Lu Z, Hadjri Z, Shi L, Wu Z, Xu W, Yao Y, Cotten A A, Sharifi Sedeh O, Weldeyesus H, Yang J, Seo J, Ye S, Zhou M, Liu H, Shi G, Hua Z, Watanabe K, Taniguchi T, Xiong P, Zumbühl D M, Fu L and Ju L 2025 Signatures of chiral superconductivity in rhombohedral graphene *Nature* **643** 654

[14] Xu F, Sun Z, Li J, Zheng C, Xu C, Gao J, Jia T, Watanabe K, Taniguchi T, Tong B, Lu L, Jia J, Shi Z, Jiang S, Zhang Y, Zhang Y, Lei S, Liu X and Li T 2025 Signatures of unconventional superconductivity near reentrant and fractional quantum anomalous Hall insulators *arXiv:2504.06972*

[15] Lin J-X, Siriviboon P, Scammell H D, Liu S, Rhodes D, Watanabe K, Taniguchi T, Hone J, Scheurer M S and Li J I A 2022 Zero-field superconducting diode effect in small-twist-angle trilayer graphene *Nat. Phys.* **18** 1221

[16] Yu J, Bernevig B A, Queiroz R, Rossi E, Törmä P and Yang B-J 2025 Quantum Geometry in Quantum Materials *arXiv:2501.00098*

[17] Liu Z and Bergholtz E J 2024 Recent Developments in Fractional Chern Insulators *Encyclopedia of Condensed Matter Physics* **1** 515

[18] Călugăru D, Jiang Y, Hu H, Pi H, Yu J, Vergniory M G, Shan J, Felser C, Schoop L M, Efetov D K, Mak K F and Bernevig B A 2025 Moiré materials based on M-point twisting *Nature* **643** 376

[19] Lei C, Mahon P T and MacDonald A H 2024 Moiré band theory for M-valley twisted transition metal dichalcogenides *arXiv:2411.18828*

[20] Zhang X-W, Wang C, Liu X, Fan Y, Cao T and Xiao D 2024 Polarization-driven band topology evolution in twisted $MoTe_2$ and WSe2 *Nat. Commun.* **15** 4223

# 10. Theory of spin-valley qubits

**Guido Burkard**[1]

[1] Department of Physics and IQST, University of Konstanz, D-78457 Konstanz, Germany

E-mail: guido.burkard@uni-konstanz.de

### Status

Various solid-state systems are being actively investigated as potential platforms for quantum information processing, leveraging quantum dynamics to perform computations that are intractable for classical computers. Among these, two main categories can be identified: macroscopic and microscopic qubit implementations. Macroscopic qubits are typically realized using superconducting electric circuits, where the quantum state corresponds to macroscopic current or voltage oscillations in a superconducting loop. These systems have demonstrated impressive progress in recent years, achieving high-fidelity quantum gates and multi-qubit entanglement. On the other hand, microscopic qubits are based on the quantum states of individual particles, such as the spin of electrons or holes confined in semiconductor quantum dots [1]. This latter approach is considered particularly promising due to the long spin coherence times, compatibility with existing semiconductor fabrication technologies, and scalability.

However, scaling quantum processors to large numbers of qubits remains one of the greatest challenges in the field. Two main issues must be addressed to enable large-scale quantum computation. The first is the need to protect qubits from external noise sources that cause decoherence. Decoherence leads to the loss of quantum information and introduces errors in quantum gate operations, fundamentally limiting performance. The second challenge lies in designing scalable and low-power control electronics capable of operating at cryogenic temperatures, where many solid-state qubits must function. Efficient integration of control and readout systems without introducing additional noise remains a significant engineering hurdle.

Decoherence arises from interactions with the environment and affects both the lifetime of qubit states (relaxation) and their quantum phase coherence (dephasing). In the case of electron spin qubits, the main sources include charged defects, spin defects such as nuclear spins in the host material, and phonons. Each of these interactions can disturb the delicate spin state, reducing lifetimes, coherence times, and gate fidelities. Considerable efforts are being directed toward developing materials and device architectures that minimize these effects, for instance, by isotopic purification or careful defect control.

Two-dimensional (2D) materials such as graphene and transition-metal dichalcogenides (TMDs) have emerged as a promising direction for qubits [2, 3], providing an alternative to three-dimensional semiconductor materials such as silicon and germanium in which qubit functionalities have already been demonstrated. The atomically thin nature of 2D materials allows for superior electrostatic control, reduced defect densities, and the possibility of engineering the spin-orbit interaction. Moreover, 2D solids and van der Waals materials may provide unique scaling advantages, enabling the fabrication of densely packed arrays of qubits while maintaining low-noise operation. These properties make them an exciting avenue for the next generation of scalable quantum information systems. On the other hand, 2D materials also offer several challenges: in monolayer graphene, the absence of a band gap prevents electrostatic electron confinement. The presence of a valley degeneracy in virtually all relevant 2D materials and the strong spin-valley coupling calls for new concepts in the design of classical [4] and quantum [2] bits in these materials. Finally, not much is known so far about the quantum phase coherence of the spin and valley degrees of freedom in quantum dots formed in 2D materials.

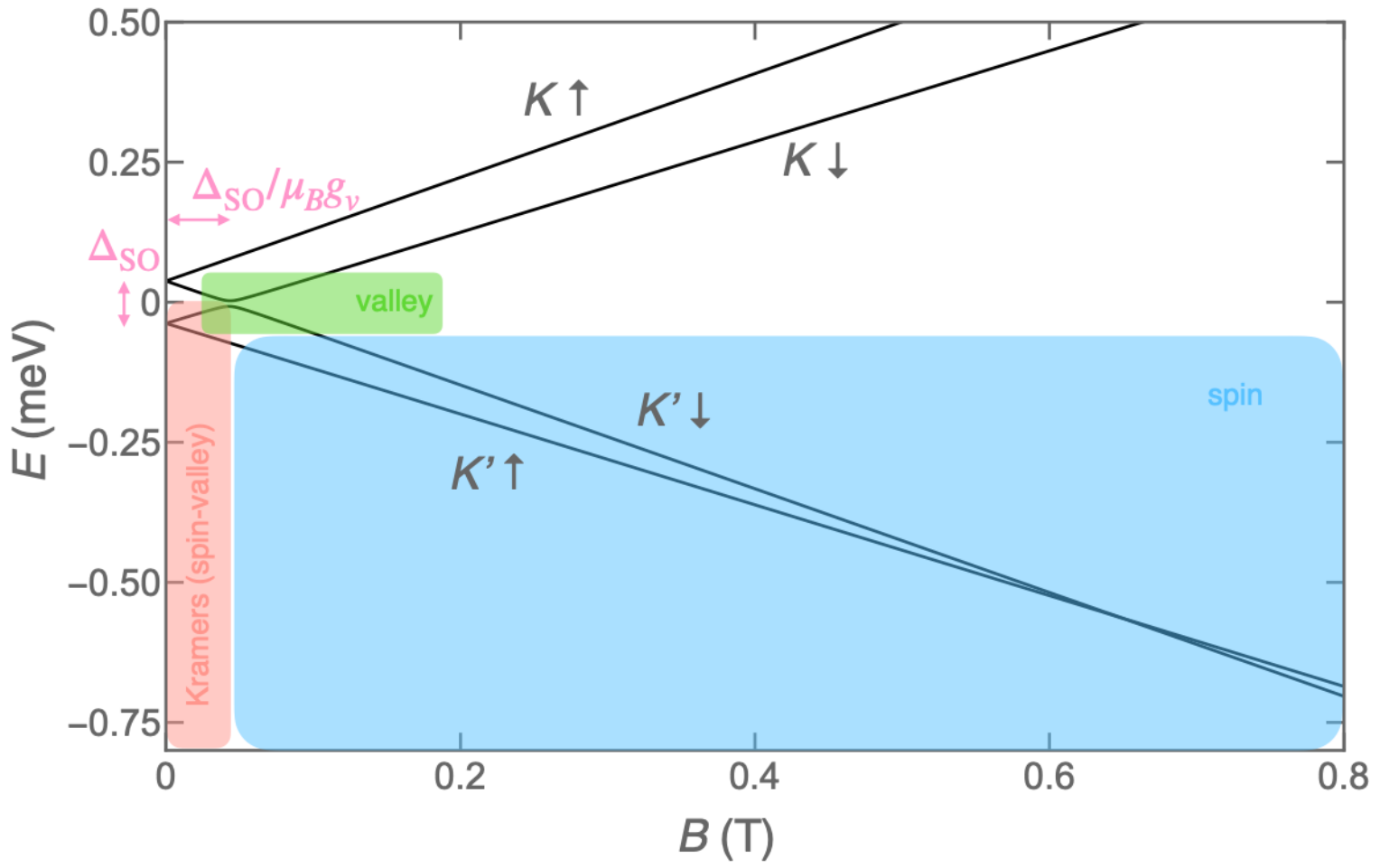

**Figure 1.** Spin and valley energy levels of a BLG or TMD quantum dot in the orbital ground state, as a function of perpendicular magnetic field $B$. Here, $K$ and $K'$ denote the two valley states, while spin up (↑) and spin down (↓) refers to the electron spin quantum number. Parameters are chosen for a typical BLG system, with spin-orbit coupling $\Delta_{SO} = 75\ \mu eV$ and valley g-factor $g_v = 30$. At low magnetic fields $B < \Delta_{SO}/\mu_B g_v$, the two lowest-energy states originate from the Kramers pair at $B = 0$, and give rise to a spin-valley (Kramers) qubit (red box) with opposite spin and valley quantum numbers ($K'$↑ and $K$↓ in the shown example). At larger fields, the lowest two energy levels form a spin qubit (blue box). The two intermediate energy levels form a valley qubit (green box); however, this qubit consists of two excited states which can therefore relax into lower-energy non-computational states.

### Current and future challenges

Implementing qubits in two-dimensional (2D) materials presents a number of fundamental and technical challenges that stem from their unique electronic and structural properties. One major obstacle arises from the absence of a band gap in monolayer graphene. While graphene's exceptional carrier mobility and long coherence lengths make it attractive for quantum applications, the lack of an intrinsic band gap hinders the ability to localize electrons—an essential requirement for defining and controlling qubit states. Without a gap, electron confinement requires methods such as nanopatterning or the formation of heterostructures, which often introduce disorder and decoherence, thus reducing qubit fidelity.

Another important challenge concerns valley degeneracy, a characteristic feature of many 2D materials where electrons can occupy equivalent energy minima (valleys) in the electronic band structure. In materials with relatively strong spin–orbit coupling, such as transition metal dichalcogenides (TMDs), this degeneracy is accompanied by spin–valley locking, where the spin and valley degrees of freedom combine to form low-energy Kramers doublets (see Figure 1). While this coupling can, in principle, be exploited to encode quantum information in hybrid spin–valley qubits, it also complicates qubit initialization and manipulation. Distinguishing and addressing specific valley states with high precision remains an experimental difficulty, and inter-valley scattering can cause rapid dephasing, leading to loss of quantum coherence.

Finally, the lack of experimental data on spin and valley phase coherence further limits the development of robust qubit implementations in 2D materials. Coherence times, spin relaxation mechanisms, and valley lifetimes are not yet fully understood in many systems, particularly under realistic device conditions. Without detailed measurements and reproducible data, it remains difficult to optimize material quality, interface engineering, and external control fields for stable qubit operation.

### Advances in science and technology to meet challenges

Significant progress has been made in recent years in both experimental and theoretical research on spin, valley, and spin-valley (Kramers) qubits in quantum dots based on two-dimensional (2D) materials. These developments have established a strong foundation for future quantum information technologies that leverage the unique electronic and spin-valley characteristics of atomically thin systems.

On the experimental front, major advances have come from high-quality quantum dots and double quantum dots in bilayer graphene (BLG). Research groups at ETH Zurich and RWTH Aachen have successfully fabricated BLG quantum dots with single-electron control, a crucial step toward realizing reliable spin and valley qubits [5, 6]. The electrostatic tunability of bilayer graphene, combined with its relatively low disorder and adjustable band gap under a perpendicular electric field, has enabled precise manipulation of charge states and the observation of discrete energy levels at the single-electron level.

Furthermore, long spin and even longer valley relaxation times have been measured in these BLG quantum dots, indicating that the system can sustain quantum states for relatively extended periods [7, 8]. These long lifetimes are essential for practical qubit operation, as they allow for multiple gate operations within the coherence window. On the theoretical side, both spin and valley relaxation mechanisms are now well understood [9]. They are attributed primarily to electron–phonon interactions, which occur via two distinct mechanisms: the deformation potential, linked to lattice strain, and bond-length changes, associated with atomic displacement. This understanding provides a solid theoretical framework for optimizing material and device design to minimize decoherence. However, for coherent qubit operation, also the spin or valley phase coherence time needs to be sufficiently long. While spin and valley relaxation times in BLG quantum dots have been measured, the corresponding coherence times have yet to be determined.

Parallel research efforts have explored valley control and manipulation in other systems. For instance, carbon nanotubes (CNTs) have been investigated theoretically for valley control. It is challenging to coherently couple the two valley states in 2D materials and CNTs, since they lie at opposite corners of the hexagonal Brillouin zone in reciprocal space. But it turns out that the atomically sharp potential of defects can provide the necessary momentum and allow for controlled and coherent valley rotations when combined with an oscillating electric field [10]. Transport through CNT double quantum dots could give rise to the combined spin-valley blockade, a generalization of the well-known Pauli spin blockade in materials without valley degeneracy [11]. Notably, the realization of a spin–valley qubit in a carbon nanotube has provided a proof of concept for encoding quantum information in combined spin and valley degrees of freedom, offering new pathways for hybrid qubit design [12].

In cases where the spin-orbit coupling dominates the Zeeman splitting, e.g., at low fields in BLG, the low-energy pairs of states that could make up a qubit are (quasi-) Kramers doublets consisting of two states with opposite spin and valley (e.g., spin-up/valley-K and spin-down/valley-K') [13, 14]. This is illustrated in Figure 1, where the four spin and valley states of a quantum dot orbital ground state are plotted as a function of the applied perpendicular magnetic field. The Kramers regime is realized most prominently in transition metal dichalcogenides (TMDs) with their strong spin–orbit coupling and intrinsic spin–valley locking [15], but also in BLG at low field strengths. Theoretical work has proposed various schemes for spin, valley, and spin–valley qubits in these materials [3, 16, 17, 18], as well as universal quantum computing architectures that utilize both spin and valley states as quantum resources [19, 20]. Collectively, these advances mark a critical step toward the realization of scalable, high-coherence quantum computing platforms based on 2D materials.

### Concluding remarks

Spin–valley (Kramers) qubits constitute a promising platform for quantum information processing, emerging naturally in materials where spin and valley degrees of freedom are locked by sufficiently strong spin-orbit coupling. Such qubits arise in the low-magnetic-field regime in graphene, carbon nanotubes, and silicon, and over a broad field range in materials with sufficiently strong intrinsic spin–orbit coupling, most notably the transition-metal dichalcogenides (TMDs). The possibility of low- or zero-field operation appears particularly promising for hybrid systems comprising superconducting elements. However, the simultaneous involvement of both spin and valley quantum numbers renders qubit control particularly demanding, as even single-qubit gates require precise manipulation of coupled spin–valley states. Several theoretical schemes have been proposed to achieve coherent control through electric, magnetic, or strain-mediated interactions, providing valuable conceptual frameworks for future experimental implementations.

Recent advances, particularly the observation of exceptionally long spin, valley, and spin–valley lifetimes in bilayer graphene quantum dots, have strengthened the case for spin–valley qubits as viable candidates for robust qubits. Despite this progress, a complete picture of coherence remains elusive. While the dominant relaxation mechanisms are now relatively well understood, quantitative knowledge of phase coherence times—both for individual spin and valley states and for the combined spin–valley degree of freedom—is still lacking. Identifying and mitigating the microscopic sources of decoherence thus remain central challenges for the field.

### Acknowledgements

We acknowledge support from Deutsche Forschungsgemeinschaft (DFG, German Research Foundation)—Project No. 425217212—SFB 1432.

### References

[1] Loss D and DiVincenzo D P 1998 Quantum computation with quantum dots Phys. Rev. A **57** 120
[2] Trauzettel B, Bulaev D V, Loss D and Burkard G 2007 Spin qubits in graphene quantum dots Nat. Phys. **3** 192
[3] Kormányos A, Zólyomi V, Drummond N D and Burkard G 2014 Spin-orbit coupling, quantum dots, and qubits in monolayer transition metal dichalcogenides Phys. Rev. X **4** 011034
[4] Rycerz A, Tworzydło J and Beenakker C W J 2007 Valley filter and valley valve in graphene, Nat. Phys. **3**, 172
[5] Eich M, Herman F, Pisoni R, Overweg H, Kurzmann A, Lee Y, Rickhaus P, Watanabe K, Taniguchi T, Sigrist M, Ihn T and Ensslin K 2018 Spin and valley states in gate-defined bilayer graphene quantum dots Phys. Rev. X **8** 031023
[6] Banszerus L, Möller S, Steiner C, Icking E, Trellenkamp S, Lentz F, Watanabe K, Taniguchi T, Volk C and Stampfer C 2021 Spin-valley coupling in single-electron bilayer graphene quantum dots Nat. Commun. **12** 5250
[7] Banszerus L, Hecker K, Möller S., Icking, E, Watanabe K, Taniguchi T, Volk C and Stampfer C 2022 Spin relaxation in a single-electron graphene quantum dot Nat. Commun. **13**, 3637
[8] Garreis R, Tong C, Terle J, Ruckriegel M J, Gerber J D, Gächter L M, Watanabe K, Taniguchi T, Ihn T, Ensslin K and Huang W W 2024 Long-lived valley states in bilayer graphene quantum dots Nat. Phys. **20** 428

[9] Wang L and Burkard G 2024 Valley relaxation in a single-electron bilayer graphene quantum dot Phys. Rev. B **110** 035409
[10] Pályi A and Burkard G 2011 Disorder-Mediated Electron Valley Resonance in Carbon Nanotube Quantum Dots 2011 Phys. Rev. Lett. **106** 086801
[11] Pályi A and Burkard G 2010 Spin-valley blockade in carbon nanotube double quantum dots Phys. Rev. B **82**, 155424
[12] Laird E, Pei F and Kouwenhoven L 2013 A valley–spin qubit in a carbon nanotube Nat. Nanotech. **8** 565
[13] Denisov A O, Reckova V, Cances S, Ruckriegel M J, Masseroni M, Adam C, Tong C, Gerber J D, Huang W W, Watanabe K, Taniguchi T, Ihn T, Ensslin K and Duprez H 2025 Spin–valley protected Kramers pair in bilayer graphene Nat. Nanotechnol. **20** 494
[14] Banszerus L, Möller S, Hecker K, Icking E, Watanabe K, Taniguchi T, Hassler F, Volk C and Stampfer C 2023 Particle-hole symmetry protects spin-valley blockade in graphene quantum dots Nature **618** 51
[15] Krishnan R, Biswas S, Hsueh Y-L, Ma H, Rahman R, and Weber B 2023 Spin-Valley Locking for In-Gap Quantum Dots in a MoS2 Transistor Nano Lett. **23** 6171
[16] Pawłowski J, Żebrowski D and Bednarek S 2018 Valley qubit in a gated MoS2 monolayer quantum dot Phys. Rev. B **97** 155412
[17] Széchenyi G, Chirolli L and Pályi A 2018 Impurity-assisted electric control of spin-valley qubits in monolayer MoS2, 2D Materials **5** 035004
[18] Altıntaş A, Bieniek M, Dusko A, Korkusiński M, Pawłowski J and Hawrylak P 2021 Spin-valley qubits in gated quantum dots in a single layer of transition metal dichalcogenides Phys. Rev. B **104** 195412
[19] Rohling N, Russ M and Burkard G 2014 Hybrid Spin and Valley Quantum Computing with Singlet-Triplet Qubits Phys. Rev. Lett. **113** 176801
[20] Rohling N and Burkard G 2012 Universal quantum computing with spin and valley states New J. Phys. **14** 083008

# 11. Spin-valley qubits in graphene

**Artem Denisov[1], Thomas Ihn[1] and Klaus Ensslin[1]**

[1]Laboratory for Solid State Physics, ETH Zurich, Switzerland

E-mail: ensslin@phys.ethz.ch

## Status

The intrinsic valley degree of freedom makes bilayer graphene (BLG) an exceptional platform for next-generation semiconductor qubits [1]. Whereas spins couple to nearby magnetic moments and phonons through spin–orbit coupling (SOC) or hyperfine interactions, flipping the valley index demands a scattering event involving an extremely large momentum swing comparable to a reciprocal-lattice vector, which can arise only from short-range disorder such as atomic-scale defects. Owing to the high material quality achievable in encapsulated graphene-based quantum dot devices, valley relaxation times have been shown to reach values as high as 500 ms and exceed spin relaxation times by more than an order of magnitude within the same device [2].

Interestingly, the ground state of a single-carrier bilayer graphene quantum dot (QD) is intrinsically two-fold degenerate, where the two states of the lowest Kramers pair carry opposite spin and valley quantum numbers [3]. Because of the orbital Berry curvature present in BLG [3], an out-of-plane magnetic field can lift this degeneracy, giving rise to a spin–valley qubit [4]. This "Kramers qubit" [5] is naturally protected from conventional spin- or valley-mixing processes, since any relaxation pathway must flip both quantum numbers simultaneously.

The idea of using spin–valley–locked states to form a semiconductor qubit was originally developed in the context of carbon-nanotube (CNT) quantum dots [6]. Early demonstrations of coherent oscillations in CNTs spin–valley qubits showed, however, relatively short relaxation times $< 10\,\mu s$ [6], and coherence times of only tens of nanoseconds, most likely limited by substantial charge noise and static disorder which allows valley mixing. Encapsulated bilayer graphene, by contrast, offers an atomically flat environment with extremely weak intervalley scattering [2], making it a far more promising host material. Recent experiments [5] demonstrated that the combined spin-valley protection in BLG yields relaxation times up to 38 s for Kramers states which is over two orders of magnitude longer than for spin-only blocked configurations (0.4 s) within the same device as shown in Fig. 1 (c). These results position BLG as an emerging platform for exceptionally long-lived semiconductor qubits, extending beyond the traditional spin-qubit paradigm.

## Current and future challenges

One of the central challenges in realizing coherent control of Kramers qubits arises, ironically, from the robustness of the valley degree of freedom in BLG. Because disorder-induced valley mixing is immeasurably small, the standard Loss–DiVincenzo spin-qubit protocol [7], which relies on electron spin resonance (ESR) to drive spin rotations, cannot be applied. In other words, the valley degree of freedom in BLG couples only to out-of-plane magnetic field, as illustrated in Fig. 1(c) inset, while for ESR two orthogonal field components are required. Additionally, intrinsically weak [8] momentum-dependent spin–orbit coupling in BLG restricts spin and valley manipulation. A more natural strategy is therefore to exploit spin–valley exchange interactions in a double–quantum-dot geometry and operate the Kramers qubit in the singlet–triplet basis. This approach, however, depends heavily on the wavefunction of the two-electron ground-state in BLG, which has remained largely unresolved until very recently [9], but still requires further experimental investigation.

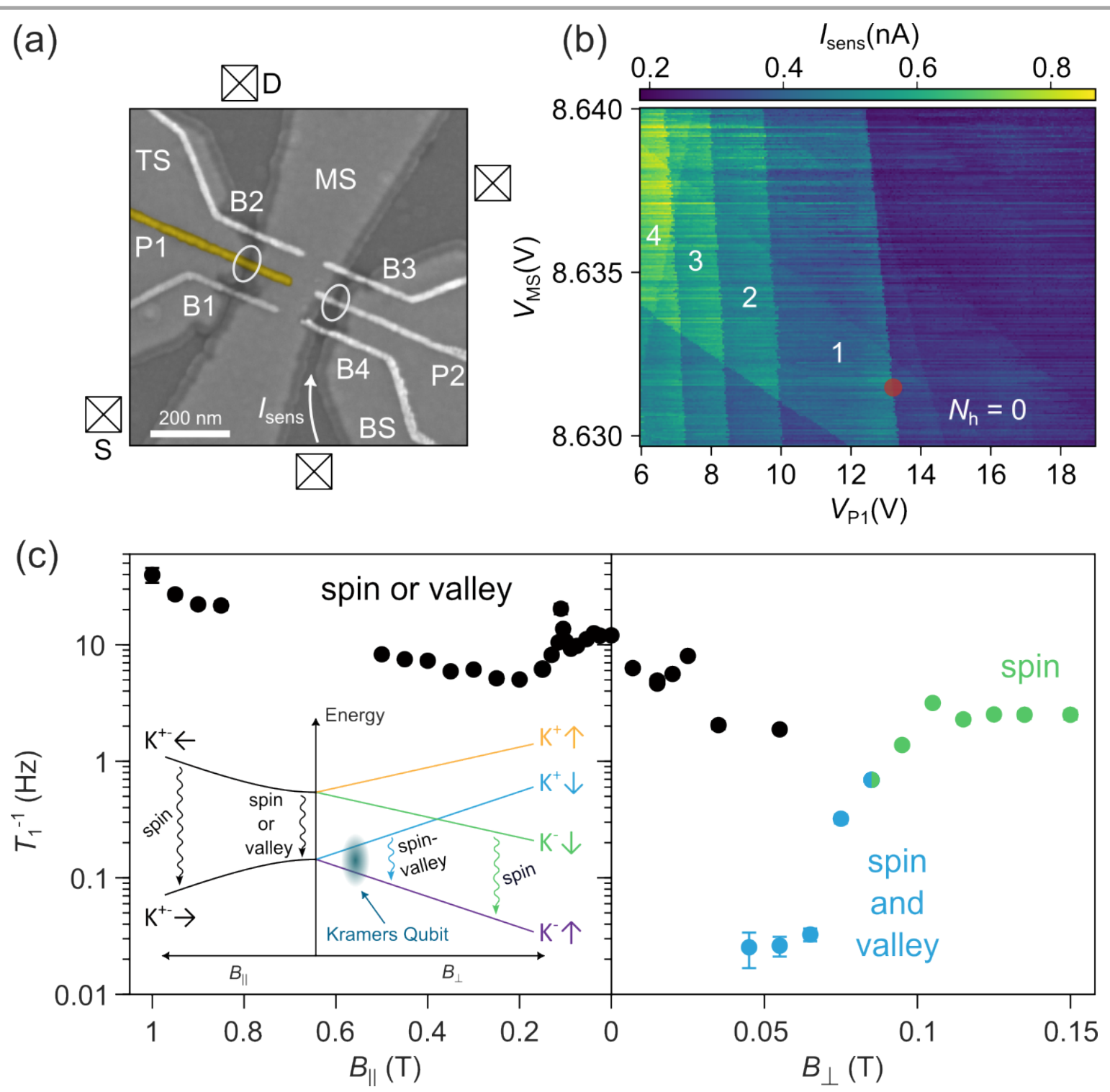

**Figure 1.** (a) SEM false-color image of the typical bilayer graphene quantum dot device with a charge sensor. (b) Charge stability diagram of the dot probed with a charge sensor current. $N_h$ indicates the number of hole carriers inside the quantum dot. The red mark indicates the point on the gate-gate diagram where relaxation time measurements were performed. (e) Measured inverse relaxation time as a function of in-plane $B_{||}$ and out-of-plane $B_\perp$ magnetic fields. Marks of different colors at the same magnetic field correspond to different relaxation. The inset shows the energy spectrum of a single carrier in the BLG QD plotted as a function of in-plane and out-of-plane magnetic fields. Figure adopted from Ref. [5].

Another challenge specific to BLG-based qubit devices is achieving sufficiently low tunnelling rates (<1 MHz), which are essential for high-fidelity qubit readout [1]. Bilayer graphene, just like monolayer graphene, has a gapless band structure. Electron confinement in channels and quantum dots therefore requires applying an out-of-plane displacement field to open a band gap [10]. However, the size of this induced gap is limited by both the achievable displacement field and the width of the confinement channel, each of which is intrinsically constrained by the thickness and quality of the hBN encapsulation and metallic gates.

A broader challenge shared across all semiconductor qubit platforms—but particularly pronounced for graphene-based quantum dot devices—is scalability and reproducibility. Each BLG quantum dot so far is fabricated on a van der Waals heterostructure composed of several individually exfoliated and hand-assembled flakes, typically four or five per device [11]. This manual and highly variable process stands in sharp contrast to the wafer-scale or foundry-level fabrication technologies already available for more established semiconductor platforms such as silicon [12]. As a result, BLG quantum dot devices currently face significant limitations in terms of large-scale integration, device uniformity, and ultimately their potential for scalable quantum information processing.

### Advances in science and technology to meet challenges

One promising route to overcoming the challenge of coherently manipulating spin–valley qubits in bilayer graphene is to engineer spin–orbit coupling through proximity to a transition-metal dichalcogenide. In particular, placing BLG on $WSe_2$ and $MoS_2$ provides a powerful way to induce SOC in pristine BLG [8]. Experiments have already shown [8] that $WSe_2$ proximity induces a substantial Ising-type SOC in BLG, nearly an order of magnitude larger than the analogous intrinsic Kane–Mele SOC in pristine BLG. However, Kane–Mele or Ising SOC alone do not provide the spin–valley mixing required for electric-dipole qubit control, since it is momentum independent. The emergence of momentum-dependent SOC terms through proximity remains an open question for future experiments, as such effects are not easily accessible through standard transport measurements. Continued progress in controlling TMD/BLG alignment and stacking order will be essential for realizing these SOC-enabled manipulation schemes.

Replacing metallic gates with graphite gates in graphene quantum dot devices offers a promising route to overcoming several key limitations of current architectures, including uncontrolled tunneling rates, disorder and charge noise from amorphous metal gate dielectrics, which are largely responsible for inhomogeneous confinement and charge noise in conventional devices. Graphite gates provide an atomically flat, chemically inert, and low-disorder electrostatic environment for graphene, outperforming metallic gates in mobility, uniformity and screening [13]. Because few-layer graphite is itself a van der Waals material, it integrates naturally into hBN/graphene stacks without requiring deposited dielectrics. The advantages of graphite gates have been demonstrated through the resolution of fragile fractional quantum Hall states [14, 15] and through the achievement of highly uniform displacement fields in bilayer graphene [16]. Extending these advantages to quantum dots is expected to significantly improve confinement quality, produce cleaner and more uniform gate-induced band gaps, and enable more reproducible graphene qubits.

Finally, the scaling challenge of graphene quantum dot devices can be addressed by leveraging large-area chemical vapour deposition (CVD) synthesis of graphene in combination with hBN. Recently, it has been demonstrated that graphene can be grown over nearly centimeter-scale areas and transferred onto substrates using a dry, Ni-assisted technique to form high-quality van der Waals heterostructures [17]. This advance paves the way for reproducible 2D quantum device fabrications without relying on mechanical exfoliation. However, a major bottleneck remains: the highest-quality hBN crystals are still available only as sub-millimeter flakes. With large-area, high-quality hBN, the full potential of CVD-graphene-based devices could be realized, or an alternative monolithic dielectric with similar qualities is found.

### Concluding remarks

Spin–valley qubits in bilayer graphene remain primarily at a fundamental research stage, with no coherence-time measurements yet available and several open challenges. Nevertheless, the unusually long relaxation times already observed in Kramers states suggest that graphene-based qubits may ultimately outperform conventional semiconductor platforms. In established systems such as silicon, germanium or GaAs, the dominant decoherence mechanisms such as hyperfine coupling, charge noise, and spin–orbit coupling are well understood and rather difficult to improve significantly. In contrast, bilayer graphene offers an intrinsically cleaner environment, strongly suppressed hyperfine noise, and the possibility of engineering spin–orbit coupling and valley mixing through proximity which otherwise are highly suppressed. Thus, while the technical difficulties facing BLG spin-valley qubits are substantial, the potential payoff in achieving exceptionally long coherence times justify the effort and could eventually shift the balance in favor of this emerging platform.

## Acknowledgements

We acknowledge financial support by the European Graphene Flagship Core3 Project, H2020 European Research Council(ERC) Synergy Grant under Grant Agreement 951541, the European Innovation Council under grant agreement number 101046231/FantastiCOF, NCCR QSIT (Swiss National Science Foundation, grant number 51NF40- 185902).

## References

[1] Burkard G, Ladd T D, Pan A, Nichol J M and Petta J R 2023 Semiconductor spin qubits Rev. Mod. Phys. **95** 025003

[2] Garreis R, Tong C, Terle J, Ruckriegel M J, Gerber J D, Gächter L M, Watanabe K, Taniguchi T, Ihn T, Ensslin K and Huang W W 2024 Long lived valley states in bilayer graphene quantum dots Nat. Phys. **20** 428–434

[3] Knothe A, Glazman L I and Fal'ko V I 2022 Tunneling theory for a bilayer graphene quantum dot's single and two electron states New J. Phys. **24** 043003

[4]Duprez H, Cances S, Omahen A, Masseroni M, Ruckriegel M J, Adam C, Tong C, Garreis R, Gerber J D, Huang W, Gächter L M, Watanabe K, Taniguchi T, Ihn T and Ensslin K 2024 Spin valley locked excited states spectroscopy in a one particle bilayer graphene quantum dot Nat. Commun. **15** 9717

[5]Denisov A O, Reckova V, Cances S, Ruckriegel M J, Masseroni M, Adam C, Tong C, Gerber J D, Huang W W, Watanabe K, Ta niguchi T, Ihn T and Ensslin K 2025 Spin valley protected Kramers pair in bilayer graphene Nat. Nanotechnol. **20** 494 499

[6] Laird E A, Kuemmeth F, Steele G A, Grove Rasmussen K, Nygård J, Flensberg K and Kouwenhoven L P 2015 Quantum transport in carbon nanotubes Rev. Mod. Phys. **87** 703

[7] Flensberg K and Marcus C M 2010 Bends in nanotubes allow electric spin control and coupling Phys. Rev. B 81 195418

[8]Gerber J D., Ersoy E., Masseroni M., Niese M., Laumer M., Denisov A.O., Duprez H., Huang W.W., Adam C., Ostertag L., Ton g C., Taniguchi T., Watanabe K., Fal'ko V.I., Ihn T. and Ensslin K. 2025 Tunable Spin-Orbit Splitting in Bilayer Graphene/$WSe_2$ Quantum Devices Nano Lett. **25** 12480–12486.

[9]Adam C, Duprez H, Lehmann N, Yglesias A, Denisov A O, Cances S, Ruckriegel M J, Masseroni M, Tong C, Huang W, Kealho fer D, Garreis R, Watanabe K, Taniguchi T, Ensslin K,and Ihn T 2025 Entropy Spectroscopy of a Bilayer Graphene Quant um Dot Phys. Rev. Lett. **135** 126202

[10]Icking E, Banszerus L, Wörtche F, Volmer F, Schmidt P, Steiner C, Engels S, Hesselmann J, Goldsche M, Watanabe K, Tan iguchi T, Volk C, Beschoten B and Stampfer C 2022 Transport Spectroscopy of Ultraclean Tunable Band Gaps in Bilayer Graphene Adv. Electron. Mater. **8** 2200510.

[11]Dean C. R., Young A. F., Meric I., Lee C., Wang L., Sorgenfrei S., Watanabe K., Taniguchi T., Kim P., Shepard K. L. and Hone J. 2010 Boron nitride substrates for high quality graphene electronics Nat. Nanotechnol. **5** 722 726.

[12]Steinacker P, Dumoulin Stuyck N, Lim W H, Tanttu T, Feng M, Serrano S, Nickl A, Candido M, Cifuentes J D, Vahapoglu E, Bartee S K, Hudson F E, Chan K W, Kubicek S, Jussot J, Canvel Y, Beyne S, Shimura Y, Loo R, Godfrin C, Raes B, Baudot S, Wan D, Laucht A, Yang C H, Saraiva A, Escott C C, De Greve K and Dzurak A S 2025 Industry compatible silicon spin qubit unit cells exceeding 99% fidelity Nature **646** 81 87

[13]Domaretskiy D, Wu Z, Nguyen V H, Hayward N, Babich I, Li X, Nguyen E, Barrier J, Indykiewicz K, Wang W, Gorbachev R V, Xin N, Watanabe K, Taniguchi T, Hague L, Fal'ko V I, Grigorieva I V, Ponomarenko L A, Berdyugin A I and Geim A K 2025 Proximity screening greatly enhances electronic quality of graphene Nature **644** 646 651

[14] Zibrov A A, Kometter C, Zhou H, Spanton E M, Taniguchi T, Watanabe K, Zaletel M P and Young A F 2017 Tunable interacting composite fermion phases in a half filled bilayer graphene Landau level Nature **549** 360 364.

[15] Cohen L A., Samuelson N. L., Wang T., Taniguchi T., Watanabe K., Zaletel M. P. and Young A. F. 2023 Universal chiral Luttinger liquid behaviour in a graphene fractional quantum Hall point contact Science **382** 542 547

[16] Zhang Y, Polski R, Thomson A, Lantagne Hurtubise E, Lewandowski C, Zhou H, Watanabe K, Taniguchi T, Alicea J and Nadj Perge S 2023 Enhanced superconductivity in spin–orbit proximitized bilayer graphene Nature **613** 268 273

[17] Amontree J, Marchese A, Pack J, Chung T, Yan X, Borisenkov Y, Yang B, Davis K, Watanabe K, Taniguchi T, Dean C and Hone J 2025 Moving Beyond Scotch Tape: Scalable Transfer of Research Grade CVD Graphene Nano Lett. **25** 15572 15578.

# 12. Spin-valley effects in TMD quantum dots

**Louis Gaudreau[1] and Justin Boddison-Chouinard[1,2]**

[1] Quantum and Nanotechnologies Research Centre, National Research Council Canada, Ottawa, Canada
[2] Department of Physics, University of Ottawa, Ottawa, Canada

E-mail: Louis.Gaudreau@nrc-cnrc.gc.ca

## Status

Atomically thin transition metal dichalcogenides present a unique platform to study spin and valley effects as well as their interplay in condensed matter physics. In particular, spin and valley degrees of freedom are intimately coupled in these materials due to broken inversion symmetry together with time-reversal symmetry, leading to the spin-valley locking effect, where carriers occupying the K and K′ valleys of the reciprocal space must have opposite spins. This effect, combined with strong spin-orbit coupling, make TMDs-based spin-valley qubits ideal candidates for future quantum technologies because they can have long coherence times and can be controlled and readout both electrically and optically. To this end, quantum dots made of TMD semiconductors are being studied to isolate and control carriers (electrons, holes or excitons) capable of encoding quantum information in their valley and spin degrees of freedom.

Currently, the most successful approaches towards the realization of TMD quantum dots have been electrostatic confinement and defect engineering [1-11]. An example of a gated quantum dot device architecture is presented in fig. 1a,b [12]. The first attempts at electrostatic confinement in TMDs consisted of creating quantum point contacts, in which the electronic transport of charge carriers is restricted to one dimension. 1D quantum confinement leads to conductance quantization, whereby adjusting the confinement gates of the 1D channel, the conductance of the device jumps in regular steps as each additional transport channel is added. This has been demonstrated by different groups [13-17] and is depicted in fig. 1 c-f. Under these confinement conditions, spin-valley effects show as anomalous conductance quantization, where instead of the expected conductance steps in units of $2e^2/h$ (e is the electron charge and h the Planck constant), experiments show $e^2/h$ steps due to strong valley-spin interactions and demonstrating the lifting of spin-valley degeneracy without an applied magnetic field [17]. Further experiments have shown successful electrostatic confinement to create quantum dots in TMDs [1-9], where the number of carriers in the quantum dots could be controlled. Different quantum dot device architectures are presented in fig. 2 (a-j) as well as the transport measurements demonstrating Coulomb blockade through the characteristic diamond features in the single and double dot regimes. Coulomb diamonds and charge stability diagrams indicate control of the number of carriers in the quantum dots. Additional transport measurements with an applied magnetic field (fig. 2 k-m) have confirmed the spin-valley locking effect by monitoring the spin-valley ground states of 2-charge transitions of in-gap quantum dots in few layer $MoS_2$ [8]. The large out of plane g-factor anisotropy extracted for the 2 transitions (fig. 2 k-l) suggest the presence of both spin and valley Zeeman effects. However, in order to extract the respective contributions of spin and valley g-factors, knowledge about the precise charge occupation is required which was unknown in the study. Furthermore, by tilting the B-field, the spin-orbit splitting of the conduction band in few-layer $MoS_2$ was estimated to be in the order of 100 μeV (fig. 2m). Optical measurements of atomic defects in TMDs [10, 11] indicate that they show similar effects, such as 0D quantum confinement. Polarization-resolved magneto-optical spectroscopy shows that in some cases two photoluminescence split peaks are observed, even at zero B-field, again indicating a lifting of the spin-valley degeneracy, where in other cases this doesn't happen. Further studies need to be performed to better understand the spin-valley physics of TMD defects to be able to use them as qubits in future quantum technologies.

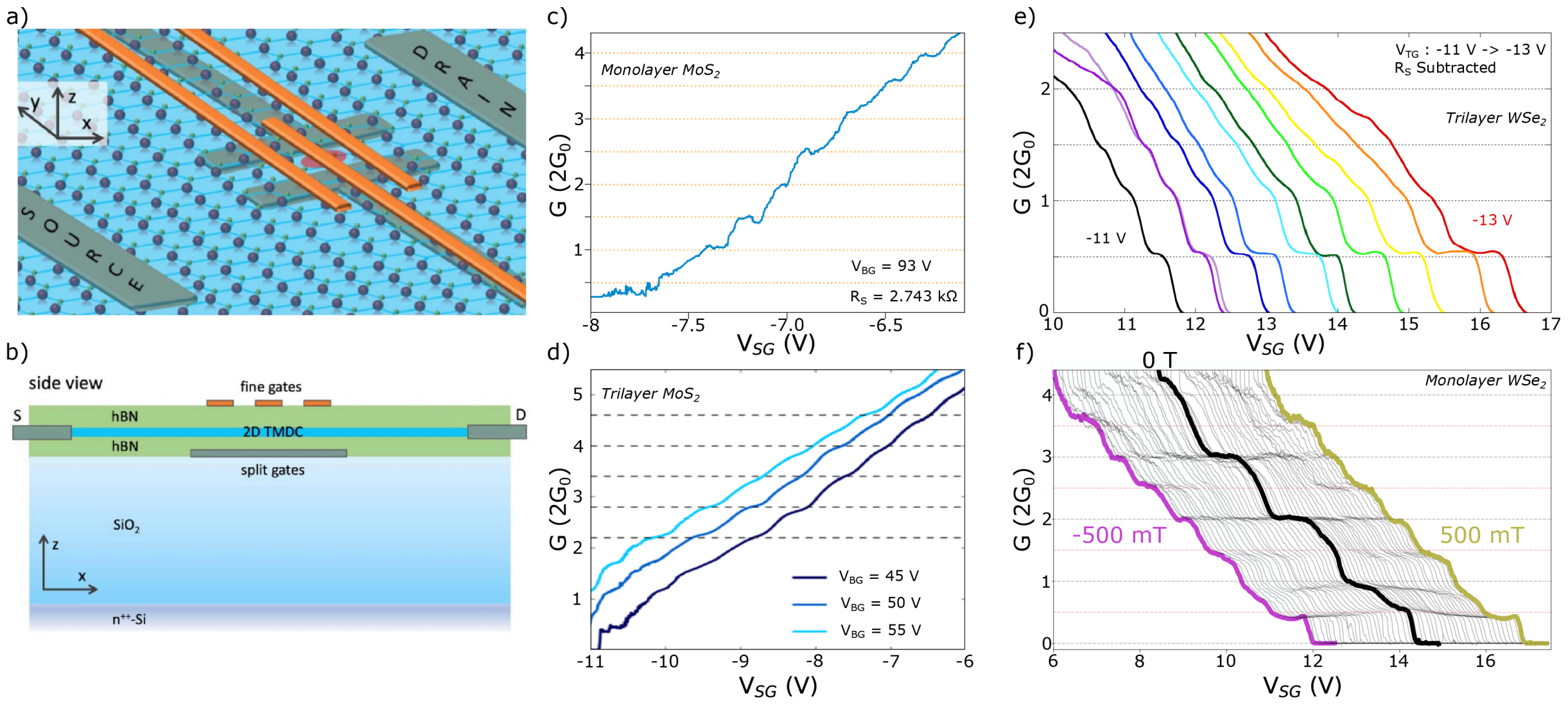

**Figure 1.** Quantum confinement in TMDs. Schematic of a proposed gated quantum dot device architecture in monolayer TMDs as seen from the top (a) and from the side (b). Adapted from [12]. Transport 1D traces as a function of the voltages applied to the split gates for (c) monolayer $MoS_2$ adapted from [13], (d) tri-layer $MoS_2$ adapted from [14], (e) tri-layer $WSe_2$ adapted from [15], (f) monolayer $WSe_2$ adapted from [17], showing conductance plateaus aligning with integer values of $G_0 = e^2/h$ at zero magnetic field, a result of spin-valley effects in TMDs.

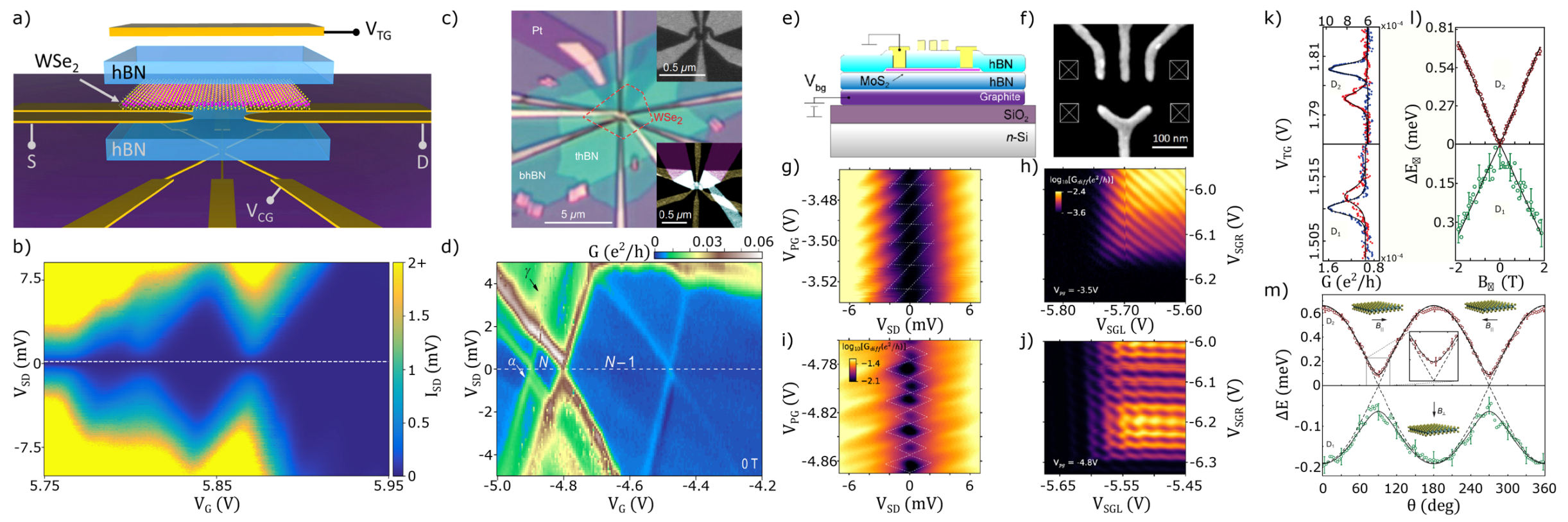

**Figure 2.** Quantum dots in TMDs. (a) Monolayer $WSe_2$ quantum dot device demonstrating charge control through Coulomb blockade in bias spectroscopy (b). Adapted from [7]. (c) Bilayer $WSe_2$ quantum dot device demonstrating charge control through Coulomb blockade in bias spectroscopy (d). Adapted from [6]. Monolayer $MoS_2$ quantum dot device (e, f) demonstrating charge control through Coulomb blockade in bias spectroscopy as well as charge stability diagrams in the single dot (g, h) and double dot regime (i, j). (k, l) Perpendicular B- field dependence of two Coulomb Blockade peaks in a few-layer $MoS_2$ quantum dot, indicating spin-valley splitting. (m) Zeeman shifts of the ground state transitions of the same Coulomb blockade peaks as a function of the magnetic field orientation (θ). Adapted from [8].

### Current and future challenges

Spin-valley physics in quantum-confined TMD devices, particularly quantum dots, holds strong potential for quantum technologies. Yet, experimental progress in these systems has been limited by significant materials and device fabrication challenges.

(i) **Electrical Contacts.** Forming high-quality ohmic contacts to TMDs remains a major challenge in transport experiments across all device types, and this limitation becomes especially pronounced in quantum dot systems. The requirement to operate at low temperatures, in order to resolve individual energy level spacings, and at low carrier densities often results in highly resistive contacts, which introduce significant noise and instability. These effects are particularly detrimental in the few- or single-carrier regime, where charge transport is extremely sensitive to local potential fluctuations.

(ii) **Material Quality.** The presence of intrinsic and extrinsic defects in TMD devices remains a key obstacle to the controlled formation of quantum-confined structures. In most current devices, quantum dots tend to form at defect sites or in disordered regions rather than being electrostatically defined by gate potentials. While this has enabled early observations of spin-valley physics, it significantly limits tunability, reproducibility, and scalability. Beyond confinement, defects also reduce carrier mobility and introduce charge noise, posing significant challenges for coherent quantum transport in the few-carrier regime.

(iii) **Scalability and Reproducibility.** Achieving scalable and reproducible TMD quantum dot devices is crucial for advancing their use in quantum technologies, yet both intrinsic material defects and fabrication-related issues pose significant hurdles. During the stacking and fabrication of heterostructures, contaminants become trapped at material interfaces and tend to coalesce into isolated pockets, creating large bubbles and leaving behind small pristine regions. Although these pristine areas benefit initial experiments and small-scale studies, the lack of uniformity severely limits reproducibility and makes scaling up difficult.

Addressing these challenges is essential to unlock the full potential of spin-valley effects in TMD-based quantum dots, as they currently limit the development of reliable, scalable quantum devices and highlight the need for targeted advances in materials and device engineering.

### Advances in science and technology to meet challenges

While these critical limitations are currently impeding TMD-based quantum dot devices, recent progress in material growth, device engineering, and fabrication techniques has focused on addressing them.

(i) **Electrical Contacts.** Innovative device architectures now allow independent gating of contact regions separate from the active TMD region [8, 17], effectively reducing the contact resistance while preserving the low carrier densities required for quantum confinement. This approach mitigates the trade-off between contact transparency and channel depletion, which is critical for accessing the single-carrier regimes. However, large electric fields are required in the contact regions in order to effectively reduce the contact resistance, which may compromise the gate dielectric over time. Alternative approaches involve the use of charge-transfer layers to induce high carrier densities locally at the contact regions [18]. This method has resulted in efficient contacts, particularly in $WSe_2$, although its application remains largely limited to p-type devices. Additionally, atomic force microscopy brooming [19] has been employed to remove residues and contaminants at the contacts, leading to cleaner interfaces and improved device performance. While effective, this technique is inherently unscalable. Despite these remaining challenges, ongoing developments in both contact engineering and interface cleaning techniques are steadily improving the reliability and performance of electrical contacts in TMD quantum devices.

(ii) **Material Quality.** Recent advances in flux-growth techniques have produced TMD crystals with low defect densities and improved carrier mobility [20]. Incorporating these high-quality materials into quantum dot devices would facilitate the controlled formation of gate-defined quantum dots, enabling cleaner and more tunable studies of spin-valley effects in confined systems.

(iii) **Scalability and Reproducibility.** Progress toward scalable quantum technologies on TMD quantum dots depends critically on improving the uniformity and reproducibility of both material growth and heterostructure assembly. Advances in the synthesis of large-area TMD films using metal organic chemical vapor deposition are progressing rapidly, providing a pathway toward wafer-scale integration. However, further progress is still required to match the purity of exfoliated crystals. At the single device level, cleaner transfer techniques that employ inorganic supports, such as flexible silicon nitride membranes instead of polymeric supports, have demonstrated a reduction in interface contamination and minimal bubble formation, resulting in improved electronic performance and reproducibility. Robotic and automated stacking platforms [21] further enhance reproducibility by enabling precise control over layer placement and alignment. In tandem, these advances represent important steps toward the reproducible fabrication of large-scale, gate-defined quantum dot arrays suitable for quantum technologies. Alternatively, exploiting the research of single defects, efforts have been made to deterministically place defects in 2D materials, with focused ion beams [11], to achieve the goal of scalable and reproducible devices.

### Concluding remarks

Quantum dots in TMDs offer a promising platform in the field of quantum technologies, driven by their distinctive spin-valley effects. The inherent spin-valley locking and strong spin-orbit coupling, together with a sizable direct bandgap, make it possible to control quantum states both electrically and optically. Although these properties make TMDs highly desirable for quantum devices, overcoming current challenges in materials growth, fabrication, and device engineering remains essential to realize their full potential in next-generation quantum technologies. Significant progress toward addressing these challenges, such as low-defect flux-grown crystals, robotic assembly in controlled environments, and charge-transfer layers for improved contact resistance, has been achieved in recent years. Yet further advances are needed to deepen our understanding of spin-valley effects in well-controlled TMD quantum dots. The development of high-quality, reproducible devices will not only enable the study of spin-valley effects in the quantum-confined regime but also provide the experimental validation needed to refine and advance current theoretical models [22, 23]. As experimental capabilities and theoretical frameworks continue to converge, TMD quantum dots will evolve from exploratory platforms into functional building blocks for spin-valley-based quantum technologies.

## Acknowledgements

We acknowledge financial support from the High Throughput and Secure Networks Challenge Program and the Quantum Sensors Challenge Program at the National Research Council of Canada, and from the Natural Sciences and Engineering Research Council of Canada (NSERC) under the funding reference number ALLRP/578466-2022.

## References

[1] Song, Liu, Mosallanejad, You, Han, Chen, Li, Cao, Xiao, Guo and Guo 2015 A gate defined quantum dot on the two-dimensional transition metal dichalcogenide semiconductor $WSe_2$ *Nanoscale* **7** 16867

[2] Song, Zhang, You, Liu, Li, Cao, Xiao and Guo 2015 Temperature dependence of Coulomb oscillations in a few-layer two-dimensional $WS_2$ quantum dot *Scientific Reports* **5** 16113

[3] Zhang, Song, Luo, Deng, Mosallanejad, Taniguchi, Watanabe, Li, Cao, Guo, Nori and Guo 2017 Electrotunable artificial molecules based on van der Waals heterostructures *Science Advances* **3** e1701699

[4] Wang, De Greve, Jauregui, Sushko, High, Zhou, Scuri, Taniguchi, Watanabe, Lukin, Park and Kim 2018 Electrical control of charged carriers and excitions in atomically thin materials *Nature Nanotechnology* **13** 128-132

[5] Pisoni, Lei, Back, Eich, Overweg, Lee, Watanabe, Taniguchi, Ihn and Ensslin 2018 Gate-tunable quantum dot in a high quality single layer $MoS_2$ van der Waals heterostructure *Appl. Phys. Lett.* **112** 123101

[6] Davari, Stacy, Mercado, Tull, Basnet, Pandey, Watanabe, Taniguchi, Hu and Churchill 2020 Gate-Defined Accumulation-Mode Quantum Dots in Monolayer and Bilayer $WSe_2$ *Phys. Rev. Applied* **13** 054058

[7] Boddison-Chouinard, Bogan, Fong, Watanabe, Taniguchi, Studenikin, Sachrajda, Korkusinski, Altintas, Bieniek, Hawrylak, Luican-Mayer and Gaudreau 2021 Gate-controlled quantum dots in monolayer $WSe_2$ *Appl. Phys. Lett.* **119** 133104

[8] Krishnan, Biswas, Hsueh, Ma, Rahman and Weber 2023 Spin-Valley Locking for In-Gap Quantum Dots in a $MoS_2$ Transistor *Nano Lett.* **23** 6171-6177

[9] Kumar, Kim, Tripathy, Watanabe, Taniguchi, Novoselov and Kotekar-Patil 2023 Excited state spectroscopy and spin splitting in single layer $MoS_2$ quantum dots *Nanoscale* **15** 18203

[10] Srivastava, Sidler, Allain, Lembke, Kis and Imamoğlu 2015 Optically active quantum dots in monolayer $WSe_2$ *Nature Nanotechnology* **10** 491-496

[11] Hötger, Amit, Klein, Barthelmi, Pelini, Delhomme, Rey, Potemski, Faugeras, Cohen, Hernangómez-Pérez, Taniguchi, Watanabe, Kastl, Finley, Refaely-Abramson, Holleitner and Stier 2023 Spin-defect characteristics of single sulfur vacancies in monolayer $MoS_2$ *npj 2D Mater. And Appl.* **7** 30

[12] Pawlowski, Kumar, Watanabe, Taniguchi, Novoselov, Churchill and Kotekar-Patil 2025 Single electron quantum dot in two-dimensional transition metal dichalcogenides *Nanotechnology* **36** 195001

[13] Marinov, Avsar, Watanabe, Taniguchi and Kis 2017 Resolving the spin splitting in the conduction band of monolayer $MoS_2$ *Nature Communications* **8** 1938

[14] Pisoni, Lee, Overweg, Eich, Simonet, Watanabe, Taniguchi, Gorbachev, Ihn and Ensslin 2017 Gate-Defined One-Dimensional Channel and Broken Symmetry States in $MoS_2$ van der Waals Heterostructures **17** 5008-5011

[15] Sakanashi, Krüger, Watanabe, Taniguchi, Kim, Ferry, Bird and Aoki 2021 Signature of Spin-Resolved Quantum Point Contact in p-Type Trilayer $WSe_2$ van der Waals Heterostructure *Nano Lett.* **21** 7534-7541

[16] Boddison-Chouinard, Bogan, Barrios, Lapointe, Watanabe, Taniguchi, Pawlowski, Miravet, Bieniek, Hawrylak, Luican-Mayer and Gaudreau 2023 Anomalous conductance quantization of a one-dimensional channel in monolayer $WSe_2$ *npj 2D Mater. and Appl.* **7** 50

[17] Boddison-Chouinard, Korkusinski, Bogan, Barrios, Waldron, Watanabe, Taniguchi, Pawlowski, Miravet, Hawrylak, Luican-Mayer and Gaudreau 2025 Valley-spin polarization at zero magnetic field induced by strong hole-hole interactions in monolayer $WSe_2$ *Science Advances* **11** eadu4696

[18] Pack, Guo, Liu, Jessen, Holtzman, Liu, Cothrine, Watanabe, Taniguchi, Mandrus, Barmak, Hone and Dean 2024 Charge-transfer contacts for the measurment of correlated states in high-mobility $WSe_2$ *Nature Nanotechnology* **19** 948-954

[19] Rosenberger, Chuang, McCreary, Hanbicki, Sivaram and Jonker 2012 Nano-"Squeegee" for the Creation of Clean 2D Material Interfaces *ACS Appl. Mater. Interfaces* **10** 10379-10387

[20] Liu, Liu, Holtzman, Li, Holbrook, Pack, Taniguchi, Watanabe, Dean, Pasupathy, Barmak, Rhodes and Hone 2023 Two-Step Flux Synthesis of Ultrapure Transition-Metal Dichalcogenides *ACS Nano* **17** 16587-16596

[21] Mannix, Ye, Sung, Ray, Muji, Park, Lee, Kang, Shreiner, High, Muller, Hovden and Park 2022 Robotic four-dimensional pixel assembly of van der Waals solids *Nature Nanotechnology* **17** 361-366
[22] Altintas, Bieniek, Dusko, Korkusinski, Pawlowski and Hawrylak 2021 Spin-valley qubits in gated quantum dots in a single layer of transition metal dichalcogenides *Physical Review B* **104** 195412
[23] Pawlowski, Bieniek and Wozniak 2021 Valley Two-Qubit System in a $MoS_2$-Monolayer Gated Double Quantum Dot *Physical Review Applied* **15** 054025

# 13. Nanophotonic control

**Zlata Fedorova[1,2,3] and Isabelle Staude[1,2,3]**

[1] Institute of Solid State Physics, Friedrich Schiller University Jena, Max-Wien-Platz 1, 07743 Jena, Germany
[2] Institute of Applied Physics, Friedrich Schiller University Jena, Albert-Einstein-Str. 15, 07745 Jena, Germany
[3]Abbe Center of Photonics, Friedrich Schiller University Jena, Albert-Einstein-Str. 6, 07745, Jena, Germany

E-mail: zlata.fedorova@uni-jena.de

## Status

With CMOS technology approaching fundamental size and energy limits, exploiting carrier degrees of freedom beyond charge, especially spin and valley, has become an attractive direction for next-generation information technology. The discovery of valley-selective optical responses in 2D materials with broken inversion symmetry, including electrically biased graphene bilayers [1] and monolayers of transition-metal dichalcogenides (TMDs) [2] established a platform where valleys K and K' act as a binary degree of freedom addressable with light. Because the K and K′ valleys couple to opposite circular components of the optical field, nanophotonic structures such as resonant plasmonic and dielectric nanoantennas, Bragg and photonic-crystal cavities, waveguides, and metasurfaces offer direct control of valley-selective processes at the nanoscale by tailoring the local photonic spin, density of states, and in-plane momentum. Over the past decade, enabled by this nanophotonic toolbox, valleytronics in 2D materials has evolved from polarization-selective pumping [3] to hybrid devices that generate, read out, modulate, and route valley information on chip and in free space. An overview of the nanophotonic control in valleytronics and the main directions for future progress are summarized schematically in Fig. 1.

Early studies of valley-nanophotonic hybrids largely focused on phenomenological observables such as enhanced circular polarization or directional emission, often employing concepts like a “chiral Purcell effect” without a rigorous microscopic treatment of the emitter-field interaction (see [4] and references therein). More recent work has shifted toward models in which valley excitons are described as circularly polarized in-plane dipoles coupled to realistic nanostructure modes, supported by Maxwell-Bloch theory [5] and by full-wave simulations combined with experiments that clarify how nanoresonators reshape far-field polarization and apparent valley contrast [6, 7]. Complementary studies of quantum emitters in $WSe_2$ coupled to chiral nanocavities have shown that the emitted photoluminescence helicity can be set deterministically via photonic mode coupling, without implying genuine valley-dependent radiative lifetime [8], showing that modified far-field polarization alone is not sufficient evidence of valley-selective emission.

In parallel, several classes of nanophotonic platforms have demonstrated functional control of valley-linked optical signals. Electron-beam excitation of TMDs coupled to nanoantennas has revealed the importance of subwavelength excitation profiles for free-space valley routing [9], while chiral plasmonic metasurfaces combined with valley Hall readout demonstrate room-temperature valley imbalance attributed to valley-selective hot electron injection [10]. Recent implementations of strongly coupled exciton-polaritons in microcavities and metasurfaces go beyond the mere inheritance of valley contrast, showing valley Hall effects [12], selective coupling of valley excitons to intrinsically chiral bound states in the continuum [13], and valley-dependent vortex emission from TMD-based polaritonic metasurfaces [11]. Device-scale implementations now include atomically thin tunable chiral cavities [12], valley-addressable monolayer lasers [13], and valley-Hall-based chiral light detectors [14]. Together, these advances establish nanophotonics as a powerful platform for initializing, manipulating, and reading out valley information.

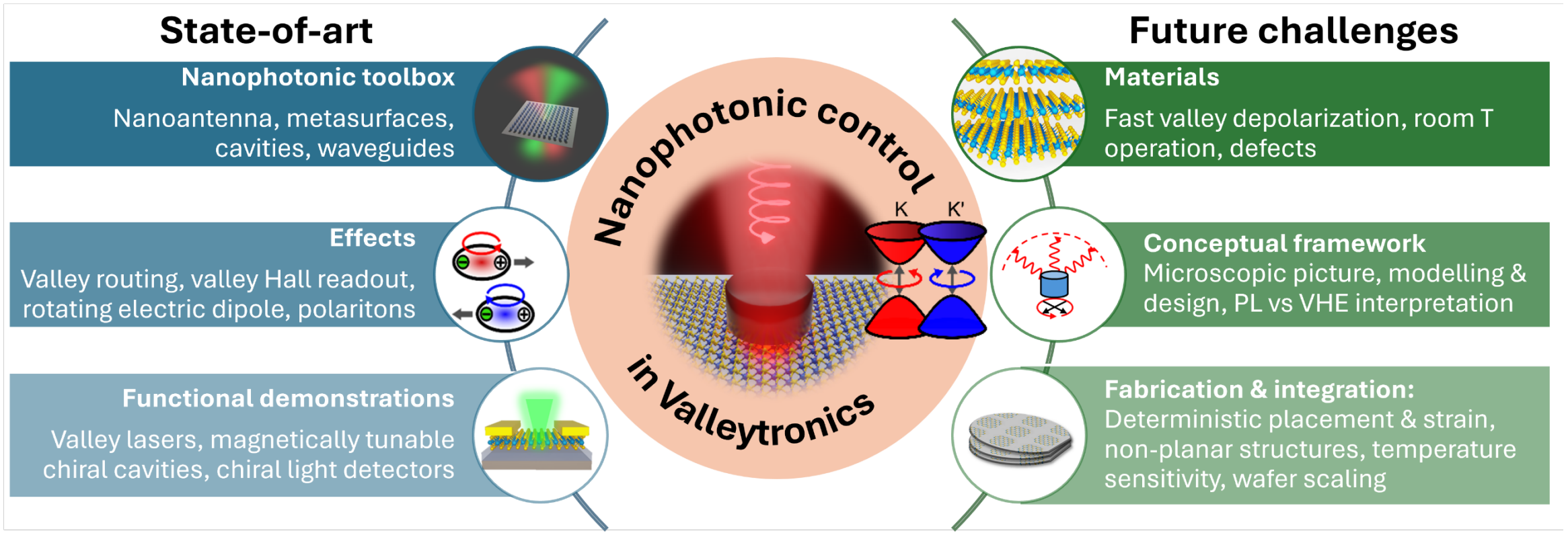

**Figure 1.** Central panel: schematic illustration of nanophotonic control of valley degree of freedom in a 2D material using a resonant nanostructure. Left: current state of the art, highlighting (top) the nanophotonic toolbox, (middle) demonstrated effects, and (bottom) functional device demonstrations. Right: key future challenges, grouped into (top) material developments, (middle) conceptual framework, and (bottom) fabrication and integration.

### Current and future challenges

Despite rapid progress, several fundamental and practical challenges still limit nanophotonic control of the valley degree of freedom in 2D materials. On the materials side, valley polarization and coherence are strongly constrained by electron-hole exchange, phonon and defect scattering, as well as charge noise (see [15] and references therein). As a result, large valley contrasts are typically confined to cryogenic temperatures, narrow excitation windows, or specially engineered heterostructures. Achieving robust, high-contrast valley signals at room temperature while maintaining strong light-matter coupling and low loss, particularly in the presence of metallic contacts and non-crystalline substrates, remains a central material challenge.

A second challenge is conceptual: there is still no unified microscopic picture of valley emitters in complex nanophotonic environments. In the literature, interpretations based on chiral emitters and a chiral Purcell effect [16, 17] coexist with models that treat valley excitons as rotating in-plane dipoles controlled by the degree of circular polarization of the local in-plane field components [6]. Since a rotating electric dipole is not a chiral object [18], these viewpoints are not fully compatible and lead to different symmetry expectations and design rules for valley-selective emission [19]. Disentangling genuine valley-selective dynamics from photonic polarization conversion and near-field artefacts will require consistent theory, as well as precise space- and time-resolved experiments that reconcile far-field helicity measurements with electronic transport measurements such as the valley Hall effect and clarify the microscopic pathways for valley-selective carrier injection at metal/2D interfaces [10].

Third, there are substantial technological and fabrication challenges regarding the scalable integration of 2D materials with nanophotonic structures. Although some progress has been made, deterministic placement of 2D emitters with nanometer precision, while preserving their optical quality, remains difficult, especially on non-planar structures where transfer-induced strain and wrinkles can strongly modify excitonic resonances [20]. Fabrication steps can introduce damage and contamination, and the strong sensitivity of excitons to the local and temperature-dependent dielectric environment complicates the predictable design of cavity and antenna geometries. Robust design rules that tolerate these variations and are transferable between different 2D materials, Van der Waals heterostructures, and alloyed systems are still lacking.

Finally, there is still a gap between proof-of-principle demonstrations and application-ready devices. Most experiments rely on optical pumping at visible wavelengths and often at cryogenic temperatures [12, 13]. Realizing low-threshold, electrically driven, valley-selective sources, modulators, and detectors that operate at room temperature, remains a major goal. Scaling up from single structures to integrated photonic circuits or stacked metasurfaces that route, store, and process valley information in a programmable fashion will demand careful management of loss, crosstalk, and decoherence across many coupled nanophotonic elements.

### Advances in science and technology to meet challenges

Progress on the materials side will be crucial for the further development of valley nanophotonics. Continued improvements in crystal quality and encapsulation for substrate passivation are expected to reduce inhomogeneous broadening and non-radiative decay and to suppress fast valley depolarization and decoherence [21, 22]. Heterostructures that exploit interlayer, dark, or trions with intrinsically longer valley lifetimes, as well as ternary alloys [23], offer additional knobs to tune exchange interactions and phonon coupling. Combining these approaches with magnetic proximity effects and electrostatic band-structure control may ultimately enable robust, high-contrast valley signals at or near room temperature [24].

Addressing the conceptual inconsistency in valley nanophotonics will require effective computational tools that are based on microscopic picture yet usable by non-specialists. Compact models that could bridge ab-initio materials theory and electromagnetic simulations, will allow to ultimately connect quantum approaches (Maxwell–Bloch, master equations) to large-scale device design. On the experimental side, coordinated use of polarization- and time-resolved photoluminescence microscopy and spectroscopy together with valley Hall and related transport measurements will be crucial to disentangling genuine valley dynamics from purely photonic effects.

Technological advances in fabrication and integration are equally important. Deterministic placement techniques, low-damage patterning, and direct growth of 2D materials on prepatterned photonic substrates are all promising routes to scalable, high-yield hybrid devices. Inverse design and topology-optimization methods, combined with temperature-dependent material models, can produce nanostructures that are robust against fabrication tolerances and variations in exciton resonance.

Finally, moving from proof-of-principle experiments to application-ready devices will require closer integration with electronic and photonic circuit technologies [25]. Electrical contacts combined with on-chip waveguides and metasurfaces for routing and demultiplexing could form the building blocks of valley-based interconnects and sensors. Reconfigurable platforms using electro-optic tuning, phase-change materials, or MEMS actuators may enable programmable valley-photonics circuits. Demonstrating such systems at scale is a central longer-term goal for the field.

### Concluding remarks

Optical control of the valley degree of freedom has moved far beyond simple demonstrations of helicity-selective photoluminescence. We now have a diverse toolbox of cavities, metasurfaces, nanoantennas, and waveguides that can generate, detect, and route valley-selective optical signals. Proof-of-principle realizations of valley-responsive lasing, magnetically tunable mirrors, and chiral light sensors evidence broad application potential of valley nanophotonics. At the same time, fast depolarization, limited coherence, and strong sensitivity to the local environment continue to constrain performance, especially at room temperature and under realistic device conditions.

Conceptually, the coexistence of "chiral emitter" and "rotating dipole" pictures, and the frequent use of far-field polarization alone as evidence for valley dynamics, highlight the need for clearer microscopic models and more discriminating measurements. Technologically, scalable and gentle integration of 2D materials with complex nanophotonic structures remains challenging, and most implementations are still at the proof-of-principle stage.

Looking ahead, progress will depend on combining improved materials, robust design methodologies, and valley-sensitive characterization with advances in fabrication and circuit integration. If successful, nanophotonic valley control could evolve from a playground for fundamental physics into a practical platform for information processing, communication, and sensing based on 2D materials.

## Acknowledgements

This work was funded by the Deutsche Forschungsgemeinschaft (DFG, German Research Foundation), Project-ID 437527638 ‐ IRTG 2675 (Meta-Active) and Project-ID 398816777 – SFB 1375 (NOA).

## References

[1] W. Yao, D. Xiao, and Q. Niu, 2008, Valley-dependent optoelectronics from inversion symmetry breaking, *Phy.s Rev. B* **77**, 23.

[2] D. Xiao, G. Bin Liu, W. Feng, X. Xu, and W. Yao, 2012, Coupled spin and valley physics in monolayers of MoS 2 and other group-VI dichalcogenides, *Phy.s Rev. Lett.* **108**, 19.

[3] K. F. Mak, K. He, J. Shan, and T. F. Heinz, 2012, Control of valley polarization in monolayer MoS2 by optical helicity, *Nat. Nanotechnol.* **7**, 8, 494–498.

[4] S. Li, H. Wang, J. Wang, H. Chen, and L. Shao, 2021, Control of light-valley interactions in 2D transition metal dichalcogenides with nanophotonic structures, *Nanoscale* **13**, 13, 6357–6372.

[5] R. Salzwedel, L. Greten, S. Schmidt, S. Hughes, A. Knorr, and M. Selig, 2024, Spatial exciton localization at interfaces of metal nanoparticles and atomically thin semiconductors, *Phys Rev B* **109**, 3.

[6] T. Bucher, Z. Fedorova, M. Abasifard, R. Mupparapu, M. J. Wurdack, E. Najafidehaghani, Z. Gan, H. Knopf, A. George, F. Eilenberger, T. Pertsch, A. Turchanin, and I. Staude, 2024, Influence of resonant plasmonic nanoparticles on optically accessing the valley degree of freedom in 2D semiconductors, *Nat. Commun.* **15**, 1.

[7] T. Bucher, J. Yan, J. Sperrhake, Z. Fedorova, M. Abasifard, R. Mupparapu, H. Chen, E. Najafidehaghani, K. Z. Kamali, A. George, M. Rahmani, T. Pertsch, A. Turchanin, D. N. Neshev, and I. Staude, 2025, Valley-dependent emission patterns enabled by plasmonic nanoantennas. [Online]. Available: http://arxiv.org/abs/2509.21023

[8] L. Yang, Y. Yuan, B. Fu, J. Yang, D. Dai, S. Shi, S. Yan, R. Zhu, X. Han, H. Li, Z. Zuo, C. Wang, Y. Huang, K. Jin, Q. Gong, and X. Xu, 2023, Revealing broken valley symmetry of quantum emitters in WSe2 with chiral nanocavities, *Nat. Commun.* **14**, 1.

[9] L. Zheng, Z. Dang, D. Ding, Z. Liu, Y. Dai, J. Lu, and Z. Fang, 2023, Electron-Induced Chirality-Selective Routing of Valley Photons via Metallic Nanostructure, *Adv.d Mater.* **35**, 34.

[10] L. Li, L. Shao, X. Liu, A. Gao, H. Wang, B. Zheng, G. Hou, K. Shehzad, L. Yu, F. Miao, Y. Shi, Y. Xu, and X. Wang, 2020, Room-temperature valleytronic transistor, *Nat. Nanotechnol.* **15**, 9, 743–749.

[11] C. J. Lee, H. C. Pan, F. HadavandMirzaee, L. S. Lu, F. Cheng, T. H. Her, C. K. Shih, and W. H. Chang, 2025, Exciton-Polariton Valley Hall Effect in Monolayer Semiconductors on Plasmonic Metasurface, *ACS Photon.* **12**, 3, 1351–1358.

[12] D. G. Suárez-Forero, R. Ni, S. Sarkar, M. Jalali Mehrabad, E. Mechtel, V. Simonyan, A. Grankin, K. Watanabe, T. Taniguchi, S. Park, H. Jang, M. Hafezi, and Y. Zhou, 2024, Chiral flat-band optical cavity with atomically thin mirrors.

[13] X. Duan, B. Wang, K. Rong, C. Liu, V. Gorovoy, S. Mukherjee, V. Kleiner, E. Koren, and E. Hasman, Valley-addressable monolayer lasing through spin-controlled Berry phase photonic cavities.

[14] H. Jiang, Y. Zhang, L. An, Q. Tan, X. Dai, Y. Chen, W. Chen, H. Cai, J. Fu, J. Zúñiga-Pérez, Z. Li, J. Teng, Y. Chen, C. W. Qiu, and W. Gao, 2025, Chiral light detection with centrosymmetric-metamaterial-assisted valleytronics, *Nat. Mater.* **24**, 6, 861–867.

[15] J. R. Schaibley, H. Yu, G. Clark, P. Rivera, J. S. Ross, K. L. Seyler, W. Yao, and X. Xu, 2016, Valleytronics in 2D materials, *Nat. Rev. Mater.* **1**, 11.

[16] Z. Chen, F. Shen, Z. Zhang, K. Wu, Y. Jin, M. Long, S. Wang, and J. Xu, 2024, Synergistic Effect of Chiral Metasurface and Hot Carrier Injection Enabling Manipulation of Valley Polarization of WSe 2 at Room Temperature , *Adv. Phys. Res.* **3**, 1.

[17] Z. Wu, J. Li, X. Zhang, J. M. Redwing, and Y. Zheng, 2019, Room-Temperature Active Modulation of Valley Dynamics in a Monolayer Semiconductor through Chiral Purcell Effects, *Adv. Mater.* **31**, 49.

[18] I. Fernandez-Corbaton, M. Fruhnert, and C. Rockstuhl, 2015, Dual and chiral objects for optical activity in general scattering directions, *ACS Photon.* **2**, 3, 376–384.

[19] Y. Lee, H. Kim, and C. Kim, 2023, Rigorous Optical Modeling of Circularly Polarized Light-Emitting Devices: Interaction of Emitters with Device Geometries, *ACS Photon.* **10**, 9, 3283–3290.

[20] K. Parto, S. I. Azzam, N. Lewis, S. D. Patel, S. Umezawa, K. Watanabe, T. Taniguchi, and G. Moody, 2022, Cavity-Enhanced 2D Material Quantum Emitters Deterministically Integrated with Silicon Nitride Microresonators, *Nano. Lett.* **22**, 23, 9748–9756.

[21] G. Gupta, K. Watanabe, T. Taniguchi, and K. Majumdar, 2023, Observation of ~100% valley-coherent excitons in monolayer MoS2 through giant enhancement of valley coherence time, *Light Sci. Appl*. **12**, 1.

[22] S. B. Kalkan, E. Najafidehaghani, Z. Gan, J. Drewniok, M. F. Lichtenegger, U. Hübner, A. S. Urban, A. George, A. Turchanin, and B. Nickel, 2023, High-Performance Monolayer MoS2 Field-Effect Transistors on Cyclic Olefin Copolymer-Passivated SiO2 Gate Dielectric, *Adv. Opt. Mater*. **11**, 2.

[23] S. Liu, A. Granados Del Águila, X. Liu, Y. Zhu, Y. Han, A. Chaturvedi, P. Gong, H. Yu, H. Zhang, W. Yao, and Q. Xiong, 2020, Room-Temperature Valley Polarization in Atomically Thin Semiconductors via Chalcogenide Alloying, *ACS Nano* **14**, 8, 9873–9883.

[24] T. Norden, C. Zhao, P. Zhang, R. Sabirianov, A. Petrou, and H. Zeng, 2019, Giant valley splitting in monolayer WS2 by magnetic proximity effect, *Nat. Commun*. **10**, 1.

[25] C. Li, K. Xing, W. Zhai, L. Sortino, A. Tittl, I. Aharonovich, M. S. Fuhrer, K. Watanabe, T. Taniguchi, Q. Ou, Z. Dong, S. A. Maier, and H. Ren, 2025, Valley optoelectronics based on meta-waveguide photodetectors. https://arxiv.org/abs/2503.19565

# 14. Scaling up valleytronic materials

**Kuan Eng Johnson Goh**[1, 2, 3, 4]

[1] Institute of Materials Research & Engineering (IMRE), Agency for Science, Technology, and Research (A*STAR), 2 Fusionopolis Way, #08-03 Innovis, Singapore 138634, Republic of Singapore
[2] Centre for Quantum Technologies, National University of Singapore, 3 Science Drive 2, Singapore 117543, Singapore
[3] Department of Physics, National University of Singapore, Republic of Singapore 2 Science Drive 3, Singapore 117551, Singapore
[4] Division of Physics and Applied Physics, School of Physical and Mathematical Sciences, Nanyang Technological University 21 Nanyang Link, 637371, Singapore

E-mail: kejgoh@yahoo.com

## Status

Following the landmark isolation of graphene by the "scotch-tape" technique, research activities to isolate and investigate other two-dimensional (2D) materials were fervent, especially for those that predict new properties that might be unique or hard to elicit in conventional bulk materials. An example is the field of valleytronics. This article focuses on the scalability of materials for valleytronics. While material scalability may not impede initial investigation of material physics or chemistry, scalability is a crucial prerequisite for those seeking to develop reliable processes and manufacturable devices.

A brief motivation and status update for valleytronics in 2D semiconductors is given here. More comprehensive reviews [1-3] are available in literature. Since the landmark papers by Xiao *et al.* [4] and Rycerz *et al.* [5] that predicted the unique coupling of spin to valley in non-centrosymmetric 2D crystals, significant experimental efforts have demonstrated the viability of valleytronics. Notably, monolayer semiconducting transition metal dichalcogenides (TMDs) such as $MoS_2$, $WS_2$, $MoSe_2$ and $WSe_2$ intrinsically possess the requisite inversion asymmetry and strong spin-orbit splitting for distinct spin-valley states, and researchers have demonstrated the possibility to address a specific valley state via the spin which is well-known to interact with electromagnetic fields. This coupling avails many techniques to manipulate and read-out the valley state – the opposite parity of the Berry curvature in the K and –K valleys of 2D TMDs results in optical circular dichroism, valley-Hall and valley-Zeeman manifestations, all experimentally verified [6]. The ready access to valleys in 2D TMD crystals using optical or electrical techniques thus provides renewed impetus for valleytronics. We shall therefore focus on the scalability of these 2D TMD crystals (Figure 1).

Currently, many of the novel valleytronic manifestations remain demonstrated on small mechanically exfoliated 2D flakes carefully transferred and stacked by hand to assure minimal contamination [3, 7]. Scalable materials and processes for the production and handling of 2D crystals remain scarce as they typically require significant reinvention of processes. In the roadmap for valleytronics, the author places current developments around a Technology Readiness Level of about 2 to 3 [8].

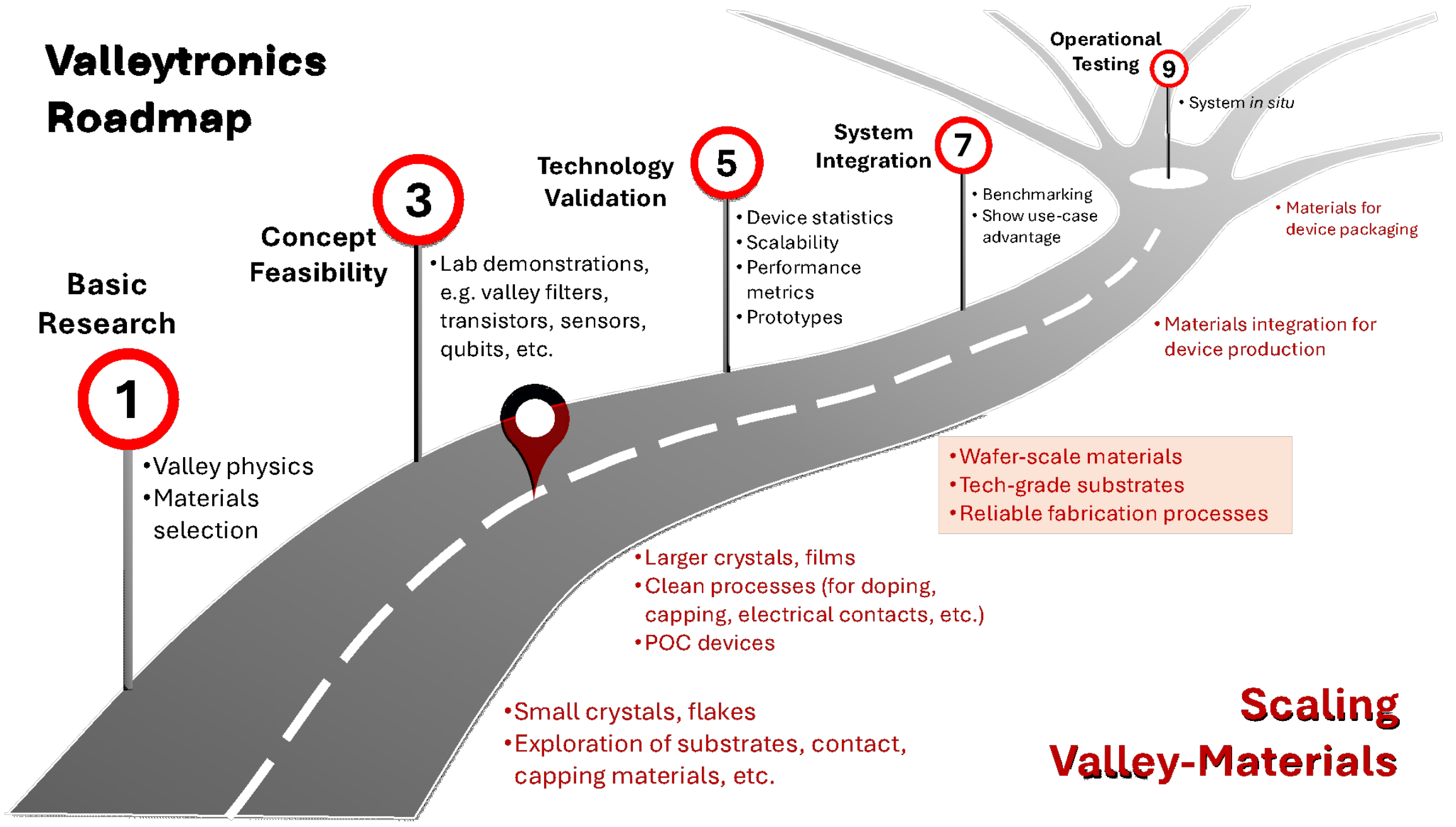

**Figure 1.** Scaling of valley-materials and related materials technologies required to support the developmental roadmap of valleytronics. The orange box highlights the next Technology Readiness Level to target beyond current proof-of-concept (POC) devices. The red map pin indicates current position of developments on the roadmap.

## Current and future challenges

We discuss 3 key current challenges here which are also relevant to 2D materials.

*1. Requirement for high quality large area materials*

Just as the microelectronics industry is dependent on the availability of high-quality silicon ingots from which wafers are sliced and processed into microprocessors, the feasibility of valleytronics would require the production of high-quality homogeneous 2D TMD monolayers. "Scotch-tape" exfoliation typically produces small micron-sized flakes which are not suitable for scale-up due to variability of the flakes and limited area for device fabrication. Thus, researchers have worked to achieve large area high quality 2D semiconductors (TMDs) using chemical vapour deposition [9, 10, 11], physical vapour deposition [12], and molecular beam epitaxy [13]. Initial efforts with wafer-scale coverage typically have multiple grain-orientation. However, recent efforts are targeting single orientation domains via step-engineering of the growth substrate [14, 15], or single 2D crystals via a Czochralski-like method [16]. Most reports highlight characteristics such as carrier mobility, on-off ratio, contact resistance, etc. which are proxy indicators for the material and process quality. Key material parameters like stoichiometry, defect densities, grain orientations and full statistical variations across the wafers may not be fully reported. Depending on the use-cases (e.g. transistors, sensors, memristors, etc.), the metrics may differ. Efforts to qualify and benchmark wafer-scale materials with a common standard remain nascent.

*2. Appropriate technology-grade substrate*

The monolithic fabrication of integrated circuits on silicon motivates a similar route for 2D semiconducting TMD devices. Naively, this starts with a substrate wafer for 2D TMD growth, followed by direct lithography and processing steps e.g. for metal contacts, dielectric insulation, gate contacts, and finally encapsulation to seal and stabilise the device. As the 2D TMD is atomically thin, the choice of a substrate and their interface is critical. This substrate should enable and survive the high-quality TMD growth. It should support the TMD monolayer yet not destroy the inherent physical property advantageous for valleytronics, and be compatible with subsequent processes. A common substrate used is the degenerately-doped silicon wafer with a terminating oxide layer (typically about 90 or 270 nm) to enable global electrical back-gating. However, high CVD temperatures typically degrade the dielectric integrity of silicon oxide [17], as also observed by the author when using such Si substrates for 2D TMD CVD growth at >900 °C. Notably, a recent report saw 2D $MoS_2$ devices fabricated directly on the growth substrate [18]. This group used sapphire wafer for TMD growth by CVD and demonstrated reliable wafer-scale logic circuits for a 32-bit RISC-V microprocessor [18]. This achievement together with the ability to grow single-oriented TMD monolayers on the sapphire substrate [14-16] bodes well for sapphire as both a growth and device substrate for 2D TMDs.

*3. Material integration for production*

Paradoxically, the isolation of monolayers of van der Waals materials (e.g. graphite, TMDs, etc.) heralded the era of 2D materials, but the careful re-integration of such monolayers with other materials is needed for valleytronic devices. Integration of 2D materials via top-down design is a recent endeavour requiring new processes. The community would require (i) effective material placement via transfer or direct growth and fabrication, and (ii) controlled processing environment. Suspending a TMD monolayer in vacuum to preserve its pristine properties is likely impractical. If 2D TMD needs to be transferred to a different substrate, there must be a feasible technique which preserves valley physics as demonstrated by various groups [7, 15]. However, possible contaminations from the transfer processes can impact device quality. Hence continued fabrication directly on the growth substrate is preferred. For electrical devices, good ohmic contacts to the TMD monolayer are needed by tailoring the work function of the metal contacts and the TMD doping [7]. The contact deposition and anneal processes must also be compatible with the other materials in the device. Encapsulation is typically required to stabilise the materials in the device stack, and the cap-layers also often act as dielectrics for metal gates deposited on top. There are thus competing demands placed on the choices of substrate, contact and insulation layers when we consider the integration of 2D materials into devices [7]. This remains a challenge for the 2D material community.

**Advances in science and technology to meet challenges**

While there are encouraging progress in wafer-scale oriented growth of 2D TMD crystals [14, 15] and direct device fabrication with high yield [18] on the sapphire substrate, these processes now need to be optimised into industrial production for reliable starting substrates. The need to encapsulate a valleytronic device is dependent on the use-case – may not be needed for a sensor where the TMD needs to interact with molecules from the environment, but essential for a reliable optical filter or transistor. Encapsulation strategies are non-trivial and require significant developmental cycles to ensure reliability over the device life-time. To date, the most common encapsulation for 2D TMDs is to sandwich them between layers of large bandgap hBN (hexagonal boron nitride) or bury them under high-K dielectrics by atomic layer deposition (ALD) [7]. The latter may cause ALD induced damage in the 2D TMD resulting in degraded devices. Alternatives such as transfer printing of ultra-thin oxides [19] might alleviate such issue, but more work is needed to assess the compatibility and scalability of such newer methods. Controlling the environments within which fabricators handle the 2D TMDs is also important for reducing contaminations (unintended doping, adsorbed molecules, or residues from lithographic resists) [7, 20]. Over-engineering the environment can be prohibitively costly, while achieving a sufficiently controlled processing environment would require deeper understanding of how key environmental factors impact the device performance [15,20]. We are in the infant stages of acquiring such acumen to determine the best compromise.

### Concluding remarks

It is exciting that the main theoretical predictions of valley physics in 2D TMDs have been demonstrated experimentally [1-3, 6, 7]. Practical valleytronics now requires materials engineering to tackle three key challenges: (i) high-quality large area materials, (ii) technology-grade substrate, and (iii) material integration. This chapter summarises the state-of-the-art and lays the next milestones in the roadmap required for valleytronics to come to fruition. Recent developments in single crystal growth by substrate engineering to facilitate single orientation growth [14, 15] or the Czochralski method [16], and direct device fabrication on the growth substrate [18], all represent significant inroads. The promise of valleytronics appears to draw nearer, but the milestones for scaling valleytronic materials are yet to be met and there remain opportunities for continued research and breakthroughs.

### Acknowledgements

K. E. J. G. acknowledges the funding from the National Research Foundation, Singapore through the National Quantum Office, hosted in A*STAR, under its Centre for Quantum Technologies Funding Initiative (S24Q2d0009).

### References

[1] Xu X, Yao W, Xiao D and Heinz T F 2014 Spin and pseudospins in layered transition metal dichalcogenides *Nat. Phys.* **10** 343
[2] Schaibley J R, Yu H, Clark G, Rivera P, Ross J S, Seyler K L, Yao W and Xu X 2016 Valleytronics in 2D materials *Nat. Rev. Mater.* **1** 16055
[3] Bussolotti F, Kawai H, Ooi Z E, Chellappan V, Thian D, Pang A L C and Goh K E J 2018 Roadmap on finding chiral valleys: screening 2D materials for valleytronics *Nano Futures* **2** 032001
[4] Xiao D, Yao W and Niu Q 2007 Valley-contrasting physics in graphene: magnetic moment and topological transport *Phys. Rev. Lett.* **99** 236809
[5] Rycerz A, Tworzydło J and Beenakker C W J 2007 Valley filter and valley valve in graphene *Nat. Phys.* **3** 172
[6] Mak K F and Shan J 2016 Photonics and optoelectronics of 2D semiconductor transition metal dichalcogenides *Nat. Photonics* **10** 216
[7] Goh K E J, Wong C P Y and Wang T (eds) 2023 *Valleytronics in 2D Materials* (Singapore: World Scientific Publishing)
[8] Kimmel W M, Beauchamp P M, Frerking M A, Kline T R, Vassigh K K, Willard D E, Johnson M A and Trenkle T G 2020 Technology Readiness Assessment Best Practices Guide NASA/SP-20205003605 (Washington, DC: NASA Office of the Chief Technologist)
[9] Lim Y F, Priyadarshi K, Bussolotti F, Gogoi P K, Cui X, Yang M, Pan J, Tong S W, Wang S, Pennycook S J, Goh K E J, Wee A T S, Wong S L and Chi D 2018 Interface engineering of monolayer $MoS_2$–$SiO_2$ junctions for high performance field effect transistors *ACS Nano* **12** 1339
[10] Wang Q, Shi R, Zhao Y, Huang R, Wang Z, Amini A and Cheng C 2021 Recent progress on kinetic control of chemical vapor deposition growth of high-quality wafer-scale transition metal dichalcogenides *Nanoscale Adv.* **3** 3430
[11] Tee J Y, John M, Fu W, Maddumapatabandi T D, Bussolotti F, Wong C P Y and Goh K E J 2023 Valley-polarized photocurrent in monolayer $MoS_2$ via chiral light excitation *Adv. Phys. Res.* **2** 202300146

[12] Yang W, Kawai H, Bosman M, Tang B, Chai J, Tay W L, Yang J, Seng H L, Zhu H, Gong H, Liu H, Goh K E J, Wang S and Chi D 2018 Interlayer interactions in 2D $WS_2/MoS_2$ heterostructures monolithically grown by in situ physical vapor deposition *Nanoscale* **10** 22927

[13] Pacuski W, Grzeszczyk M, Nogajewski K, Bogucki A, Oreszczuk K, Kucharek J, Połczyńska K E, Seredyński B, Rodek A, Bożek R, Taniguchi T, Watanabe K, Kret S, Sadowski J, Kazimierczuk T, Potemski M and Kossacki P 2020 Narrow excitonic lines and large-scale homogeneity of transition-metal dichalcogenide monolayers grown by molecular beam epitaxy on hexagonal boron nitride *Nano Lett.* **20** 3058

[14] Fu J H, Min J, Chang C K, Tseng C C, Wang Q, Sugisaki H, Li C, Chang Y M, Alnami I, Syong W R, Lin C, Fang F, Zhao L, Lo T H, Lai C S, Chiu W S, Jian Z S, Chang W H, Lu Y J, Shih K, Li L J, Wan Y, Shi Y and Tung V 2023 Oriented lateral growth of two-dimensional materials on c-plane sapphire *Nat. Nanotechnol.* **18** 1289

[15] Fu W, Chai J, Kawai H, Maddumapatabandi T D, Bussolotti F, Huang D, Lee R, Teo S L, Tan H R, Wong C P Y, Sng A, Chen Y, Lau C S, Zhang M, Medina H, Lin M, Bosman M and Goh K E J 2025 Evidence of air-induced surface transformation of atomic step-engineered sapphire in relation to epitaxial growth of 2D semiconductors *Nat. Commun.* **16** 8488

[16] Jiang H, Zhang X, Chen K, He X, Liu Y, Yu H, Gao L, Hong M, Wang Y, Zhang Z and Zhang Y 2025 Two-dimensional Czochralski growth of single-crystal $MoS_2$ *Nat. Mater.* **24** 188

[17] Knyazev A V, Guseva M V, Kolesnikov N N, Kolesnikov A V, Gusev A A and Guseva N V 2023 Electrical properties of silicon oxide layers subjected to high-temperature treatment reproducing the growth conditions for thin carbon films *J. Electron. Mater.* **52** 10498

[18] Zhou P, Ao Z, Zhang Y, Kong X, Gou S, Chen S, Dong X, Zhu Y, Sun Q, Zhang Z, Zhang J, Zhang Q, Hu Y, Sheng C, Wang K, Wang S, Wan J, Han J, Bao W and Peng Z 2025 A RISC-V 32-bit microprocessor based on two-dimensional semiconductors *Nature* **619** 105

[19] Dasari V, Mishra A, Tarn Y, Lee R, Verzhbitskiy I A, Huang D, Bussolotti F, Maddumapatabandi T D, Teh Y W, Ang Y S, Goh K E J and Lau C S 2024 Liquid metal oxide-assisted integration of high-k dielectrics and metal contacts for two-dimensional electronics *ACS Nano* **18** 26911

[20] Mukherjee S, Wang S, Dasari V, Tarn Y, Talha-Dean T, Lee R, Verzhbitskiy I A, Huang D, Mishra A, John J W, Das S, Bussolotti F, Maddumapatabandi T D, Teh Y W, Ang Y S, Goh K E J and Lau C S 2025 Toward phonon-limited transport in two-dimensional transition metal dichalcogenides by oxygen-free fabrication *ACS Nano* **19** 9327

# 15. Novel valleytronic platforms beyond TMDs and graphene

**Zhichao Zhou[1] and Xiao Li[1]**

[1] Center for Quantum Transport and Thermal Energy Science, Institute of Physics Frontiers and Interdisciplinary Sciences, School of Physics and Technology, Nanjing Normal University, Nanjing 210023, China

E-mail: lixiao@njnu.edu.cn

## Status

Beyond TMDs and graphene, recent studies have also proposed a variety of new two-dimensional (2D) valleytronic materials, which serve as novel platforms for investigating valley physics and associated applications. These new materials include conventional valley-degenerate materials [1-3], magnetic valley materials [4-9], topological valley materials [10-13], and so on. In the following, we will introduce them respectively.

Firstly, degenerate but inequivalent valleys have been discovered in many 2D nonmagnetic materials. In these materials, the valley degeneracy is protected by time-reversal symmetry, and distinct valleys exhibit contrasting optoelectronic and Berry-curvature-driven transport properties (see Figure 1a and 1b), similar to TMDs and graphene. Typical instances include Si-N passivated transition-metal nitrides, $MoSi_2N_4$, and other monolayer materials with the same structure [1, 2], as well as the all-inorganic lead-free perovskite derivative, $Cs_3Bi_2I_9$ [3].

Taking one step further from nonmagnetic valley materials, researchers have recently focused on 2D magnetic valley materials, where exchange interactions can break the time-reversal symmetry and associated valley degeneracy (see Figure 1c). The valley degeneracy splitting results in exotic physical properties, such as static valley polarization and anomalous valley Hall effect. By coupling valleys with Néel antiferromagnetic order, an emergent spin-valley coupled degree of freedom has been discovered in manganese chalcogenophosphates monolayers [4]. Beyond the magnetic proximity effect [5], hundreds of intrinsic ferromagnetic valley materials with spontaneous valley splitting have been proposed, such as monolayers of $VSe_2$ [6], $VSi_2N_4$ [7], and $Cr_2Se_3$ [8], which are referred to as ferrovalley materials [6]. Further considering noncollinear magnetic order, the scalar-spin chirality also brings about multiple valley modulations [9].

The valley degree of freedom is also closely related to band topology, and correspondingly a number of topological valley materials have been designed. In such materials, band gaps of different valleys can be selectively regulated, thereby giving rise to rich topological physics. For example, the valley-polarized quantum anomalous Hall effect has been realized in silicene, where both Chern number and valley Chern number are nonzero with valley-dependent topological edge states (see Figure 1d) [10]; through the couplings of valley-related surface states, various topological phases, including strong and weak topological insulators, and crystalline topological insulator, can be created in SnTe superlattice [11, 12]; In $ScI_2$ bilayer, hybrid-order topological phase transitions also occur by the valley engineering [13].

In addition to the aforementioned materials, researchers have also designed multifunctional valley materials, like valley-related ferroelectric [13] and piezoelectric materials [14]. Besides the spin degree of freedom, valleys are also coupled to such other degrees of freedom as layer number, stacking configuration, and twist angle. Furthermore, from the perspective of crystal lattice, 2D valley materials have evolved from the initial honeycomb lattice to a variety of lattice types, including square lattices [15] and Kagome lattices [9]. The aforementioned diverse valley materials provide powerful platforms for exploring valley physics and advancing its practical applications, thereby boosting the development of valleytronics.

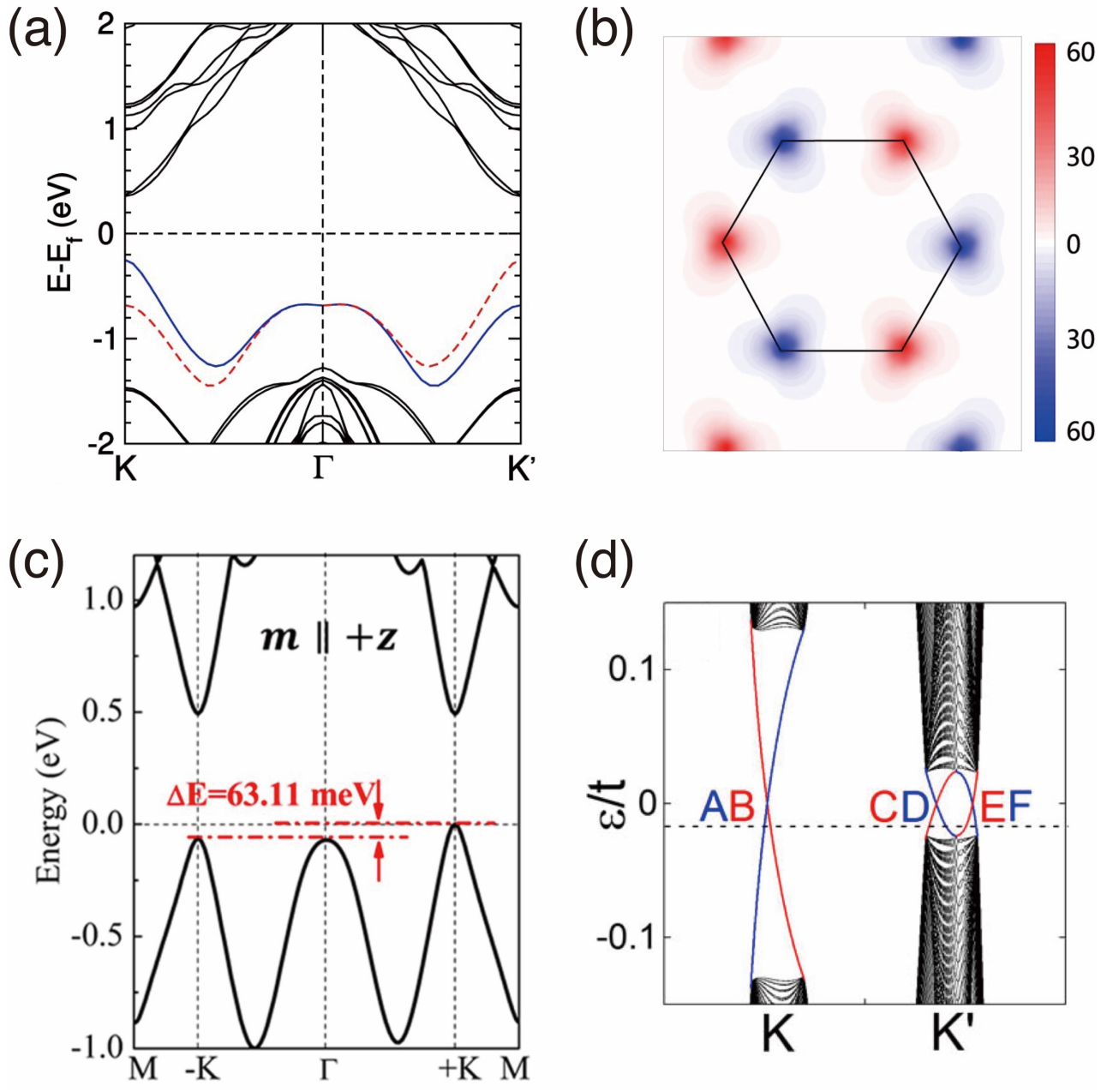

**Figure 1.** Novel valleytronic platforms. (a) The band structure of nonmagnetic $WSi_2P_4$ monolayer, where degenerate but inequivalent valleys appear at the corners of the first Brillouin zone. Adopted from [2]. CC BY 4.0. (b) The valley-contrasting Berry curvature in the monolayer, which leads to valley Hall effect. Adopted from [2]. CC BY 4.0. (c) The band structure of ferromagnetic $VSi_2N_4$ monolayer, where the valley degeneracy splittings occur. Adopted from [7]. (d) The band structure of zigzag-terminated silicene exhibiting the valley-polarized quantum anomalous Hall effect with valley-contrasting topological edge states. Adopted from [10].

## Current and future challenges

As mentioned above, an increasing number of valley materials have been discovered, exhibiting rich valley-contrasting physics. However, the aforementioned research on emergent valley materials is mainly performed via theoretical calculations, with relatively limited experimental progress. For well-known valley materials like $MoS_2$ monolayer, breakthroughs have been made across the board, spanning from experimental measurements to the development of device prototypes [16]. For example, room-temperature valley transistors have been fabricated with an on-off ratio of $10^2$–$10^3$, realizing a full sequence of the generation, propagation, and manipulation of valley information [17]. Nevertheless, most of the newly proposed 2D valley materials remain in the stage of theoretical prediction. The experimental growth of these valley materials and detection of their valley properties are urgently needed. There is thus no experimental basis for further verifying control precision and stability of the valley state that are required for the device fabrication of these valley materials.

Moreover, although a multitude of emergent valley materials have been put forward, the design of such materials lacks novel mechanisms, and the selection range of materials is limited. For example, the valley Hall effect is one of the experimental hallmarks of valley materials, and its realization conventionally relies on the breaking of the spatial inversion symmetry, which excludes a large number of centrosymmetric materials from the design of valley Hall semiconductors. While the interplays between conventional magnetic orders and valley degree of freedom have been extensively studied in magnetic valley materials, novel valley-spin coupling modes are relatively scarce.

In addition, there are also several constraints on realizing the application of valley materials. For instance, ferrovalley materials require a relatively high Curie temperature and a considerable valley splitting; the maintenance of valley polarization necessitates the effective suppression of intervalley scatterings from phonons, impurities, etc.; valley Hall transport requires high carrier mobility in valley materials. They are common scientific and technological challenges faced by both theoretical and experimental researchers.

### Advances in science and technology to meet challenges

The progression of emergent valley materials, from their theoretical proposal, experimental verification to device application, requires close and active collaborations between theoretical and experimental researchers. For these emergent valley materials, in addition to focusing on their physical properties such as electronic band structure and optoelectronic responses, the factors affecting the experiment, including the feasibility of material synthesis and the effects of intervalley scatterings on valley polarization, also deserves the attention of theoretical researchers. By high-performance computations with considering such experimental factors and high-throughput materials screening, theoretical studies are expected to construct the database of valley materials and provide targeted guidance for experimental exploration on emergent valley materials. On the other hand, with the help of current advanced material growth techniques, e.g., chemical vapor deposition and molecular beam epitaxy, large-size, low-defect valley materials are anticipated to be fabricated in experiment, thereby minimizing intervalley scattering. Moreover, large-scale valley materials can be utilized for realizing the mass fabrication and integration of valleytronic devices, by micro-nano processing technologies.

In recent years, there have been numerous promising advancements in the field of condensed matter physics, such as nonlinear transports [18] and unconventional magnetic orders [19], which offer abundant opportunities for designing novel valley materials with intriguing valley properties. Unlike the conventional linear valley Hall effect, the nonlinear valley Hall effect and other valley-related nonlinear transport have distinct symmetry requirements. They are likely to be not restricted by the breaking of the spatial inversion symmetry and occur in centrosymmetric materials. Unconventional magnetic orders not only provide a new landscape for the study of magnetism, but also are apt to give birth to more novel mechanisms of the valley-spin coupling and material realizations [20]. Moreover, extending from electronic systems to quasiparticle systems, valley-contrasting physical properties can also be exhibited in phonons and magnons, and the interactions between electrons and phonons or magnons in novel valley materials are worthy of further exploration. These valley-related new research directions are in its infancy, and they are expected to significantly expand the selection range of valley materials and enrich mechanisms of the expression and manipulation of valley degree of freedom, thereby further promoting the development of valleytronics.

### Concluding remarks

Beyond TMDs and graphene, a large number of new 2D valley materials have been designed, including magnetic valley materials, topological valley materials, and so on, broadening the pool of valley materials. These emergent valley materials have compelling physical properties such as static valley polarization, anomalous valley Hall effect, and valley-polarized quantum anomalous Hall effect, and manifest couplings between multiple electronic degrees of freedom, including valley, spin, layer, and charge. These intriguing characteristics make 2D valley materials promising for the design of new valleytronic devices. Furthermore, the combination of theory and experiment is expected to advance valleytronics through the design of new materials, experimental verification, and the fabrication and application of practical devices. Additionally, new developments in condensed matter physics also bring enormous opportunities for the advancement of 2D valley materials.

### Acknowledgements

We are supported by the National Natural Science Foundation of China (Grants No. 12374044, No. 12004186, and No. 11904173).

### References

[1] Hong Y-L, Liu Z, Wang L, Zhou T, Ma W, Xu C, Feng S, Chen L, Chen M-L, Sun D-M, Chen X-Q, Cheng H-M and Ren W 2020 Chemical vapor deposition of layered two-dimensional $MoSi_2N_4$ materials *Science* **369** 670

[2] Wang L, Shi Y, Liu M, Zhang A, Hong Y-L, Li R, Gao Q, Chen M, Ren W, Cheng H-M, Li Y and Chen X-Q 2021 Intercalated architecture of $MA_2Z_4$ family layered van der Waals materials with emerging topological, magnetic and superconducting properties *Nat. Commun.* **12** 2361

[3] Liang J, Fang Q, Wang H, Xu R, Jia S, Guan Y, Ai Q, Gao G, Guo H, Shen K, Wen X, Terlier T, Wiederrecht G P, Qian X, Zhu H and Lou J 2020 Perovskite-derivative valleytronics *Adv. Mater.* **32** 200411

[4] Li X, Cao T, Niu Q, Shi J and Feng J 2013 Coupling the valley degree of freedom to antiferromagnetic order *Proc. Natl. Acad. Sci.* **110** 3738

[5] Qi J, Li X, Niu Q and Feng J Giant and tunable valley degeneracy splitting in $MoTe_2$ 2015 *Phys. Rev. B* **92** 121403

[6] Tong W-Y, Gong S-J, Wan X and Duan C-G 2016 Concepts of ferrovalley material and anomalous valley Hall effect *Nat. Commun.* **7** 13612

[7] Cui Q, Zhu Y, Liang J, Cui P and Yang H 2021 Spin-valley coupling in a two-dimensional $VSi_2N_4$ monolayer *Phys. Rev. B* **103** 085421

[8] Chuang C-W, Kawakami T, Sugawara K, Nakayama K, Souma S, Kitamura M, Amemiya K, Horiba K, Kumigashira H, Kremer G, Fagot-Revurat Y, Malterre D, Bigi C, Bertran F, Chang F H, Lin H J, Chen C T, Takahashi T, Chainani A. and Sato T 2025 Spin-valley coupling enhanced high-$T_c$ ferromagnetism in a non-van der Waals monolayer $Cr_2Se_3$ on graphene *Nat. Commun.* **16** 3448

[9] Zhou Z, Wang H and Li X 2024 Multiple valley modulations in noncollinear antiferromagnets *Nano Lett.* **24** 11497

[10] Pan H, Li Z, Liu C-C, Zhu G, Qiao Z and Yao Y 2014 Valley-polarized quantum anomalous Hall effect in silicene *Phys. Rev. Lett.* **112** 106802

[11] Li X, Zhang F, Niu Q and Feng J. 2014 Superlattice valley engineering for designer topological insualtors *Sci. Rep.* **4** 6397

[12] Li X and Niu Q. 2017 Topological phase transitions in thin films by tuning multivalley boundary-state coupling *Phys. Rev. B* **95** 241411

[13] Yang N-J, Zhang J-M, Li X-P, Zhang Z, Yu Z-M, Huang Z and Yao Y 2025 Sliding ferroelectrics induced hybrid-order topological phase transitions *Phys. Rev. Lett.* **134** 256602

[14] Zhang C, Nie Y, Sanvito S and Du A 2019 First-principles prediction of a room-temperature ferromagnetic Janus VSSe monolayer with piezoelectricity, ferroelasticity, and large valley polarization *Nano Lett.* **19** 1366

[15] Wan W, Yao Y, Sun L, Liu C-C and Zhang F 2017 Topological, valleytronic, and optical properties of monolayer PbS *Adv. Mater.* **29** 1604788

[16] Mak K F, McGill K L, Park J and McEuen P L 2014 The valley Hall effect in $MoS_2$ transistors *Science* **344** 1489

[17] Li L, Shao L, Liu X, Gao A, Wang H, Zhang B, Hou G, Shehzad K, Yu L, Miao F, Shi Y, Xu Y and Wang X 2020 Room-temperature valleytronic transistor *Nat. Nanotech.* **15** 743

[18] Du Z Z, Lu H-Z and Xie X C 2021 Nonlinear Hall effects *Nat. Rev. Phys.* **3** 744

[19] Liu Q, Dai X and Blügel S 2025 Different facets of unconventional magnetism *Nat. Phys.* **21** 329

[20] Ma H-Y, Hu M, Li N, Liu J, Yao W, Jia J-F and Liu J 2021 Multifunctional antiferromagnetic materials with giant piezomagnetism and noncollinear spin current *Nat. Commun.* **12** 2846